\documentclass[twocolumn]{aastex6}

\pdfoutput=1

\usepackage{amsmath,amssymb,amstext}
\usepackage{aas_macros} %
\usepackage{hyperref}  
\usepackage{graphicx} 
\usepackage{bm}

\graphicspath{ {figures/} }

\shorttitle{GMC Collisions as Triggers of Star Formation. II.}
\shortauthors{Wu et al.}

\begin{document}


\title{GMC Collisions as Triggers of Star Formation. II. \\
3D Turbulent, Magnetized Simulations}

\author{Benjamin Wu}
\affiliation{Department of Physics, University of Florida, Gainesville, FL 32611, USA}
\affiliation{National Astronomical Observatory, Mitaka, Tokyo 181-8588, Japan}
\author{Jonathan C. Tan}
\affiliation{Department of Physics, University of Florida, Gainesville, FL 32611, USA}
\affiliation{Department of Astronomy, University of Florida, Gainesville, FL 32611, USA}
\author{Fumitaka Nakamura}
\affiliation{National Astronomical Observatory, Mitaka, Tokyo 181-8588, Japan}
\author{Sven Van Loo}
\affiliation{School of Physics and Astronomy, University of Leeds, Leeds LS2 9JT, UK}
\author{Duncan Christie}
\affiliation{Department of Astronomy, University of Florida, Gainesville, FL 32611, USA}
\and
\author{David Collins}
\affiliation{Department of Physics, Florida State University, Tallahassee, FL 32306-4350, USA}

\email{ben.wu@nao.ac.jp}


\begin{abstract}
We investigate giant molecular cloud (GMCs) collisions and their
ability to induce gravitational instability and thus star
formation. This mechanism may be a major driver of star formation
activity in galactic disks.
We carry out a series of three dimensional, magnetohydrodynamics
(MHD), adaptive mesh refinement (AMR) simulations to study how cloud
collisions trigger formation of dense filaments and clumps. Heating
and cooling functions are implemented based on photo-dissociation
region (PDR) models that span the atomic to molecular transition and
can return detailed diagnostic information. The clouds are initialized
with supersonic turbulence and a range of magnetic field strengths and
orientations.  Collisions at various velocities and impact parameters
are investigated. Comparing and contrasting colliding and
non-colliding cases, we characterize morphologies of dense gas,
magnetic field structure, cloud kinematic signatures, and cloud
dynamics. We present key observational diagnostics of cloud
collisions, especially: relative orientations between magnetic fields
and density structures, like filaments; $^{13}$CO($J$=2-1),
$^{13}$CO($J$=3-2), and $^{12}$CO($J$=8-7) integrated intensity maps
and spectra; and cloud virial parameters. We compare these results to
observed Galactic clouds.
\end{abstract}





\section{Introduction}
\label{sec:intro}

Collisions between giant molecular clouds (GMCs) within the
interstellar medium have been proposed as a mechansim for triggering
star formation \citep{Loren_1976,Scoville_Sanders_Clemens_1986,Tan_2000},
potentially even setting global star formation rates
(SFRs) of disk galaxies. It is an attractive mechanism because it is a
process that is expected to create $\sim$parsec-scale dense gas clumps
that are prone to gravitational instability and are the precursors to
star clusters, while at the same time being sensitive to global
galactic dynamics, such as the shear rate 
\citep{Tan_2000,Tasker_Tan_2009,Tan_2010,Suwannajak_ea_2014}
and the presence of spiral arms \citep{Dobbs_2008}. 
Such a connection to orbital shear naturally explains the dynamical 
Kennicutt-Schmidt relation \citep{Kennicutt_1998,Leroy_ea_2008}, 
$\Sigma_{\rm SFR} \propto \Sigma_{\rm gas}\Omega$
where $\Sigma_{\rm SFR}$ and $\Sigma_{\rm gas}$ are surface densities
of star formation rate and total gas and $\Omega$ is the orbital
angular frequency. Global galactic simulations have shown that in a
flat rotation curve disk, GMC collision timescales are relatively
frequent, at $t_{\rm coll}\simeq 0.2 t_{\rm orbit}$
\citep{Tasker_Tan_2009,Dobbs_ea_2015}.

Most star formation is observed to occur within GMCs, which are
generally defined to have masses $\geq10^4\:M_\odot$, with mean mass
surface densities $\Sigma \sim 100\:M_{\odot}\:{\rm pc^{-2}}$, and
mean volume densities $n_{\rm H} \simeq 100\:{\rm cm^{-3}}$, but with
large variation and substructure in the form of filaments/clumps/cores
\citep[e.g.,][]{McKee_Ostriker_2007,Tan_Shaske_Van_Loo_2013}.  Average
radii of GMCs range from $\sim6-100$~pc, although they are typically
not well described by a simple spherical geometry. Rather, filamentary
and/or complex irregular morphologies are often observed 
\citep[e.g.,][]{Jackson_ea_2010,Roman-Duval_ea_2010,Ragan_ea_2014,Hernandez_Tan_2015}.

At typical molecular cloud temperatures of $\sim10-20$~K, thermal
pressure support is insufficient for preventing gravitational collapse
of GMCs and their protocluster gas clumps. Magnetic fields
\citep[e.g.,][]{Mouschovias_2001,Crutcher_2012,Li_ea_2014} and
turbulence
\citep[e.g.,][]{Krumholz_McKee_2005,Padoan_Nordlund_2011,Federrath_Klessen_2012,Padoan_ea_2014}
are both likely to be more important in influencing the gravitational
stability of molecular gas and thus the regulation of star formation.

Magnetic field strengths have been measured in the ISM via the Zeeman
effect, revealing a magnitude that is density-dependent.  In the
diffuse ISM, the magnetic field has been measured to be
$6\pm2~{\rm \mu G}$ locally and $10\pm3~{\rm \mu G}$ at 3~kpc Galactocentric distance \citep{Beck_2001}.
Within molecular clouds, clumps and cores with $n_{\rm H}>300\:{\rm
  cm^{-3}}$ the distribution of magnetic field strengths has been
inferred to be bounded by a relation that scales as $B_{\rm max} =
B_{0}(n_{\rm H}/300\:{\rm cm^{-3}})^{2/3}$, where $B_{0} = 10~{\rm \mu
  G}$ \citep{Crutcher_ea_2010}, while at lower densities, $B_{\rm max}
= B_{0} = 10~{\rm \mu G}$. We refer to this as the ``Crutcher
relation.''

Kinematically, GMCs have internal velocity dispersions similar to the
virial velocity, which is at least an order of magnitude larger than
the sound speed ($c_{s} \sim 0.2$~km/s for $\sim 10$~K gas) 
\citep[e.g.,][]{Solomon_ea_1987,Ossenkopf_MacLow_2002,Heyer_Brunt_2004,Roman-Duval_ea_2010,Hernandez_Tan_2015}. 
Thus GMCs are expected to be permeated by supersonic
turbulence.

Random bulk velocities of GMCs have been observed in the Galaxy to be
$\sim 5-7~{\rm km\:s^{-1}}$ \citep[e.g.,][]{Liszt_ea_1984,Stark_1984}.
However, actual interaction velocities are expected to be set by the
shear velocity at 1-2 cloud tidal radii, which may be several times
faster \citep{Gammie_ea_1991,Tan_2000}.

On scales of GMCs and clumps conversion of gas into stars has been
proposed to be a slow and inefficient process relative to local
dynamical timescales
\citep{Zuckerman_Evans_1974,Krumholz_Tan_2007,DaRio_ea_2014}. Faster
conversion rates have been proposed for some of the most active
star-forming regions in the Galaxy \citep{Murray_ea_2010,Lee_ea_2015}.
Star formation is seen to be highly localized in
space and time, with relatively higher overall efficiencies eventually
achieved within these clusters
\citep[e.g.,][]{Lada_Lada_2003,Gutermuth_ea_2009,Federrath_Klessen_2013}.

There are a number of observational candidates for triggering of star
formation by cloud collisions. The most common criteria for
identifying candidates is the presence of two distinct velocity
components of molecular gas (traced by CO rotational line spectra),
surrounding populations of dense cores or young stars.
Potential examples include NGC133 \citep{Loren_1976}, ${\rm LkH\alpha
  198}$ \citep{Loren_1977}, W75-DR 21 \citep{Dickel_ea_1978}, GR110-13
\citep{Odenwald_ea_1992}, Westerlund 2
\citep{Furukawa_ea_2009,Ohama_ea_2010}, M20 \citep{Torii_ea_2011},
Cygnus OB 7 \citep{Dobashi_ea_2014}, and N159 West
\citep{Fukui_ea_2015}. However, problems remain in verification of
collisions. It can be difficult to rule out chance alignments of
multiple, independent velocity components that are seen in
projection. It is also very challenging to discern the overall 3D
distribution of cloud structures from position-position-velocity data.

The basic question we seek to answer is whether realistic models for
GMC-GMC collisions, i.e., converging flows of molecular gas that are
already prone to gravitational instability, result in dense gas
structures and star formation activity that can explain typical
observed star-forming regions. In our first paper in this series,
\citet[][hereafter Paper I]{Wu_ea_2015}, we presented idealized 2D
simulations of GMC collisions and their effect on a pre-existing
dense, magnetized clump. Paper I introduced many of the methods that
will be adopted here, including photo-dissociation region (PDR)-based
heating and cooling functions. These allow prediction of molecular
line diagnostics of cloud collisions: e.g., collisions lead to high
ratios of $^{12}$CO($J$=8-7) to lower $J$ line intensities.
Here in Paper II, we will extend these models to 3D, turbulent GMCs
and focus on the properties of dense gas created in GMC-GMC
collisions. Paper III will explicitly model the star formation that
may result from such collisions.

Our work is part of a growing body of numerical studies that have
investigated cloud-cloud collisions. Early simulations typically
initialized two spherical clouds and studied the physical effects of
collisions. It was shown that collisions produce bow shocks which lead
to compression of gas and gravitational instability
\citep{Habe_Ohta_1992}, bending mode instabilities and highly
inhomogeneous high-density regions \citep{Klein_Woods_1998},
thin-shell and Kelvin-Helmholtz instabilities due to shear
\citep{Anathpindika_2009}, and filament formation from a shock-
compressed layer \citep{Balfour_ea_2015}. Recent simulations of
turbulent, unmagnetized clouds showed core formation at the collision
interface with properties favorable to massive star creation
\citep{Takahira_ea_2014}, and with observational signatures
potentially found in position-velocity diagrams
\citep{Haworth_ea_2015a,Haworth_ea_2015b}.

Our work is distinguished from the above studies by modeling
magnetized, turbulent clouds, with realistic heating and cooling
functions. These enable us to focus on a number of diagnostic
signatures of cloud collisions that can be compared against observed
clouds.

Section \ref{sec:methods} describes our numerical methods and initial
setup. Section \ref{sec:results} discusses our results, which
include morphologies (\S\ref{sec:results-morphology}), magnetic
fields, (\S\ref{sec:results-Bfield}), probability distribution
functions (\S\ref{sec:results-pdf}), integrated intensity maps
(\S\ref{sec:results-intensity}), kinematics
(\S\ref{sec:results-kinematics}), and dynamics
(\S\ref{sec:results-dynamics}). We present our conclusions in Section
\ref{sec:conclusion}.

\section{Numerical Model}
\label{sec:methods}

\subsection{Initial Conditions}

We choose initial conditions to match properties of observed GMCs. We
include physical processes likely to be dominant in the formation and
evolution of structure within GMCs: self-gravity, supersonic
turbulence, and magnetic fields. We then focus on the effects of
colliding two clouds that are converging at a given velocity and with
a given initial impact parameter.

Our simulation volume is a $128\:{\rm pc}$-sided cube containing two
identical, initially spherical GMCs with uniform densities of H nuclei
of $n_{\rm H,GMC} = 100\:{\rm cm^{-3}}$ and radii $R_{\rm GMC} =
20.0$~pc, giving each GMC a mass $M_{\rm GMC} = 9.3 \times
10^4\:M_\odot$.  The clouds are embedded in ambient gas, representing
the atomic cold neutral medium (CNM). This material is set to have
$n_{\rm H,0} = 10\:{\rm cm^{-3}}$.

We introduce supersonic turbulence in order to approximate 
the velocity and density fluctuations present in observed GMCs. Our 
method borrows from turbulent-box type simulations with a few key 
differentiating features.
A random velocity field is initialized within the GMC material. 
This velocity field is chosen to be purely solenoidal in nature and is 
created via a 3D power spectrum following the relation 
$v_{k}^{2} \propto k^{-4}$, where $k=\pi/d$ is the wavenumber for an 
eddy diameter $d$. All modes within this range are excited.
We chose the minimum $k$-mode to be that spanning our cloud diameters,
i.e., setting the largest-scale turbulent velocities, and the maximum
$k$-mode to be ten times greater so that our fiducial range for both
clouds is $2 < \frac{k}{\pi/L} < 20$ for simulation volume length $L$. 
Turbulence will then cascade to smaller scales
(larger $k$ numbers), eventually limited by numerical resolution,
during the course of the simulation.

Note that we do not initialize turbulence in the surrounding ambient 
medium. We adopt this method for simplicity in order to focus on the GMCs,
and because we expect the dynamical effects of sub-sonic turbulence in the
atomic envelope to be relatively low.

The scaling of the turbulence is chosen such that the GMCs are 
initialized to be moderately super virial, i.e., with a
1D velocity dispersion of $\sigma=5.2\:{\rm km\:s^{-1}}$ so that the
virial parameter $\alpha_{\rm vir}\equiv\frac{5\sigma^2R}{GM}=6.8$. 
This corresponds with Mach numbers measured within the clouds of 
$\mathcal{M}_{s} \equiv \sigma/ c_{s}= 23$ (for $T=15$~K conditions).
Since we do not drive turbulence, the kinetic energy content decays 
within a few dynamical times due to internal shocks, leading to lower 
velocity dispersions and lower
values of $\alpha_{\rm vir}$. Note also that the GMCs are somewhat 
confined by the pressure of the ambient, uniform density medium. Observed 
GMCs appear to have smaller virial parameters, $\sim 1$ 
\citep[e.g.,][]{Roman-Duval_ea_2010, Kauffmann_ea_2013},
especially when considering their position-velocity connected, 
$^{13}$CO-emitting structures \citep{Hernandez_Tan_2015}.  Our choice of
initializing with a larger kinetic energy content is motivated by the
desire to not have rapid global collapse of the clouds within the
first few Myr, i.e., the timescale of the collision.

The simulation box is initialized with a large-scale uniform magnetic
field directed at an angle ($\theta$) relative to the collision axis
of the clouds. The fiducial direction is $\theta=60^{\circ}$, though
various orientations are explored. The fiducial magnetic field strength
is set to be $10~{\rm \mu G}$, following Zeeman measurements of GMC
field strengths \citep{Crutcher_2012}.  Additionally, we test
non-magnetized as well as more strongly magnetized ($30~{\rm \mu G}$)
cases to explore the effects of magnetic field strength. We define
magnetic criticality via the dimensionless mass-to-flux ratio
\begin{equation}
\lambda_{\rm GMC} = \frac{M/\Phi}{1/(2\pi G^{1/2})}
\end{equation}
In this case, we calculate the 
mass-to-flux ratio by averaging over the cross section of one GMC
through the volume of the box, including ambient gas. This yields
$B_{\rm crit} = 43\:{\rm \mu G}$. Thus our GMCs are magnetically
supercritical and so should be able to undergo global collapse if
their internal turbulence is at a small enough level.

The default relative collision velocity of the clouds is chosen to be
$v_{\rm rel}=10\:{\rm km\:s^{-1}}$, though values of 5 and 20 ${\rm
  km\:s^{-1}}$ are also explored. The CNM envelope of each GMC is
assumed to be co-moving with the cloud and thus is also colliding at
the same relative velocity. In terms of the simulation domain, half
the volume is initialized with a velocity $+v_{\rm rel }/2$ while the
other half moves with $-v_{\rm rel}/2$.

Generally, the simulations are run for 5 Myr. The freefall time for
the adopted initial density of the clouds is $t_{\rm
  ff}=(3\pi/[32G\rho])^{1/2}\simeq 4.35$~Myr, but $t_{\rm ff}$ for the
denser substructures created by turbulence is much less. Most of the
analysis is performed at a time 4~Myr after the beginning of the
simulations, though the time-evolution of various cloud properties is
also explored.

\begin{figure}
\centering
\includegraphics[width=1\columnwidth]{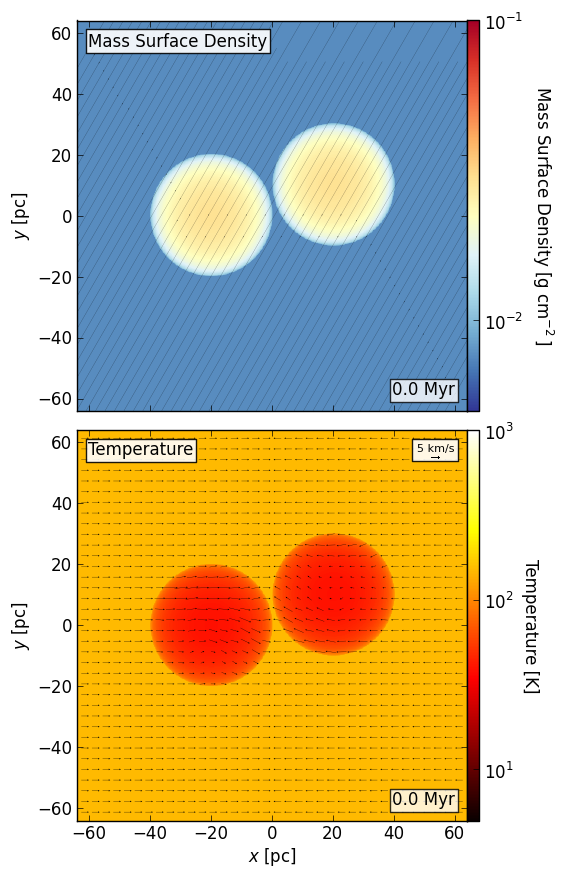}
\caption{
Fiducial initial conditions. Top panel: Mass surface density, shown
together with magnetic field structure (gray lines). Bottom panel:
Mass-weighted temperature, shown together with the velocity field
(black vectors; velocity scale in the top right).
GMCs 1 (left) and 2 (right) have identical dimensions with an initial
separation of their centers of $2R_{\rm GMC}$ in the $x$-direction 
and 0 in the $z$-direction. 
In the $y$-direction, they are offset by an impact parameter 
$b=0.5R_{\rm GMC}$.
\label{fig:initial}}
\end{figure}

The initial conditions of the set-up are shown in
Figure~\ref{fig:initial} and their properties are summarized in
Table~\ref{tab:initial}. A complete list of models, illustrating the
range of parameter space explored, is shown in
Table~\ref{tab:all_runs}. In our subsequent discussion, we shall refer
to Models 1 and 2 as the ``fiducial colliding'' and ``fiducial
non-colliding'' models, respectively, while the remaining models
(3-11) will be collectively referred to as the ``parameter models.''

\begin{table}
\caption{Initial Simulation Properties}
\label{tab:initial}
\begin{center}
\begin{tabular}{lcccc}
\hline\hline
            &                    & GMC                 & ambient \\
\hline
$n_{\rm H}$ & (${\rm cm}^{-3}$)  & 100                 & 10      \\
$R$         & (${\rm pc}$)       & 20                  &  -      \\
$M$         & ($M_{\odot}$)      & $9.3\times10^{4}$   &  -      \\
$T$         & (K)                & 15                  & 150     \\
$t_{\rm ff}$ & (Myr)             & 4.35                &  -      \\
$c_{\rm s}$ & (km/s)             & 0.23                & 0.72    \\
$v_{A}$     & (km/s)             & 1.84                & 5.83    \\
$v_{\rm vir}$ & (km/s)           & 4.9                 &  -      \\
$\sigma$    & (km/s)             & 5.2                 &  -      \\
$\mathcal{M}_{\rm s}$ &          & 23                  & -       \\ 
$\mathcal{M}_{A}$ &              & 2.82                & -       \\ 
$k$-mode    & ($k_{1},k_{2}$)    & (2,20)              &  -      \\
$v_{\rm bulk}$ & (km/s)          & $\pm5$              & $\pm5$  \\
$B$         & (${\rm \mu G}$)    & 10                  & 10      \\
$\lambda$   &                    & 4.3                 & 1.5     \\
$\beta$     &                    & 0.015               & 0.015   \\
\hline
\end{tabular}
\end{center}
\end{table}

\begin{table}
  \caption{Summary of Simulations and Explored Parameter Space}
  \label{tab:all_runs}
  \centering
  \begin{tabular}{cccccc} \hline \hline
model &  name                & $v_{\rm rel}$ & $B$             & $\theta$     & $b$              \\
      &                      & ${\rm km\:s^{-1}}$        & (${\rm \mu G}$) & ($^{\circ}$) & ($R_{\rm GMC}$)  \\
\hline
1\footnote{includes additional runs exploring lower resolutions using 1 and 2 levels of AMR}  & Colliding                   & 10  & 10  & 60  & 0.5  \\
2  & Non-Colliding               &  0  & 10  & 60  & 0.5  \\
3  & $v_{\rm rel}=5\:{\rm km/s}$  &  5  & 10  & 60  & 0.5  \\
4  & $v_{\rm rel}=20\:{\rm km/s}$ & 20  & 10  & 60  & 0.5  \\
5  & $\theta=0^{\circ}$          & 10  & 10  & 0   & 0.5  \\
6  & $\theta=30^{\circ}$         & 10  & 10  & 30  & 0.5  \\
7  & $\theta=90^{\circ}$         & 10  & 10  & 90  & 0.5  \\
8  & $b=0R_{\rm GMC}$            & 10  & 10  & 60  & 0    \\
9  & $B=30{\rm \mu G}$           & 10  & 30  & 60  & 0.5  \\
10 & $B=0{\rm \mu G}$, Col.      & 10  & 0   & -   & 0.5  \\
11 & $B=0{\rm \mu G}$, Non-Col.    & 0   & 0   & -   & 0.5  \\
\hline
\end{tabular}
\end{table}

\subsection{Numerical Code}
\label{sec:methods-code}

We use the numerical code
\texttt{Enzo}\footnote{http://enzo-project.org (v2.4-dev, changeset 845edacb82b1+)},
a magnetohydrodynamics
(MHD) adaptive mesh refinement (AMR) code \citep{Bryan_ea_2014}. This
code solves the MHD equations using the MUSCL 2nd-order Runge-Kutta
temporal update of the conserved variables with the Harten-Lax-van
Leer with Discontinuities (HLLD) method and a piecewise linear
reconstruction method (PLM).
The hyperbolic divergence cleaning method of \citet{Dedner_ea_2002} is
adopted to ensure the solenoidal constraint on the magnetic field
\citep{Wang_Abel_2008}.

For our main results, we use a top level root grid of $128^{3}$ and 3
additional levels of refinement, giving a minimum grid cell size of
0.125~pc and maximum resolution of $1024^{3}$. 
To test numerical convergence, two additional models of the fiducial 
colliding case are run at lower resolution. These models have the same 
$128^{3}$ root grid, but instead have 1 and 2 total levels of AMR, 
respectively. We perform the equivalent analysis for each resolution 
case and compare any noteworthy differences in the respective sections.

For all cases, a cell is refined
when the local Jeans length becomes smaller than 8 cells. This results
in larger volumes of highly refined regions within the GMCs when 
compared to the 4 cells typically used to avoid artificial 
fragmentation (i.e., the Truelove criterion; \citealt{Truelove_ea_1997}).
However, we note that for our magnetically supported gas the effective
``magneto-Jeans mass'' will be significantly larger than the thermal
Jeans mass.
While these refinement conditions do not necessarily capture the full 
turbulent cascade or dynamo amplification, which would require 30 cells
per Jeans length \citep{Federrath_ea_2011}, they should nonetheless 
provide approximations to real GMC structures while sufficiently avoiding 
artificial fragmentation.

We make use of the {\textquotedblleft}dual energy
formalism{\textquotedblright} that solves the internal energy equation
in addition to the total energy equation. This is necessary when
thermal energy is dominated by magnetic and kinetic energy, as it is
in our case. This method calculates the temperature from the internal
pressure when the ratio of thermal to total energy is less than 0.001,
and from the total energy otherwise.

We also use a method of limiting the Alfv\'en speed in order to avoid
extremely small timesteps set by Alfv\'en waves. This was done by 
setting a magnetic field dependent density floor, determined by a 
chosen maximum Alfv\'en velocity,
$v_{A}=B/\sqrt{\mu_{0}\rho}=1\times10^{7}\:{\rm cm \:s^{-1}}$. 
Thus, for $B\sim 10\:{\rm \mu G}$, only gas at densities below 
$n_{\rm H} \sim 0.1\:{\rm cm^{-3}}$ is affected by this limit.
In our simulations, 
this corresponds with $\ll 1\%$ of the cells and an even smaller
percentage of the total gas mass,
thus we determine the overall results to be essentially
unaffected.

\subsection{Thermal Processes}
\label{sec:methods-thermal}

We are primarily interested in the dense internal structures of
GMCs. 
This gas is almost entirely
molecular with densities $n_{\rm H} \gtrsim 10^2\:{\rm cm^{-3}}$ and
equilibrium temperatures of $\sim 15$~K. For simplicity, we use a
constant value of mean particle mass
$\mu=2.33\:m_{\rm H}$. We also choose a constant adiabatic index
$\gamma=5/3$ throughout the entire simulation domain, following
methods adopted in Paper I. While this does not account for the
excitation of rotational and vibrational modes of $\rm H_2$ that would
occur in some shocks, we consider that this is the most appropriate
single-valued choice of $\gamma$ for our simulation setup, given our
focus on the dynamics of the dense molecular gas.

We implement PDR-based heating and cooling functions that were created
and described in detail in Paper I. These functions include atomic and
molecular heating and cooling processes in nonequilibrium conditions,
taking into account extinction, density, and temperature. Again
following Paper I, we assume a FUV radiation field of $G_{0}=4$ (i.e.,
appropriate for inner Galaxy conditions, e.g., at Galactocentric
distances of $\sim 4$~kpc) and background cosmic ray ionization rate
of $\zeta = 1.0 \times 10^{-16}\:{\rm s^{-1}}$. The heating/cooling
functions span the density and temperature space of $10^{-3} \geq
n_{\rm H}/{\rm cm^{-3}} \geq 10^{10}$ and $2.7 \geq T/{\rm K} \geq 10^7$
(increasing the upper limits from $10^6\:{\rm cm^{-3}}$ and 
$10^5\:{\rm K}$, respectively, from Paper I),
encompassing our desired regime of interest and approximating a
multi-phase fluid.

The resulting heating and cooling rates are incorporated into
\texttt{Enzo} via the \texttt{Grackle} external chemistry and cooling
library\footnote{https://grackle.readthedocs.org/}
\citep{Bryan_ea_2014, Kim_ea_2014}. The information is read in via the
purely tabulated method and modifies the gas internal energy, $E_{\rm
  int} = p/\left(\gamma - 1\right)$, of a given cell with a net
heating/cooling rate calculated by
\begin{equation} \label{eq:heating}
H = n_{\rm H}[\Gamma - n_{\rm H}\Lambda] \; \rm erg\, cm^{-3}\, s^{-1},
\end{equation}
where $\Gamma$ is the heating rate and $\Lambda$ is the cooling rate.

\begin{figure*}
\centering
\includegraphics[width=2\columnwidth]{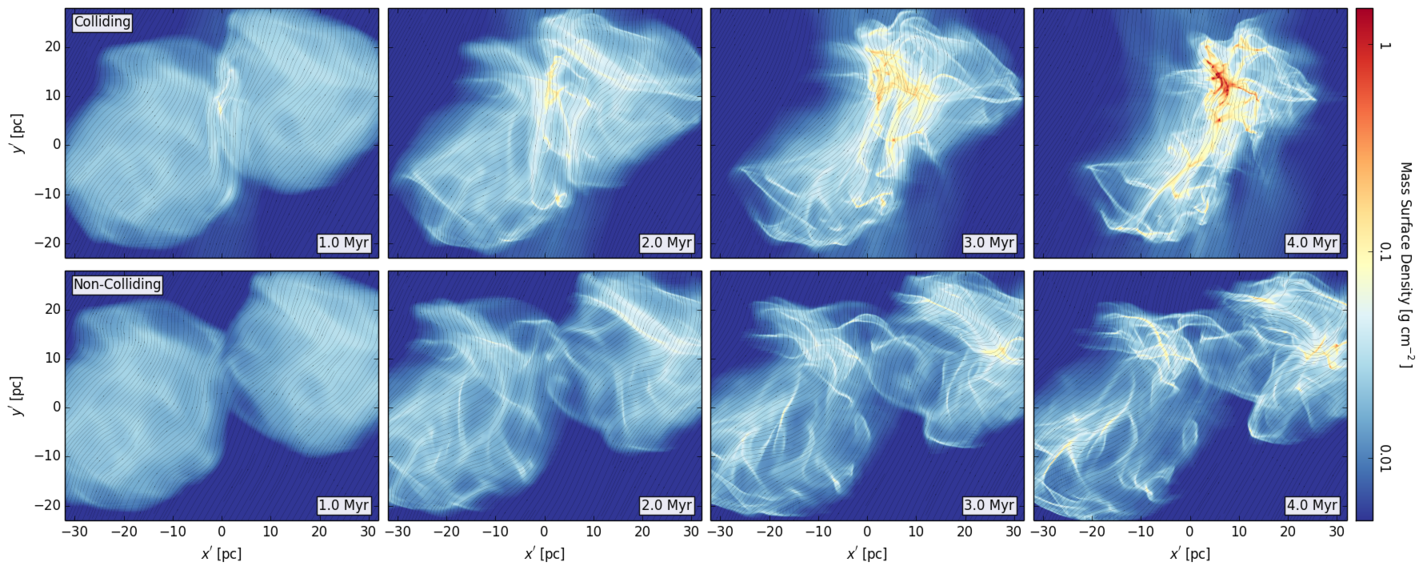}
\includegraphics[width=2\columnwidth]{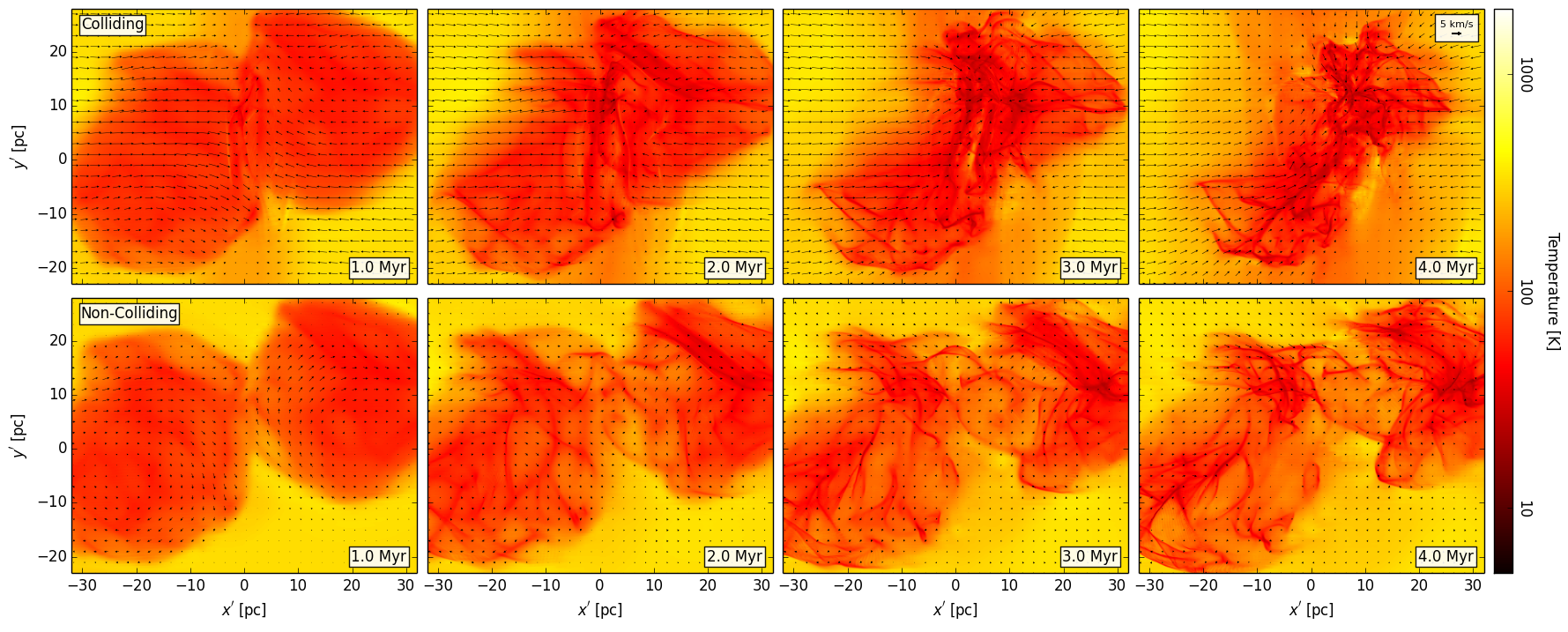}
\figcaption{
Top: Time evolution of mass surface density for the the fiducial colliding
(model 1, 1st row) and non-colliding (model 2, 2nd row) cases.
Bottom: Time evolution of mass-weighted 
temperature for the same models (model 1, 3rd row; model 2 4th
row). Snapshots at 1.0, 2.0, 3.0, and 4.0 Myr are
shown. Mass-weighted magnetic fields are shown as gray streamlines
while velocities are shown as black vectors with the
velocity scale shown in the top right.
\label{fig:morphology_fiducial}}
\end{figure*}

\subsection{Observational Diagnostics}

\label{sec:methods-diagnostics}

A key output of the aforementioned heating/cooling functions is the
detailed information of specific components that contribute to the
total heating and cooling rates (see Paper I for the full
method). Specifically, by extracting rotational line cooling rates of
$^{12}$CO and $^{13}$CO, we are able to create synthetic observations
of self-consistent CO emissivities via post-processing. Paper I
introduced a number of observational diagnostics, namely high-$J$ to
low-$J$ CO line intensity ratios and velocity spectra. The analysis in
this paper revisits these metrics, but now for 3D geometries and
initially turbulent clouds.

We note that while radiative transfer of emissivities is not
calculated during post-processing (i.e., we sum contributions along
sight lines that is valid in the optically thin limit), it is
indirectly incorporated in each cell via the heating/cooling
functions.  Self-shielding and line optical depths are accounted for
in the PDR models, which assume a one-to-one density-extinction
relation (see Paper I). Nevertheless, we choose lines in which optical
depths should be relatively small. The resulting intensities are
simply integrated directly through the simulation domain. We also note
that CO freeze-out onto dust grains is not treated in our PDR
models. A more detailed study with comparison of our approximate
functions to 3D PDR models and radiative transfer calculations is
currently in preparation (Bisbas et al., in prep.).

We will present integrated intensity maps and spectra of CO lines and
line ratios with rotational excitations $J$=2-1, 3-2, and 8-7 in
\S\ref{sec:results-intensity} and \S\ref{sec:results-kinematics},
respectively. The dynamical analysis of \S\ref{sec:results-dynamics}
is performed on synthetic $^{13}$CO($J$=1-0) maps.

\vspace{10mm}

\section{Results}
\label{sec:results}

We perform analysis of the simulations, focusing on the following
categories of interest: density and temperature morphologies
(\S\ref{sec:results-morphology}); magnetic field morphologies and
strengths (\S\ref{sec:results-Bfield}); mass surface density
distributions (\S\ref{sec:results-pdf}); CO line diagnostics
(\S\ref{sec:results-intensity}); kinematics (i.e., spectra and
velocity gradients) (\S\ref{sec:results-kinematics}); and dynamics
(i.e., virial analysis) (\S\ref{sec:results-dynamics}).

Primarily, we investigate relative differences between the fiducial
colliding and non-colliding cases, with the goal of understanding the
physical effects of GMC-GMC collisions and determining potential
differentiating observational diagnosis techniques. Additionally, the
remaining parameter models are analyzed to supplement the main results
by understanding the effects of variations in the collisional
parameters.

For visualization and analysis, we often use a rotated coordinate
system ($x'$,$y'$,$z'$) relative to the simulation axes ($x$,$y$,$z$)
such that $x'$,$y'$, and $z'$ are rotated by the polar and azimuthal
angles, respectively, $(\theta,\phi) = (15^{\circ},15^{\circ})$ about
each axis.  The purpose of this is to remove biases from an artificial
collisional plane that develops as a result of our initial conditions
of colliding flows of uniform CNM. This plane has negligible dynamical
effects on the GMCs, but a magnified observational signature when the
line-of-sight is directly aligned along this plane. In some cases, a
non-rotated coordinate system denoted by ($x$,$y$,$z$) is sufficiently
unaffected by the initial conditions and is thus used for simplicity.

\begin{figure*}
  \centering
  \includegraphics[width=1.7\columnwidth]{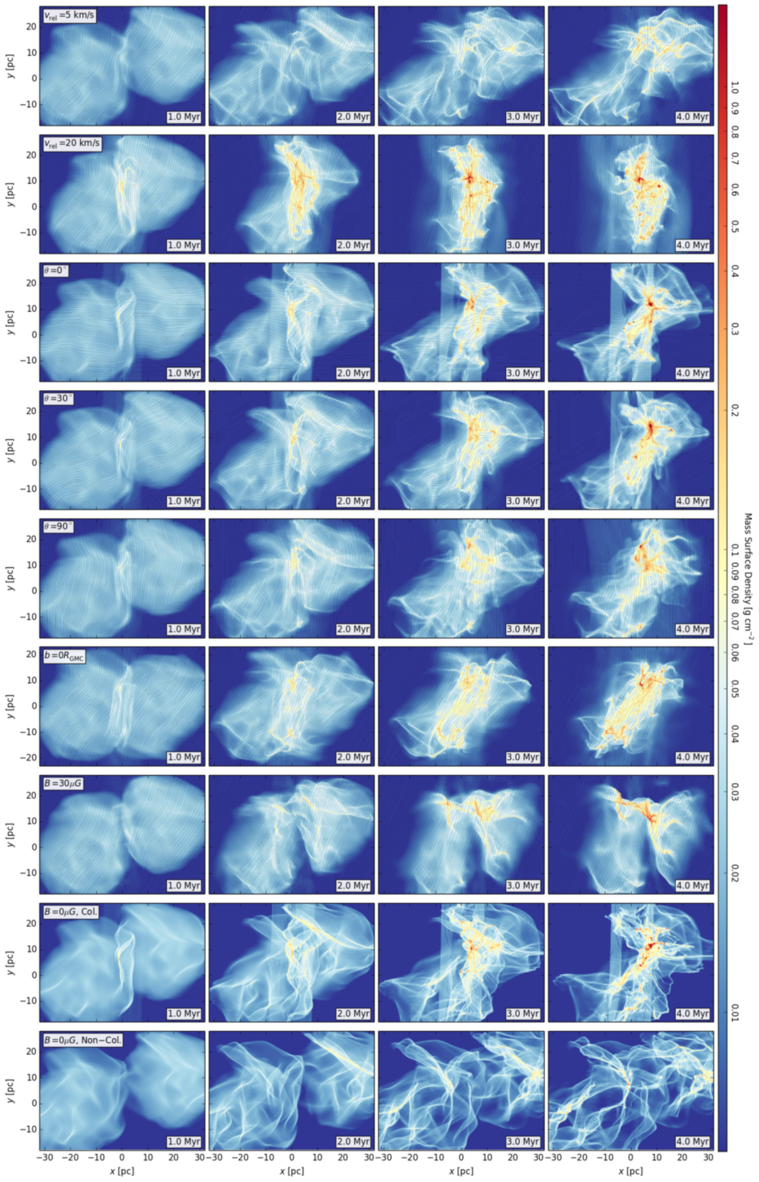}
\figcaption{
Time evolution of mass surface density for the remaining simulations
(models 3 through 11). Each row represents a specific model as
labeled, while columns are snapshots at $t$ = 1.0, 2.0, 3.0, and 
4.0~Myr are shown. 
Mass-weighted magnetic fields are represented by gray streamlines.
  \label{fig:morphology_dens_all}}
\end{figure*}

\begin{figure*}
\centering
\includegraphics[width=1.7\columnwidth]{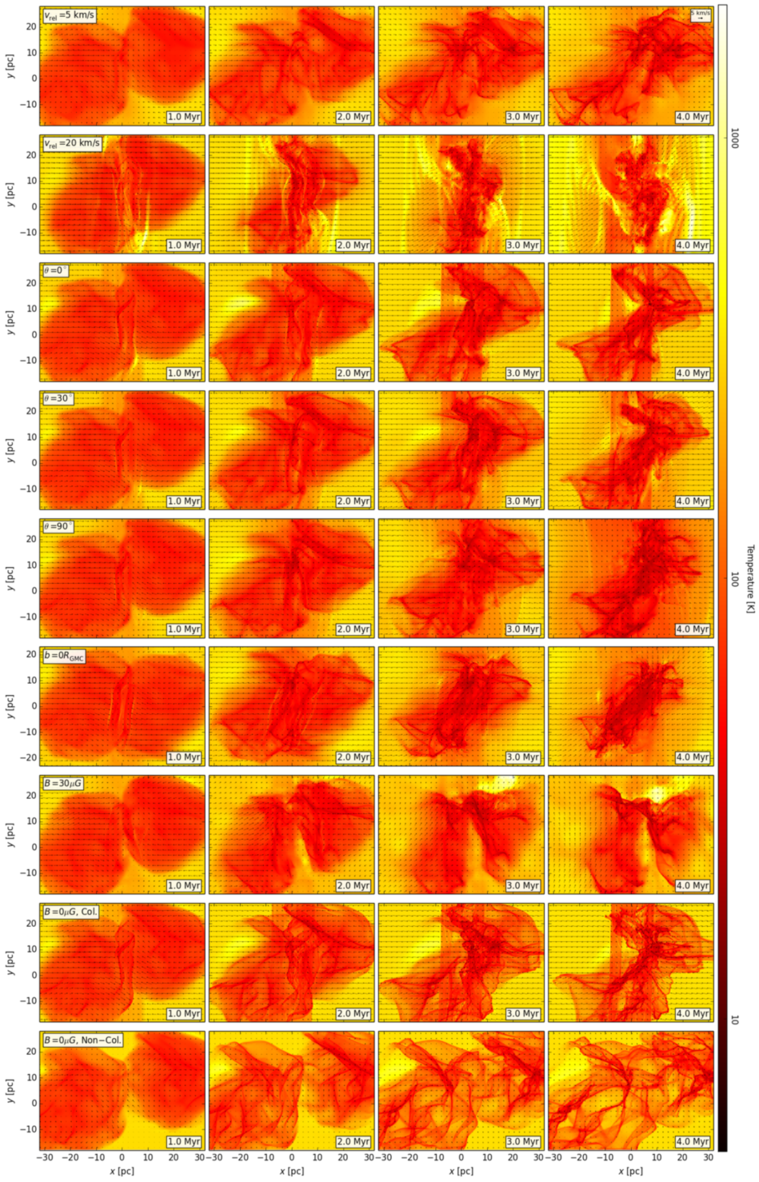}
\figcaption{
Time evolution of mass-weighted temperature for the remaining
simulations (models 3 through 11). Each row represents a specific
model as labeled, while columns are snapshots at 
$t$ = 1.0, 2.0, 3.0, and 4.0~Myr.
Mass-weighted velocities are represented by black vectors, with the
velocity scale shown in the top right.
  \label{fig:morphology_temp_all}}
\end{figure*}

\subsection{Mass Surface Density and Temperature Morphology}
\label{sec:results-morphology}

The time evolution of mass surface density (superposed with magnetic
field lines) and temperature (superposed with gas velocity vectors)
structures in the fiducial colliding and non-colliding cases are shown
in Fig.~\ref{fig:morphology_fiducial}. Similar plots for the remaining
nine parameter models are shown in Figs.~\ref{fig:morphology_dens_all}
and \ref{fig:morphology_temp_all} for density and temperature,
respectively.

\subsubsection{Fiducial Models}

Both the fiducial colliding and non-colliding cases develop
filamentary density structures within the GMCs as a result of the
turbulent velocity fields. The spatial extent of the non-colliding
GMCs is generally retained over the course of $\gtrsim 1$ free-fall
time, though the density distribution evolves from an initially
uniform density to a network of relatively slowly growing filaments
and with increasing differentiation in densities.

For the colliding case, an elongated filamentary sheet-like structure
of much higher density quickly develops near the colliding region,
with both GMC material and CNM gas being swept up in the large
shocks created by the colliding flows. A primary filamentary region
results, generally lying in the plane oriented perpendicular to the
collision axis, with smaller filaments extending outward. Structures
with mass surface densities exceeding $1\:{\rm g\:cm^{-2}}$ are more
localized and form at fractions of the original $t_{\rm ff}$, much 
more quickly relative to the non-colliding case.

The mass surface density structure and magnetic fields mutually affect
one another. In the non-colliding case, the densest filaments are
qualitatively preferentially aligned perpendicular to magnetic field
lines. Additionally, the turbulent material drags the magnetic fields
with it, creating twisted and more complex magnetic structures from an
initially uniform geometry. In the colliding case, the large-scale
flows compress the magnetic fields into the plane perpendicular to the
collision axis, effectively re-orienting the magnetic fields in a new
locally dominant direction. Relative orientations between mass surface
density structure and magnetic fields may be an observable
differentiating factor between relatively isolated turbulent GMCs and
those which have undergone a major binary collision. A more detailed
analysis quantifying these relative orientations is discussed in
\S\ref{sec:results-Bfield}. The strong coupling between magnetic field
and density in the simulations is expected from flux-freezing in ideal
MHD. Non-ideal MHD effects such as ambipolar diffusion may become
dominant in certain regimes within the GMCs and will be explored in a
subsequent paper.

The PDR-based heating/cooling functions (described in
\S\ref{sec:methods-thermal} and Paper I) enable us to approximate the
thermal behavior of gas in the atomic-to-molecular regime and model
non-equilibrium effects, specifically shocks. For both models, the
temperature is generally near the equilibrium temperature for the
particular density: $\sim$ tens of Kelvin at $n_{\rm H}>100\:{\rm
  cm^{-3}}$ and $\sim 10^{2}$~K to $10^{3}$~K for $n_{\rm
  H}\lesssim10\:{\rm cm^{-3}}$. In the non-colliding case, the
deviation of actual gas temperature from the equilibrium temperature
curve is generally small.  In the colliding case, large shock waves
are created, resulting in a high-temperature shock front that sweeps
through GMC material as it enters the post-shock region. Upon doing
so, a central region of low temperature filamentary gas develops,
again strongly correlating with density structures. This region of $T
\sim 15\:{\rm K}$ gas grows in size as more dense material
accumulates.

\subsubsection{Parameter Models}

Next, we discuss how variations in the collision parameters affect the
morphologies of mass surface density and temperature through their
subsequent evolution. Figs~\ref{fig:morphology_dens_all} and
\ref{fig:morphology_temp_all} provide a direct comparison between
these models.

Collision velocities of $v_{\rm rel}=5$ and 20 km/s are explored in
models 3 and 4, respectively. By $t=4\:{\rm Myr}$, Model 3 has not yet
produced gas of $\Sigma>1\:{\rm g\:cm^{-2}}$ but contains morphological
features somewhat in between the non-colliding and colliding fiducial
models. A relatively dense filament can be seen forming in the central
collision region, while a separate region within GMC 2 has begun to
form a second dense filament. Both regions correspond spatially with
dense structures that form in the non-colliding case, which points to
turbulence as the dominant formation mechanism, but their densities
are further enhanced at earlier times due to the collision. These
regions are also sites of the lowest temperatures, cooling to $\sim
15\:{\rm K}$. Model 4 creates a stronger shock, higher-density
collision region, and higher-density clumps at earlier times. The main
filamentary sheet appears more localized to the central collision
region, and many dense core-like structures are created along the
length of this general filament relative to the fewer, more elongated
structures created in more slowly colliding cases. The higher
collision velocity also created high-temperature ($T>1000\:{\rm K}$)
shock fronts propagating anti-parallel to the incoming flow as well as
oblique shocks created at the GMC boundaries corresponding to the
impact parameter.

Initial magnetic field orientations of $\theta=0^{\circ}$,
$30^{\circ}$, and $90^{\circ}$ are explored in models 5, 6, and 7,
respectively. As magnetic pressure acts in directions perpendicular to
the field lines, it is expected that smaller values of $\theta$ should
result in less inhibited flow and yield higher density gas. While
turbulence does stir up the magnetic field lines, the larger-scale
uniform direction and bulk flow dominate the resulting
morphology. Thus, higher density gas is formed at earlier times for
smaller $\theta$, with the extent of general GMC substructures
greatest along the direction of the large-scale magnetic fields. The
temperatures within the dense regions are near equilibrium, aside from
regions through which shocks are actively crossing. Among these
models, ambient gas near the collisional region exhibit differences in
the temperature morphology due to the density of post-shock
material. More perpendicular values of $\theta$ result in post-shock
regions with densities spread over larger extents created by built-up
magnetic pressure from the flows; this produces growing regions of
$T\sim 100\:{\rm K}$ gas surrounding the GMC material.

Model 8 explores the effects of a head-on collision ($b=0$). 
Compared to the fiducial colliding case, the head-on collision 
produces fairly similar structures in density and
temperature, though the features exhibit greater morphological symmetry:
dense, cold clumps and filaments are created at both positive and negative
y-values as opposed to predominantly positive y-values for the
$b=0.5R_{\rm GMC}$ cases.

Model 9 explores a case of stronger magnetic field, with $B=30\:{\rm
  \mu G}$, resulting in GMCs with a mass-to-flux ratio $\lambda_{\rm
  GMC}= 1.43$, only slightly magnetically supercritical, and CNM with
$\lambda_{0} = 0.5$, distinctly magnetically subcritical.  This
threefold increase in magnetic field strength, however, creates
roughly an order of magnitude increase in magnetic pressure ($P\propto
B^{2}$.) The final result is an evolution in which the clouds are
compressed by the bulk flows, but merging is inhibited. The resulting
filaments still accumulate towards the central colliding region but
are more dispersed than in the fiducial case.

\begin{figure*}
\centering
\includegraphics[width=1\columnwidth]{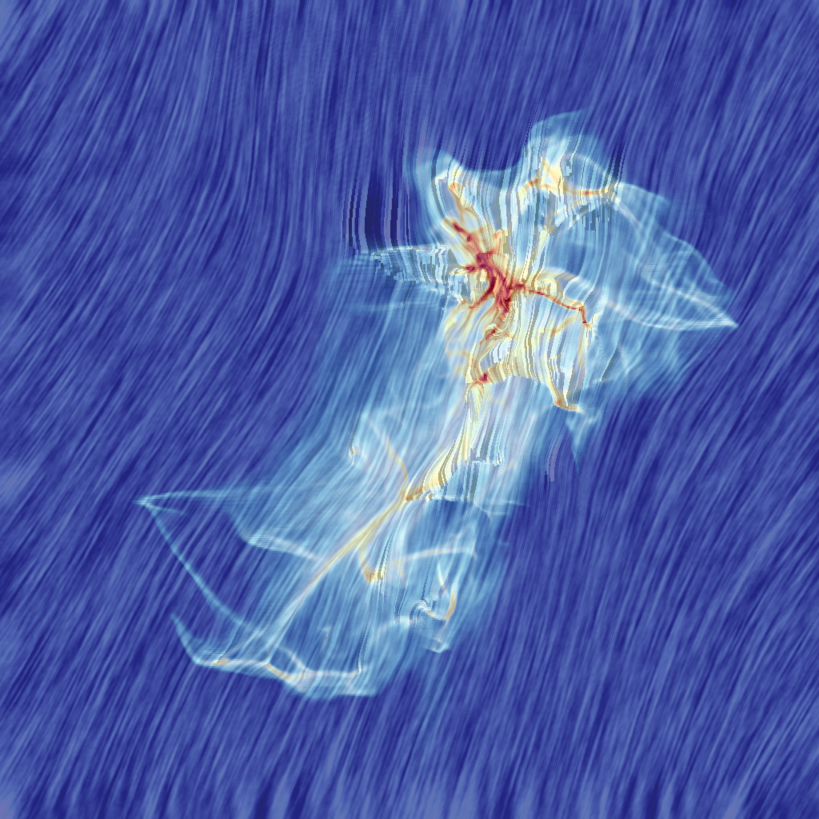}
\includegraphics[width=1\columnwidth]{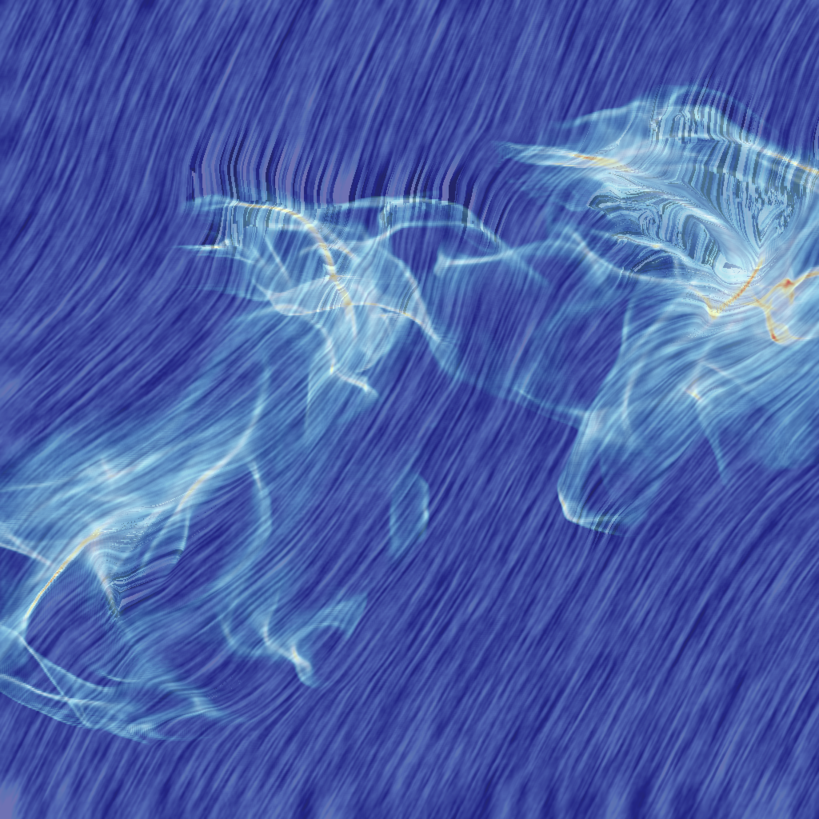}
\figcaption{
Visualization of mass surface density and projected magnetic field
polarization vectors for (left) the fiducial colliding and (right) 
non-colliding simulations. Mass surface density is represented by the
underlying colors, while the magnetic fields are "painted" along their 
polarization direction using the LIC method. The domain shown represents 
a physical projected area of $64~{\rm pc}^2$.
  \label{fig:LIC_plots}}
\end{figure*}

Unmagnetized cases are explored in models 10 and 11, the respective
colliding and non-colliding simulations. In both models, deviations in
density structures arise quickly in the evolution as there are fewer
forces inhibiting collapse. Denser filaments form more quickly, which
in turn collapse into clump-like structures on the order of $t_{\rm
  ff}$. The collision acts to localize the resulting clumps in the
central region, while the non-colliding clouds form clumps at fairly
evenly spatially distributed regions throughout each parent cloud. The
density and temperature contrasts are sharper for the non-magnetized
clouds, compared to the smoother, more connected structures of the
magnetized cases.

A more detailed quantitative analysis investigating mass surface
density distribution and evolution using probability distribution
functions (PDFs) is discussed in \S\ref{sec:results-pdf}.

\subsection{Magnetic Fields}
\label{sec:results-Bfield}

Interstellar magnetic fields and their complex interactions with both
turbulence and gravity likely play an important role in the formation
and evolution of GMCs, filaments, and eventually stars. However, their
dynamical importance is not well-determined.

Two important magnetic field parameters that influence gas dynamics
are magnetic field orientation and strength. Observationally, the
projected magnetic field orientation averaged along the line-of-sight
can be studied via dust polarization maps (assuming a particular grain
alignment model), while the line-of-sight component of the magnetic
field strength can be calculated from molecular line splitting due to
the Zeeman effect.

Recently, the ability to understand magnetic field orientations in
Galactic molecular clouds has been greatly expanded by the {\it
  Planck} space observatory, with its all-sky capability of measuring
both dust polarization and optical depth, and resolution to probe the
interiors of nearby ($d<450\:{\rm pc}$) clouds (see \citet[][hereafter
  PlanckXXXV]{PlanckXXXV_2016}).

From our simulations incorporating magnetized turbulence on the
GMC-scale, we can perform similar types of analysis in order to better
understand observable magnetic field signatures and their connections
with underlying physical processes. We first analyze magnetic field
orientation relative to mass surface density structures and then
investigate magnetic field strength relative to gas volume
density. Fig.~\ref{fig:LIC_plots} uses the line integral convolution
(LIC, first proposed by \citealt{Cabral_Leedom_1993}) method to combine
visualization of column density and projected magnetic field structure
for the fiducial colliding and non-colliding cases.

\subsubsection{Relative Orientations: $B$ vs. iso-$N_{\rm H}$}

\begin{figure}
\centering
\includegraphics[width=1\columnwidth]{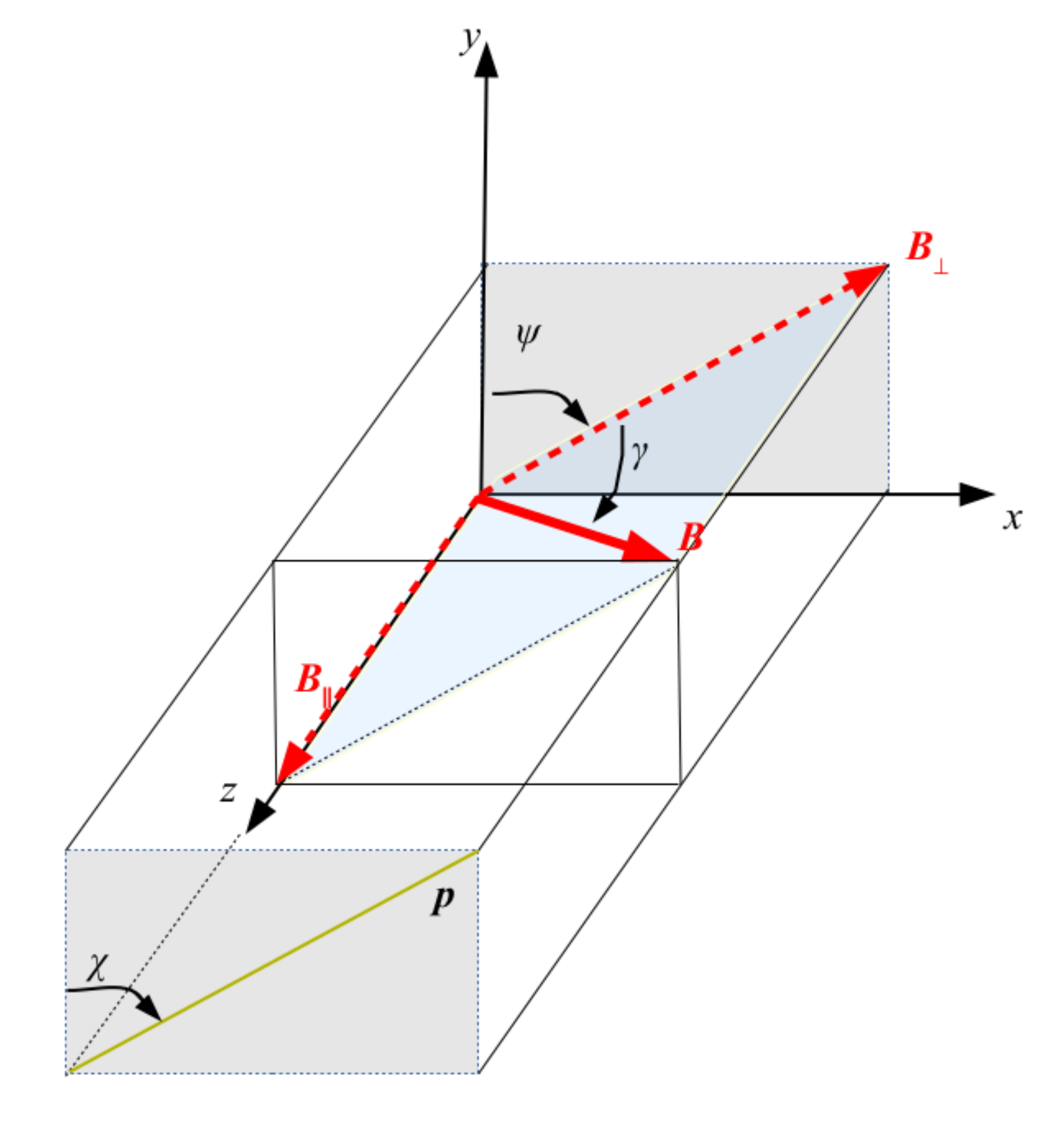}
\figcaption{
Diagram of angle definitions. For a magnetic field $\bm{B}$ and an 
observer viewing along the -$z$ axis, $\gamma$ is the inclination angle
between $\bm{B}$ and the plane-of-sky, while $\psi$ is the position angle
between $\bm{B}_{\perp}$ (the plane-of-sky magnetic field component) and 
the ``north'' direction (in this case, $y$). The integrated polarization
pseudo-vector yields an angle $\chi$.
  \label{fig:stokes_angles}}
\end{figure}

\begin{figure*}
\centering
\includegraphics[width=2\columnwidth]{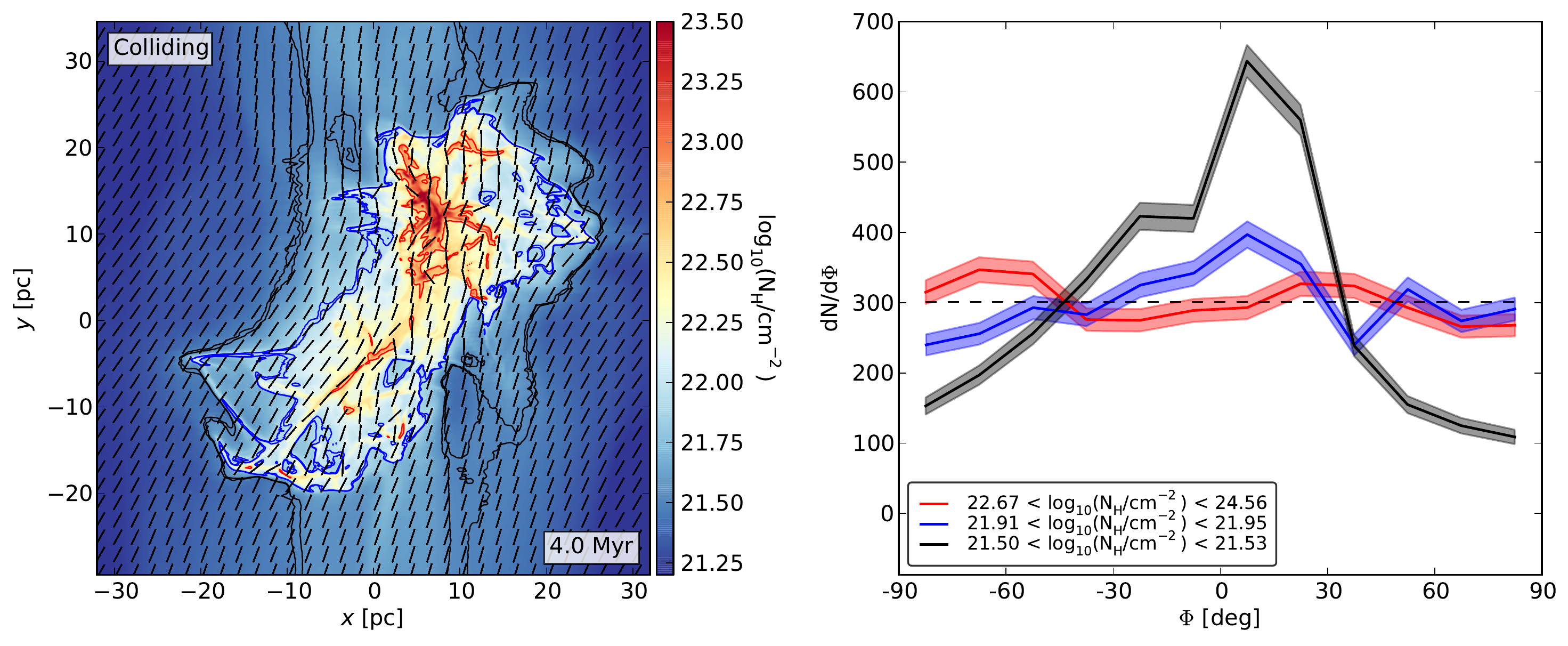}
\includegraphics[width=2\columnwidth]{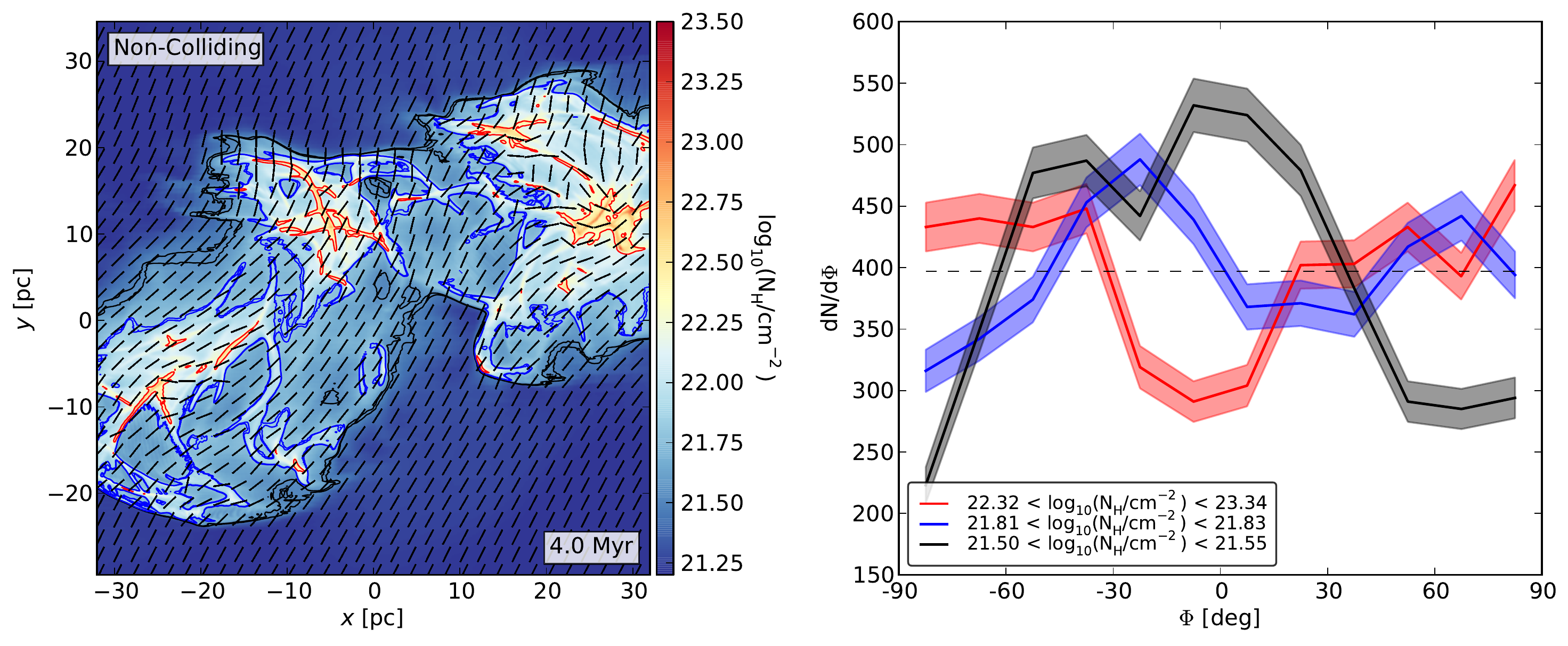}
\figcaption{
Left panels: Column density maps, $\log_{10}(N_{\rm H}/\rm{cm^{-2}})$,
with black vectors representing the normalized plane of sky
polarization field. The
colliding case is shown in the top figure, while the non-colliding
case is in the bottom figure.  Right panels: Histograms of Relative
Orientations (HROs) comparing the angle between
the polarization pseudo-vector $p$
vs. iso-$N_{\rm H}$ contours pixel-by-pixel in the fiducial colliding
(top) and non-colliding (bottom) simulations. The projected map is
divided into 25 column density bins of equal pixel count. HROs for the
lowest (1st bin; black), middle (12th bin; blue), and highest (25th
bin; red) $N_{\rm H}$ bin are shown, using angle bins of $15^{\circ}$.
The histogram color corresponds with the colored contours that bound
low (black), intermediate (blue), and high (red) column density
regions of the projection map. Histograms with peaks at $0^{\circ}$
correspond to $p$
predominantly aligned with iso-$N_{\rm H}$
contours (i.e., $B$-fields aligned along filaments). Histograms with
peaks at $\pm90^{\circ}$ correspond to $p$
predominantly
perpendicular to iso-$N_{\rm H}$ contours (i.e., $B$-fields aligned
perpendicular to filaments).
  \label{fig:HRO_fiducials}}
\end{figure*}

To study magnetic field orientations, we utilize the Histogram of
Relative Orientations (HRO, \citealt{Soler_ea_2013}). The HRO is a 
statistical tool that quantifies the magnetic field orientation 
relative to the gradient of the column density. 
It can be performed on polarization observations (e.g., PlanckXXXV) 
as well as numerical simulations \citep[e.g.,][]{PlanckXX_2015,Chen_ea_2016arXiv}
to study the mutual dependence of magnetic fields on density structures.

The HRO investigates the angle $\phi$ between the polarized emission 
$\bm{p}$ and $N_{\rm H}$ iso-contours (orthogonal to $\nabla N_{\rm H}$):
\begin{equation}
\phi = \arctan \left( \frac{\nabla N_{\rm H} \cdot
  \bm{p}}{|\nabla N_{\rm H} \times \bm{p}|} \right)
\end{equation}
$\bm{p}$ is a pseudo-vector defined by
\begin{equation}
\bm{p} = (p \sin \chi)\hat{\bm{x}} + (p \cos \chi)\hat{\bm{y}}
\end{equation}
where $p$ is the polarization fraction and $\chi$ is the polarization
angle. Thus, one can think of $\phi$ also as being the
relative angle between the magnetic field and the filamentary axis
of structures seen in mass surface density maps. Note that the 
convention we adopt for $\phi$ follows PlanckXXXV but
is shifted $\pi/2$ from that defined in
\citet{Soler_ea_2013} and \citet{Chen_ea_2016arXiv}.

We assume a constant polarization fraction $p=0.1$ (though 
\citet{PlanckXX_2015} and \citet{Chen_ea_2016arXiv}
use various grain polarization fraction models in their analysis) while
$\chi$ is the polarization angle derived from the Stokes parameters.

The relative Stokes parameters can be calculated following previous work
\citep{Lee_Draine_1985,Fiege_Pudritz_2000,Kataoka_ea_2012,Chen_ea_2016arXiv}:
\begin{equation}
q = \int n \cos 2\psi \cos^{2}\gamma ds
\end{equation}
\begin{equation}
u = \int n \sin 2\psi \cos^{2}\gamma ds
\end{equation}
where $\gamma$ is the angle between the local magnetic field relative to 
the plane of the sky, while $\chi$ is the angle of the magnetic
field on the plane of the sky relative to the ``north'' axis (see Fig.~ \ref{fig:stokes_angles}). For a
coordinate orientation where the $y$-axis can be defined as ``north''
with the line of sight directed along the $z$-axis, the relative Stokes
parameters can be written as (see \citet{Chen_ea_2016arXiv}):
\begin{equation}
q = \int n \frac{B_{y}^{2}-B_{x}^{2}}{B^{2}} ds
\end{equation}
\begin{equation}
u = \int n \frac{2B_{x}B_{y}}{B^{2}} ds
\end{equation}

Finally, we can calculate $\chi$, the polarization angle on the plane of
the sky:
\begin{equation}
\chi = \frac{1}{2} \arctan2(u,q)
\end{equation}
where $\arctan2$ is the arctangent function with two arguments, 
returning angles within $[-\pi,\pi]$ based on the quadrant of the inputs. 

To distinguish cloud structure from background structure, PlanckXXXV
selected pixels in regions where the magnitude of the column density 
gradient exceeded the mean gradient of a reference diffuse background
field. In our case, the gradient threshold was chosen to be 0.25 the
average value of the fiducial colliding case in order to better 
capture the GMC material. We apply this value for each case and 
additionally apply a cut of the lowest column density values
($N_{\rm H}<21.5\:{\rm cm^{2}}$). (Note: we assume 
$n_{\rm He}=0.1n_{\rm H}$, giving a mass per H of 
$2.34\times10^{-24}\:{\rm g}$.)
$\phi$ is then calculated for each remaining pixel in the projected 
domain. This domain is divided into
25 bins of $N_{\rm H}$ ranges, each containing an equal number of
pixels. For a given $N_{\rm H}$ range, an HRO plot can be created,
comparing the distribution of cells for each angle
$-90^{\circ}<\phi<90^{\circ}$. We create HROs for the lowest,
intermediate, and highest column density bins to investigate how the
magnetic field orientation changes as a function of column
density. This means histograms peaking at $\phi=0^{\circ}$ correspond
to $\bm{p}$ mostly aligned parallel to filamentary structure, while
peaks at $\phi=\pm 90^{\circ}$ correspond to perpendicular alignment
of magnetic fields with filaments.

The left-hand column of Fig.~\ref{fig:HRO_fiducials} shows column
density maps of the fiducial colliding and non-colliding cases
over-plotted with magnetic field vectors and colored contours defining
the three aforementioned $N_{\rm H}$ ranges. The right-hand column
shows the respective HROs, representing material within the specific
column density range. In the fiducial colliding case, the HRO peaks
near $0^{\circ}$ especially for the low column density bins, while the
intermediate and high column density bins show slight preference to
this value. This signifies a predominantly parallel alignment of
$\bm{p}$ with iso-$N_{\rm H}$ contours for the colliding case. 
Likewise, the fiducial non-colliding case exhibits strong peak near
$0^{\circ}$ for the low column density bin, but is roughly flat for
moderate column densities while peaking at $\phi=\pm 90^{\circ}$ for 
the highest column densities. This signifies a shift from predominantly
parallel alignment of $\bm{p}$ with iso-$N_{\rm H}$ contours
at low densities to a predominantly perpendicular alignment at high 
densities.

In order to distinguish trends along the entire column density range
and compare models with various collisional parameters, we further
quantify HROs using the histogram shape parameter $\xi$, which is
defined as (see \citet{Soler_ea_2013} and PlanckXXXV): 
\begin{equation}
\xi = \frac{A_{\rm c}-A_{\rm e}}{A_{\rm c}+A_{\rm e}},
\end{equation}
where $A_{c}$ is the area within the central region
($-22.5^{\circ}<\phi<22.5^{\circ}$) under the HRO, while $A_{e}$ is
the area within the extrema ($-90^{\circ}<\phi<-67.5^{\circ}$ and
$67.5^{\circ}<\phi<90^{\circ}$) of the HRO. 
Thus $\xi$ is independent of total bin
number and normalizes relative differences within the individual
histogram.  $\xi>0$ is indicative of a concave histogram ($\bm{p}$
preferentially parallel to iso-$N_{\rm H}$ contours), while $\xi<0$ is 
indicative of a convex histogram ($\bm{p}$ preferentially 
perpendicular to $N_{\rm H}$).

From PlanckXXXV, uncertainties in the HROs were found to be dominated 
by histogram binning, which we include in our analysis here.
The $k$th bin in the histogram has variance
\begin{equation}
\sigma_{k}^{2} = h_{k}\left(1-\frac{h_{k}}{h_{\rm tot}}  \right)
\end{equation}
with $h_{k}$ and $h_{\rm tot}$ being the number of samples in the 
$k$th bin and total number of samples, respectively.
The total uncertainty of $\xi$, given by $\sigma_{\xi}$, is then
calculated from
\begin{equation}
\sigma_{\xi}^{2} = \frac{4(A_{\rm e}^2 \sigma_{A_{\rm c}}^{2} + A_{\rm c}^2 \sigma_{A_{\rm e}}^{2})}{(A_{\rm c}+A_{\rm e})^{4}}.
\end{equation}

Also following PlanckXXXV, we can study trends in $\xi$ vs 
$\log_{10}(N_{\rm H}/{\rm cm^{2}})$ by fitting a linear function
\begin{equation}
\xi = C_{\rm HRO}[\log_{10}(N_{\rm H}/{\rm cm^{2}})-X_{\rm HRO}].
\end{equation}
$C_{\rm HRO}$ and $X_{\rm HRO}$ can be used as quantitative parameters
to compare general relationships between all the simulation models. A
negative slope $C_{\rm HRO}$ represents $\bm{p}$ becoming more
parallel with filaments as $N_{\rm H}$ increases, while a positive
$C_{\rm HRO}$ would signify an increasingly perpendicular relative
orientation. $X_{\rm HRO}$ represents the crossover value of $N_{\rm
  H}$ at which $\bm{p}$ switches from perpendicular to parallel to
iso-$N_{\rm H}$ contours.

\begin{figure*}
\centering
\includegraphics[width=2\columnwidth]{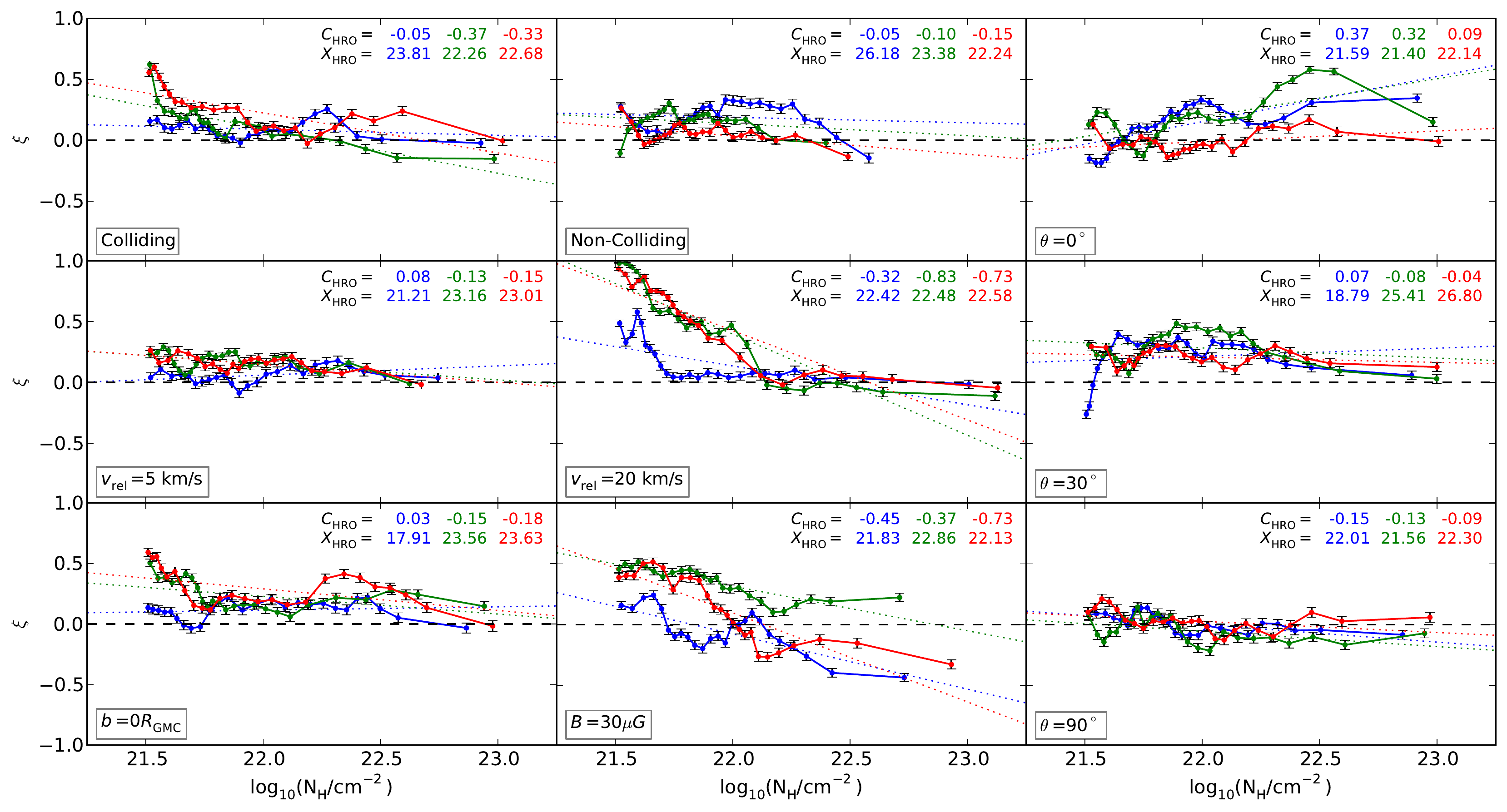}
\figcaption{
Comparison of the histogram shape parameter, $\xi$, vs column density,
$N_{\rm H}$, among the magnetized simulations. $\xi>0$ represents a
preferentially parallel orientation between magnetic field lines and
iso-$N_{\rm H}$ contours, while $\xi<0$ represents a preferentially
perpendicular orientation. The blue, green, and red lines represent
lines of sight from the $x'$,$y'$, and $z'$ axes, respectively. The
parameters $C_{\rm HRO}$ and $X_{\rm HRO}$ for the best linear fit for
each line of sight are indicated in the respective color.
\label{fig:HRO_shape_param}}
\end{figure*}

Figure~\ref{fig:HRO_shape_param} shows $\xi$ vs. $N_{\rm H}$ for our
magnetized runs (models 1-9). 
This relation does not appear to have strong dependence on line of sight,
agreeing fairly well for each model along the $y'$ and $z'$ viewing
directions. Viewing from the $x'$-direction does result in  
occasional deviations, but for the most part it is well-correlated. 
These models are generally fit with
$C_{\rm HRO}<0$ and $X_{\rm HRO}\approx 22$, which agree with the 
observational results from \citet{PlanckXXXV_2016}. From the 
10 molecular clouds in their study, mean values of $C_{\rm HRO}=
-0.41$ and $X_{\rm HRO}=22.16$ were found, with uncertainties in 
$\xi$ generally in the tens of percent range.

However, between the various models themselves, there are notable 
differences. The fiducial non-colliding case has a fairly flat slope, 
with $C_{\rm HRO}\approx-0.1$
and $X_{\rm HRO}\approx23.9$.
In comparison, the fiducial colliding case has
a steeper slope, with $C_{\rm HRO}\approx-0.3$, and 
a slightly lower intercept, $X_{\rm HRO}\approx22.9$. The linear fits
to both fiducial models are similarly consistent.
The physical interpretation is that the collision influences the 
overall preferential alignment of magnetic fields relative to 
gas filaments, specifically driving the value of $\xi$ more positive 
for lower column structures (i.e., more concave HRO; $\bm{p}$ preferentially 
parallel to low $N_{\rm H}$), and more negative for higher column structures
(i.e., more convex HRO; $\bm{p}$ preferentially perpendicular 
to high $N_{\rm H}$).

This is emphasized when varying collisional velocities are explored 
(models 3 and 4). The intermediate collision velocity ($v_{\rm rel}=
5\:{\rm km/s}$) results in a slight increase in $C_{\rm HRO}$, while
the high collision velocity ($v_{\rm rel}=20\:{\rm km/s}$) increases
the slope strongly, with $C_{\rm HRO}$ as steep as -0.83. The value
of $\xi$ seems most affected at low $N_{\rm H}$, while staying relatively
steady at $\xi \lessapprox 0$ for high $N_{\rm H}$.
The $x'$
line-of-sight in this case does not capture much of the effect of 
the collision on the magnetic field polarization.

The effects of initial magnetic field orientation (models 5, 6, and 7)
are less direct, but initial orientation appears to primarily influence
the value of $C_{\rm HRO}$, with $\theta=0^{\circ}$ resulting
in positive $C_{\rm HRO}$.

A head-on collision (model 8) appears to have a small effect on the 
overall $\xi$ vs $N_{\rm H}$ relation when compared to the fiducial 
colliding model. There is a slight upward shift in values of
$X_{\rm HRO}$, but the overall shape is generally similar. The impact 
parameter, while significant on the GMC scale, would not be expected 
to greatly influence the behavior of collisions between smaller 
individual substructures that determine the local B-field polarization.

Lastly, the stronger-field case of $B=30\:{\rm \mu G}$ (model 9) 
has notable effects on the slope, with $C_{\rm HRO}=-0.73$ in the $z'$
line of sight, as well as a moderate crossover point $X_{\rm HRO} 
\approx 22$. This model produces the most preferentially perpendicular
alignment of B-field and filamentary structure at high $N_{\rm H}$.

The resolution analysis of HRO results yielded similar values for all 
lines of sight in the fiducial colliding model, with signs of convergence 
when increasing resolution from 1-2 AMR levels to 2-3. 

HROs and subsequent histogram shape parameter analysis may be a useful
tool for differentiating between non-colliding and colliding clouds
given the strong correlations with collision velocity and B-field 
strength and, to a lesser extent, various other collisional parameters.

\subsubsection{Magnetic Field Strengths: $|B|$ vs. $n_{\rm H}$}

\begin{figure*}
\centering
\includegraphics[width=1\columnwidth]{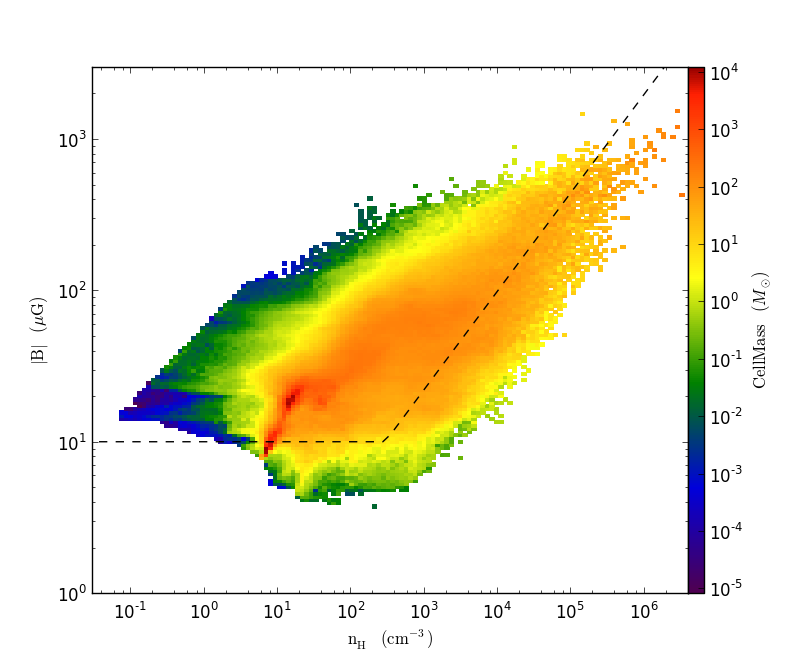}
\includegraphics[width=1\columnwidth]{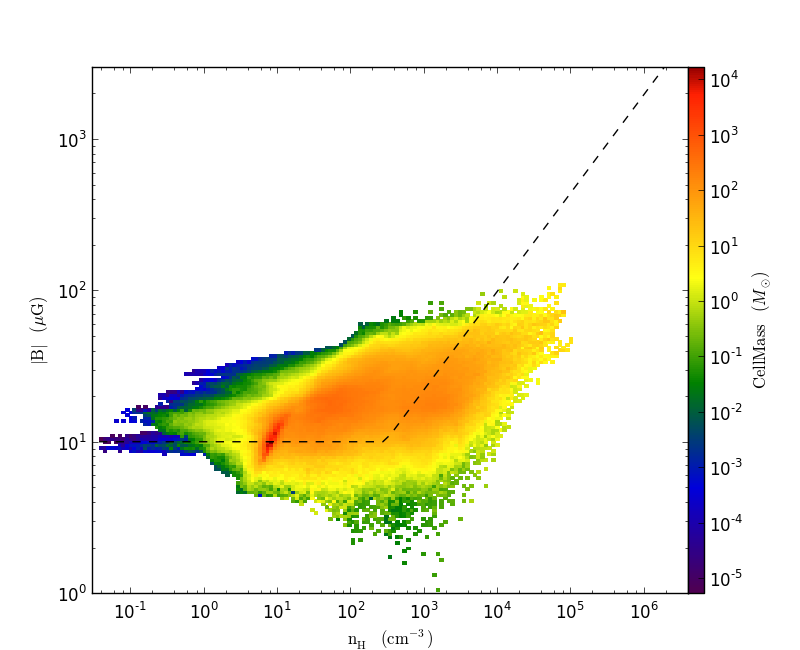}
\figcaption{
Phaseplots examining $|B|$ vs. $n_{\rm H}$ for (left) the fiducial
colliding and (right) non-colliding simulations at $t=4.0\:{\rm
Myr}$. The colorbar displays the total gas mass at each point. 
The dashed line represents the ``Crutcher relation,'' where
$B_{\rm max} = B_{0} = 10~{\rm \mu G}$ for $n_{\rm H}<300\:{\rm 
cm^{-3}}$ and $B_{\rm max} = B_{0}(n_{\rm H}/300\:{\rm cm^{-3}})^{2/3}$
otherwise. The cutoff in the low density regions is due to the
Alfv\'en limiter.
\label{fig:Bmag_vs_nH}}
\end{figure*}

The magnetic field strength as a function of density in GMCs is
another property that is potentially important for the evolution of
substructure.  Figure~\ref{fig:Bmag_vs_nH} explores the $|B|$
vs. $n_{\rm H}$ relation for the fiducial colliding and non-colliding
GMCs. The colliding case involves creation of regions of both high
density and higher magnetization than the non-colliding case. The
majority of the overall gas mass remains near the initialized values
of $B$ and $n_{\rm H}$, but the collision generally produces stronger
field strengths for a given density. The concentration of gas mass
from $10 < n_{\rm H}/{\rm cm^{-3}} < 100$ corresponds primarily with
the ambient gas accumulating in the collision region, where the
initial magnetic field is similarly compressed. In the non-colliding
case, the gas mass mostly stays concentrated near the initialized
levels, with especially the ambient, CNM gas evolving in a mostly
quiescent manner.

Although our simulations are initialized with relatively idealized
conditions, both fiducial models develop a $B$ vs. $n_{\rm H}$
behavior approximately consistent at least in general shape with the 
``Crutcher relation'' \citep{Crutcher_ea_2010} where 
$B_{\rm max} = B_{0} = 10~{\rm \mu G}$
for $n_{\rm H}<300\:{\rm cm^{-3}}$ and $B_{\rm max} = B_{0}(n_{\rm
H}/300\:{\rm cm^{-3}})^{2/3}$, for $n_{\rm H}>300\:{\rm cm^{-3}}$. 
Our models exhibit relatively stronger $|B|$ overall, exceeding the 
maximum values statistically determined by observations comprising 
the relation. The gas in the colliding case 
reaches ${\rm mG}$ strengths at $n_{\rm H}\approx10^6\:{\rm cm^{-3}}$ 
as the accumulation of gas to higher densities in turn compresses the 
magnetic fields along with it. 
The lower envelope of the phaseplot appears to exhibit a slight elbow 
near $10^{3}\:{\rm cm^{-3}}$ in both the colliding and non-colliding cases.
This is roughly consistent with the Crutcher relation, although it is 
important to note that the elbow occurs in the upper envelope of the 
Crutcher data.
The gas near this range retains roughly constant values of $|B|$ in the tens 
of ${\rm \mu G}$ range. The non-colliding case also 
exhibits a similar lower envelope relation, with a smaller overall range in 
density and $|B|$.

Deviations exist between $|B|$ found in our models and the maximum 
$|B|$ statistically predicted from observations, particularly in the 
highly magnetized, low-density gas of the colliding case. However, this 
may be attributable to the particular choice of our initial field
strengths and other simulation parameters.

\vspace{10mm}

\subsection{Mass Surface Density Probability Distribution Functions}
\label{sec:results-pdf}

PDFs of mass surface density (or $N_{\rm H}$ or $A_{\rm V}$) have been
used as tools to study the physical characteristics of observed
molecular clouds and IRDCs 
\citep[e.g.][]{Kainulainen_ea_2009,Kainulainen_Tan_2013,Butler_ea_2014}.
Mechanisms such as turbulence, self-gravity, shocks, and magnetic
fields all contribute to the resulting distribution of $\Sigma$.

For turbulent clouds, the $\Sigma$-PDF is generally
characterized as log-normal at lower $N_{\rm H}$ ranges, while at
higher $N_{\rm H}$ an additional power law tail component is often
measured and attributed to compression due to self-gravity.
The width of the log-normal component is expected to correlate with
the strength of turbulence, i.e., the Mach number of typical
shocks. The fraction of mass in the high-$\Sigma$ power law tail may
correlate with the degree of gravitational instability and the
efficiency of star formation.

We present area ($p_{A}(\Sigma)$) and mass-weighted ($p_{M}(\Sigma)$)
PDFs of $(32\:{\rm pc})^{3}$ extracted regions projected along the 
$z'$-direction through each of our models. For each distribution, we
also find the best-fit log-normal function:
\begin{equation}
p(\Sigma) = \frac{A}{(2 \pi)^{1/2}\sigma_{\ln \Sigma}}\exp{\left[ - \frac{(\ln \Sigma-\overline{\ln \Sigma})^{2}}{2\sigma_{\ln \Sigma}^{2}} \right]}
\end{equation}
where $\sigma_{\ln \Sigma}$ is
the standard deviation of $\ln \Sigma$. 
A scale factor $A$ is included to allow for adjustment
between differing PDF normalization schemes. A summary of the fit
parameters for each run is shown in Table~\ref{tab:PDFs}.

\begin{figure*}
\centering
\includegraphics[width=1\columnwidth]{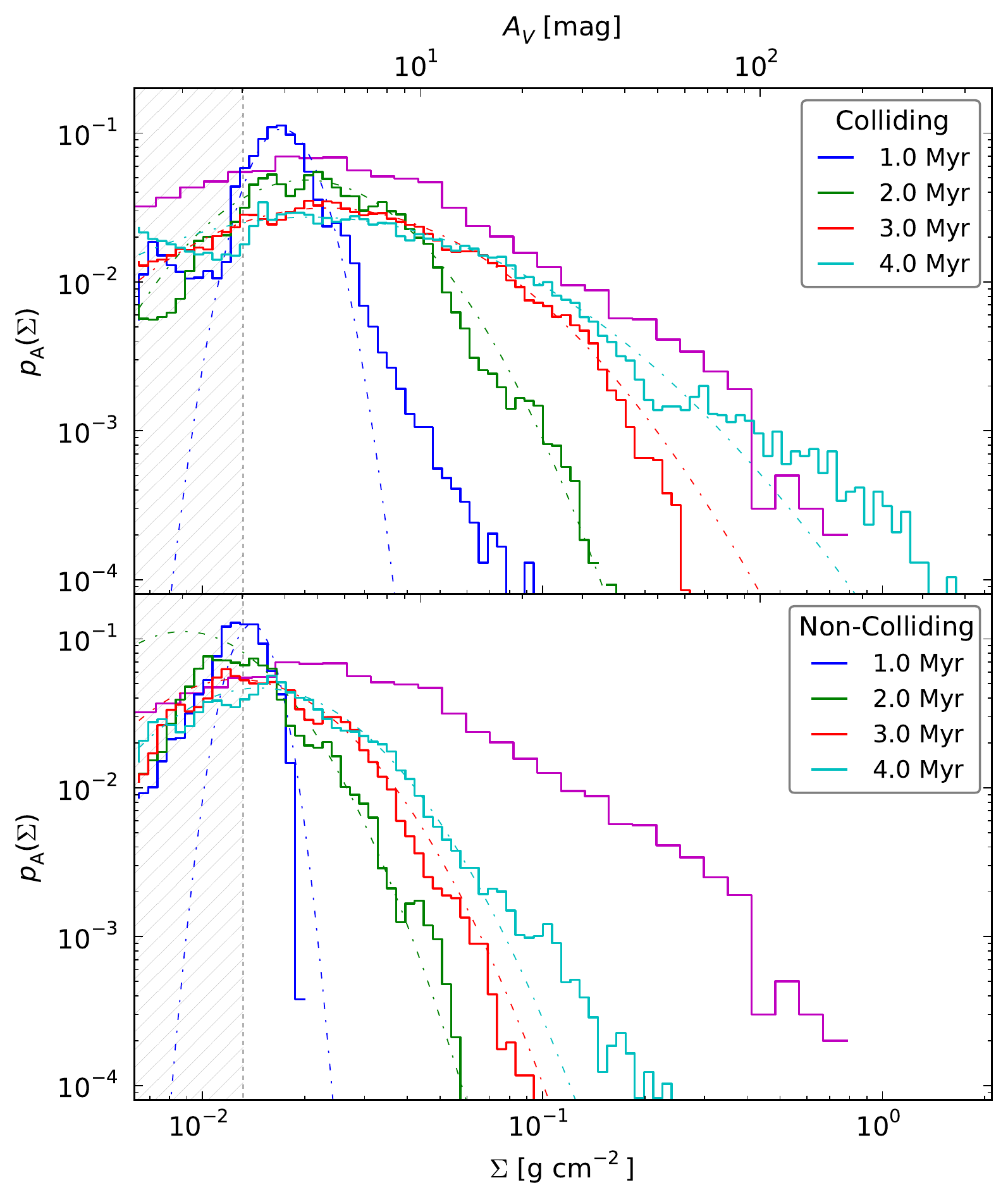}
\includegraphics[width=1\columnwidth]{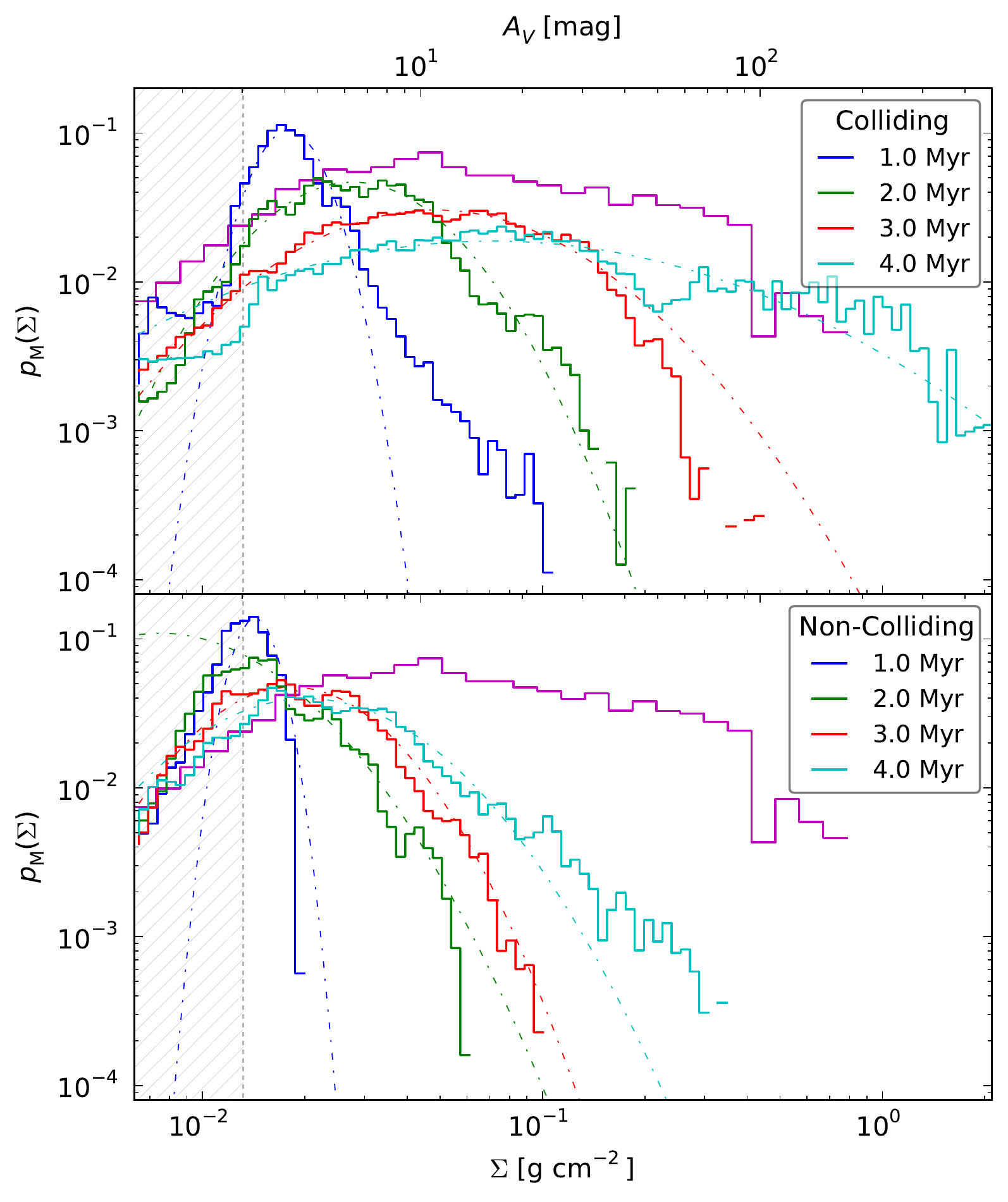}
\figcaption{
Area-weighted (left column) and mass-weighted (right column)
$\Sigma$-PDFs of $(32\:{\rm pc})^3$ regions from the fiducial
colliding (top) and non-colliding (bottom) cases as they evolve in
time. $\Sigma$-PDFs for each case at $t$=1.0, 2.0, 3.0, and 4.0~Myr are 
shown in blue, green, red, and cyan, respectively. The best log-normal 
fits for each case are plotted as dash-dotted lines of the same color.
In each panel, the $\Sigma$-PDFs from observations of a massive IRDC from 
\citet{Lim_ea_2016arXiv} is shown in magenta.
The shaded region denotes areas of $A_{V}<3\:{\rm mag}$, matching the
completeness levels the observed IRDCs.
\label{fig:PDFstime}}
\end{figure*}

Figure~\ref{fig:PDFstime} shows the time evolution of area and
mass-weighted $\Sigma$-PDFs for the fiducial colliding and non-colliding
cases. For each case, the region is centered on the position of
maximum $\rho$ at the respective 4 Myr timestep to capture the
evolution of the dense filament.

As the region evolves, both cases exhibit a broadening of the 
distribution, with $\sigma_{\ln \Sigma, A}$ increasing over 3 
Myr from 0.200 to 1.079 for colliding clouds and 0.143 to 0.600 
for non-colliding clouds. Likewise, $\sigma_{\ln \Sigma, M}$ 
increases from 0.215 to 1.413 (colliding) and 0.142 to 0.691 
(non-colliding). The values for $\overline{\Sigma}_{A}$ stay 
relatively constant, with slight increases for the colliding
case. $\overline{\Sigma}_{M}$ increases for both cases, with a
much stronger increase in the colliding case due to the high
densities created. 

Both the area and mass-weighted $\Sigma$-PDFs are generally well-fit 
with a single log-normal, though the colliding case at 1.0 Myr 
and 4.0 Myr and non-colliding case at 4.0 Myr exhibit slight 
excesses at the high-$\Sigma$ end.

\begin{figure*}
  \centering
  \includegraphics[width=1\columnwidth]{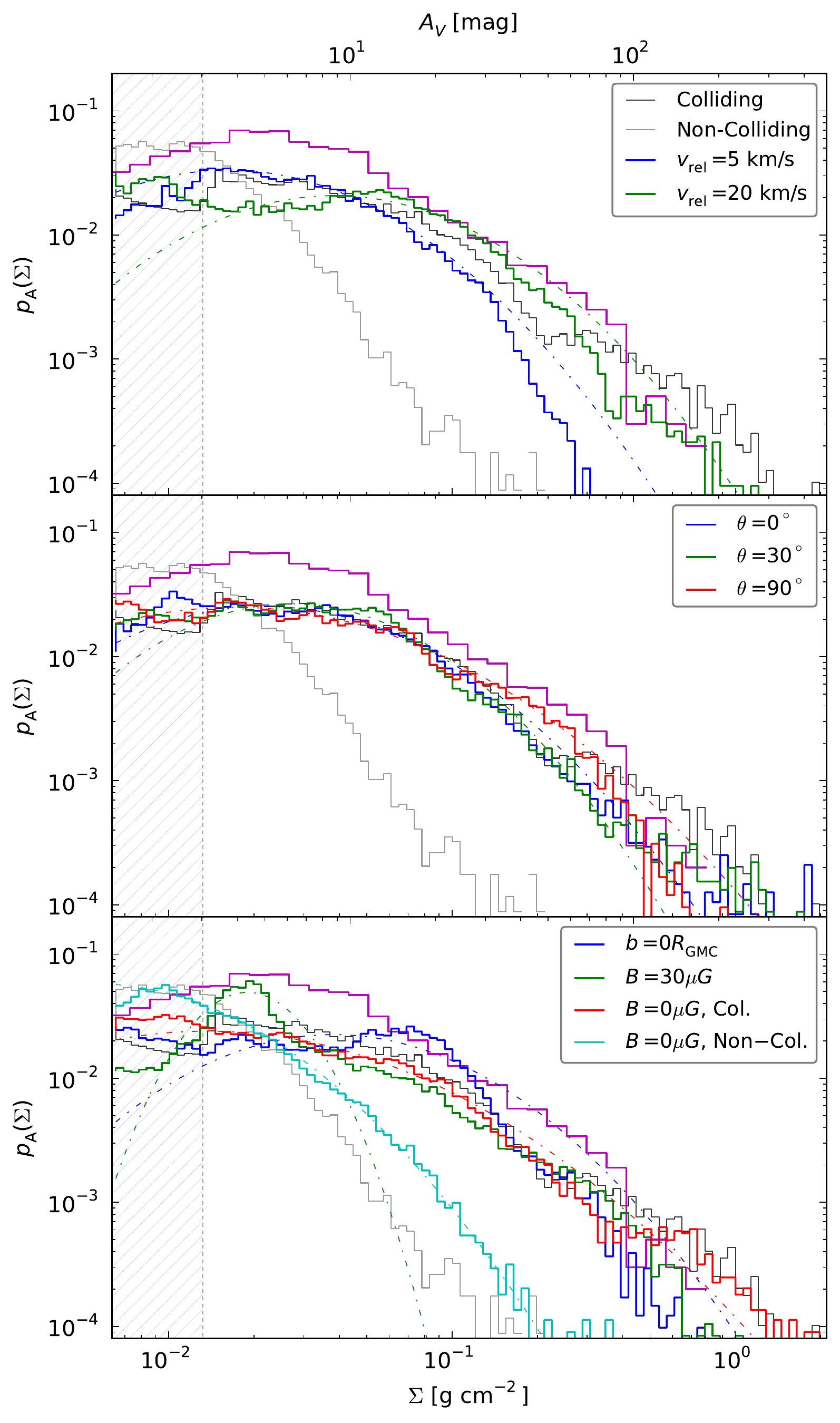}
  \includegraphics[width=1\columnwidth]{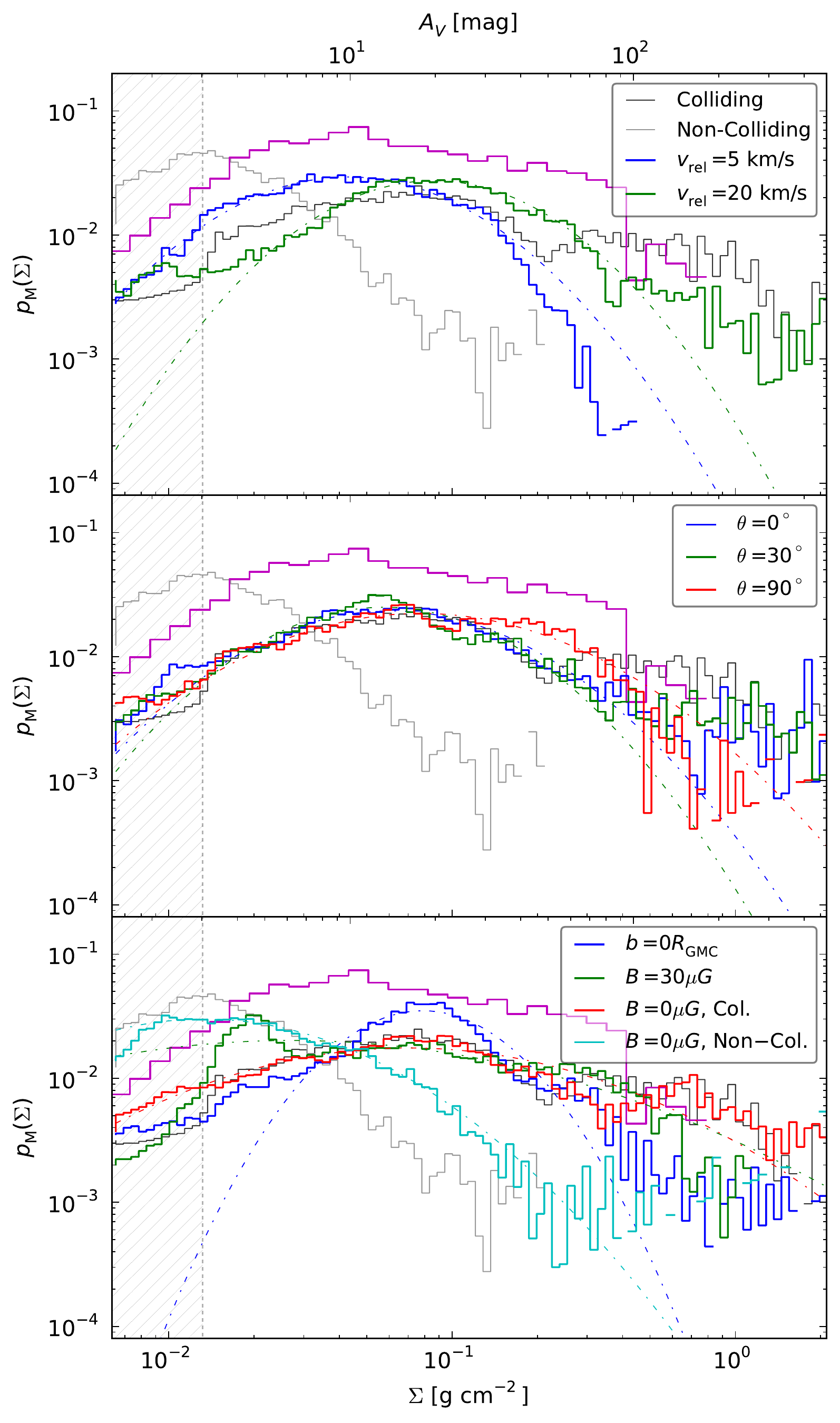}
  \figcaption{
Area-weighted (left column) and mass-weighted (right column)
$\Sigma$-PDFs of $(32\:{\rm pc})^3$ regions from the parameter models
for each category of (top) $v_{\rm rel}$, (middle) $\theta$, and
(bottom) $|B|$ and $b$. $\Sigma$-PDFs for each case at $t= 4\:{\rm Myr}$ are
shown 
The fiducial colliding and non-colliding cases at $t= 4\:{\rm Myr}$
are plotted in dark and light gray, respectively, for reference in
each figure. The best log-normal fits for each case are plotted as
dash-dotted lines of the same color. In each panel, the $\Sigma$-PDFs 
from observations of a massive IRDC from 
\citet{Lim_ea_2016arXiv} is shown in magenta and 
the shaded region denotes areas of $A_{V}<3\:{\rm mag}$, matching the
completeness levels the observed IRDCs.
\label{fig:PDFsparams}}
\end{figure*}

In Figure~\ref{fig:PDFsparams}, we calculate area and mass-weighted
$\Sigma$-PDFs for the parameter models at $t=4\:{\rm Myr}$. 
The regions are centered at the position of maximum $\rho$ in each
case. The figures are organized by models comparing collision velocity
(models 3, 4), magnetic field direction (5, 6, 7), and impact
parameter (8) and magnetic field strength (9), respectively. The
fiducial colliding and non-colliding cases are also included for
reference in each figure.

The greatest differences arise from the collision velocity. Higher
values of $v_{\rm rel}$ create greater relative amounts of gas at both
high and low mass surface densities, resulting in increasingly higher
values of $\sigma_{\ln \Sigma_{A}}$ and 
$\sigma_{\ln \Sigma_{M}}$.  $\overline{\Sigma}_{A}$ and
$\overline{\Sigma}_{M}$ also show monotonic increases with collision
velocity.

An inspection of initial magnetic field orientation yields fairly 
similar $\Sigma$-PDFs and corresponding PDF parameters for each case. Thus,
although the variation of $\theta$ leads to quite different density
and temperature morphologies, the resulting $\Sigma$-PDFs are much
less affected.

The variation of $b$ and $|B|$ resulted in insignificant changes to
the $\Sigma$-PDFs, though the unmagnetized colliding case reached the highest
mass surface densities. However, the PDF parameters for each of these
colliding cases are relatively similar. When compared with the
unmagnetized, non-colliding, case, the differences in $\Sigma$-PDFs due to
collision velocity are emphasized further.

\citet{Butler_ea_2014} and \citet{Lim_ea_2016arXiv} have presented the
$\Sigma$-PDF of a 30~pc scale region centered on a massive IRDC that
is embedded in a GMC. Near+mid infrared extinction mapping and sub-mm
dust continuum emission methods have been used to derive the PDF. The
region contains a minimum close contour of $\Sigma=0.013\:{\rm
  g\:cm^{-2}}$ ($A_V=3$~mag), so is expected to be complete for higher
values of $\Sigma$. The area-weighted PDF (weighting by the total area
of those pixels with $A_V\geq3$~mag) is well fit by a single
log-normal with $\overline{\Sigma}_{A} = 0.039\:{\rm g\:cm^{-2}}$ and
$\sigma_{\ln \Sigma, A}=1.4$. 
There is a relatively limited fraction of material at high $\Sigma$'s
in excess of the log-normal, i.e., $\epsilon_{\rm pl}\lesssim
0.1$. These features are quite similar to some of the simulated PDFs,
especially the colliding case at 4 Myr, which is well-fit with
$\overline{\Sigma}_{A} = 0.021\:{\rm g\:cm^{-2}}$ and $\sigma_{\ln
\Sigma, A}=1.1$. The $v_{\rm rel}=20\:{\rm km\:s^{-1}}$, $\theta=0^{\circ}$
and $b=0R_{\rm GMC}$ models also have similar values.
While this does not prove any particular scenario, the colliding cases
in general demonstrate strong consistency with observations.

In studying resolution effects, the $\Sigma$-PDFs are well-converged, 
with histogram noise decreasing as resolution increases and overall 
values of log-normal fit parameters in agreement within a few percent.

\begin{table}
\caption{Properties of $\Sigma$-PDFs}
\label{tab:PDFs}
\centering
\begin{tabular}{ccccc} \hline \hline
name         & $\sigma_{\ln\Sigma,A}$ & $\overline{\Sigma}_{A}$   &  $\sigma_{\ln\Sigma,M}$ & $\overline{\Sigma}_{M}$   \\
      &      & (${\rm g\:cm^{-2}}$)   &  & (${\rm g\:cm^{-2}}$)  \\
\hline

Colliding (1.0 Myr) & 0.200 & 0.017 & 0.215 & 0.018 \\
Colliding (2.0 Myr) & 0.567 & 0.020 & 0.539 & 0.028 \\ 
Colliding (3.0 Myr) & 0.850 & 0.023 & 0.835 & 0.048 \\ 
Colliding (4.0 Myr) & 1.079 & 0.021 & 1.413 & 0.071 \\
Non-Col. (1.0 Myr) & 0.143 & 0.014 & 0.142 & 0.014 \\
Non-Col. (2.0 Myr) & 0.503 & 0.009 & 0.691 & 0.008 \\
Non-Col. (3.0 Myr) & 0.586 & 0.013 & 0.545 & 0.018 \\
Non-Col. (4.0 Myr) & 0.600 & 0.015 & 0.691 & 0.020 \\

\hline

$v_{\rm rel}=5$ km/s & 1.004 & 0.016 & 0.876 & 0.043 \\
$v_{\rm rel}=20$ km/s & 0.986 & 0.038 & 0.818 & 0.086 \\
$\theta=0^{\circ}$ & 1.045 & 0.022 & 0.969 & 0.061 \\
$\theta=30^{\circ}$ & 0.893 & 0.027 & 0.883 & 0.058 \\
$\theta=90^{\circ}$ & 1.255 & 0.017 & 1.123 & 0.077 \\
$b=0R_{\rm GMC}$ & 0.982 & 0.038 & 0.610 & 0.079 \\
$B=30\mu G$ & 0.407 & 0.019 & 1.868 & 0.027 \\
$B=0\mu G$, Col. & 1.317 & 0.014 & 1.423 & 0.070 \\
$B=0\mu G$, Non-Col. & 1.007 & 0.005 & 1.119 & 0.013 \\

\hline
\hline
\end{tabular}
\end{table}

\subsection{Integrated Intensity Maps}
\label{sec:results-intensity}

From the PDR-based heating and cooling functions, we extract 
$^{12}$CO and $^{13}$CO molecular line cooling information to create
self-consistent synthetic integrated intensity maps via
post-processing. $^{12}$CO and $^{13}$CO line emissivities at
different $J$ levels are affected to various extents by density and
temperature. Generally, we expect the lower-$J$ CO lines to act as a
tracer of the bulk of the molecular gas, while higher-$J$ lines probe
higher temperature, denser gas. These mid- to high-$J$ CO lines are
often signatures of shocked regions and have been studied in GMCs and
IRDCs \citep{Pon_ea_2015}. The general strength of the shock can be
followed with increasing values of $J$.

Paper I found the $^{12}$CO($J$=8-7)/$^{13}$CO($J$=2-1) line intensity
ratio to be a good tracer of cloud collisions due to the strong shocks
created in colliding cases but not in isolated scenarios.

Using similar methods as Paper I, we assume a fiducial distance to the
GMCs of $d=3$kpc. From this, we determine flux contributions from
each cell in the simulation and calculate integrated intensities using
\begin{equation}
I = \int{I_{\nu}d\nu} = \frac{2k}{\lambda^{2}}\int{T_{\rm mb}d\nu}.
\end{equation}
where $I_{\nu}$ is the specific intensity, $\lambda$ is the line 
wavelength, and $T_{\rm mb}$ is the main beam temperature.

To calculate the temperature contribution of the cells, we use
\begin{equation}
\int{T_{\rm mb}d\nu}=\frac{\lambda^{3}}{2k}I=\frac{\lambda^{3}jV}{8\pi kd^{2}\Omega}.
\end{equation}
where $j$ is the volume emissivity, $V$ the cell volume, and $\Omega$
the solid angle subtended by the cell.

\begin{figure*}
  \centering
  \includegraphics[width=2\columnwidth]{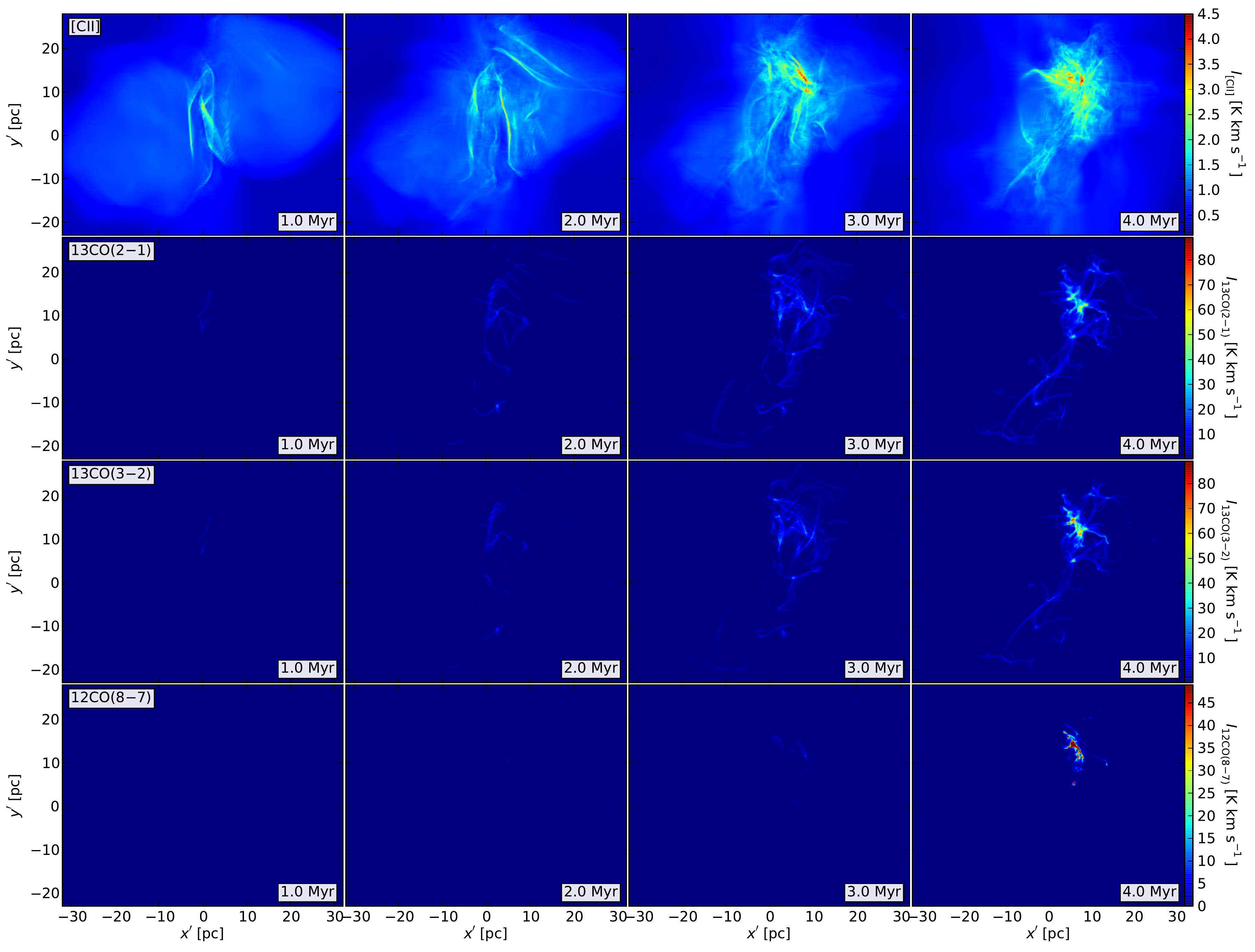}
  \figcaption{
Time evolution of the fiducial colliding GMC model, simulating various
line emissivities at 1.0, 2.0, 3.0, and 4.0 Myr. Integrated intensity
maps derived from the PDR-based cooling functions: 
Row 1: [CII].  Row 2: $^{13}$CO($J$=2-1).  Row 3: $^{13}$CO($J$=3-2).
Row 4: $^{12}$CO($J$=8-7).
  \label{fig:diagnostics_colliding}}
\end{figure*}

\begin{figure*}
  \centering
  \includegraphics[width=2\columnwidth]{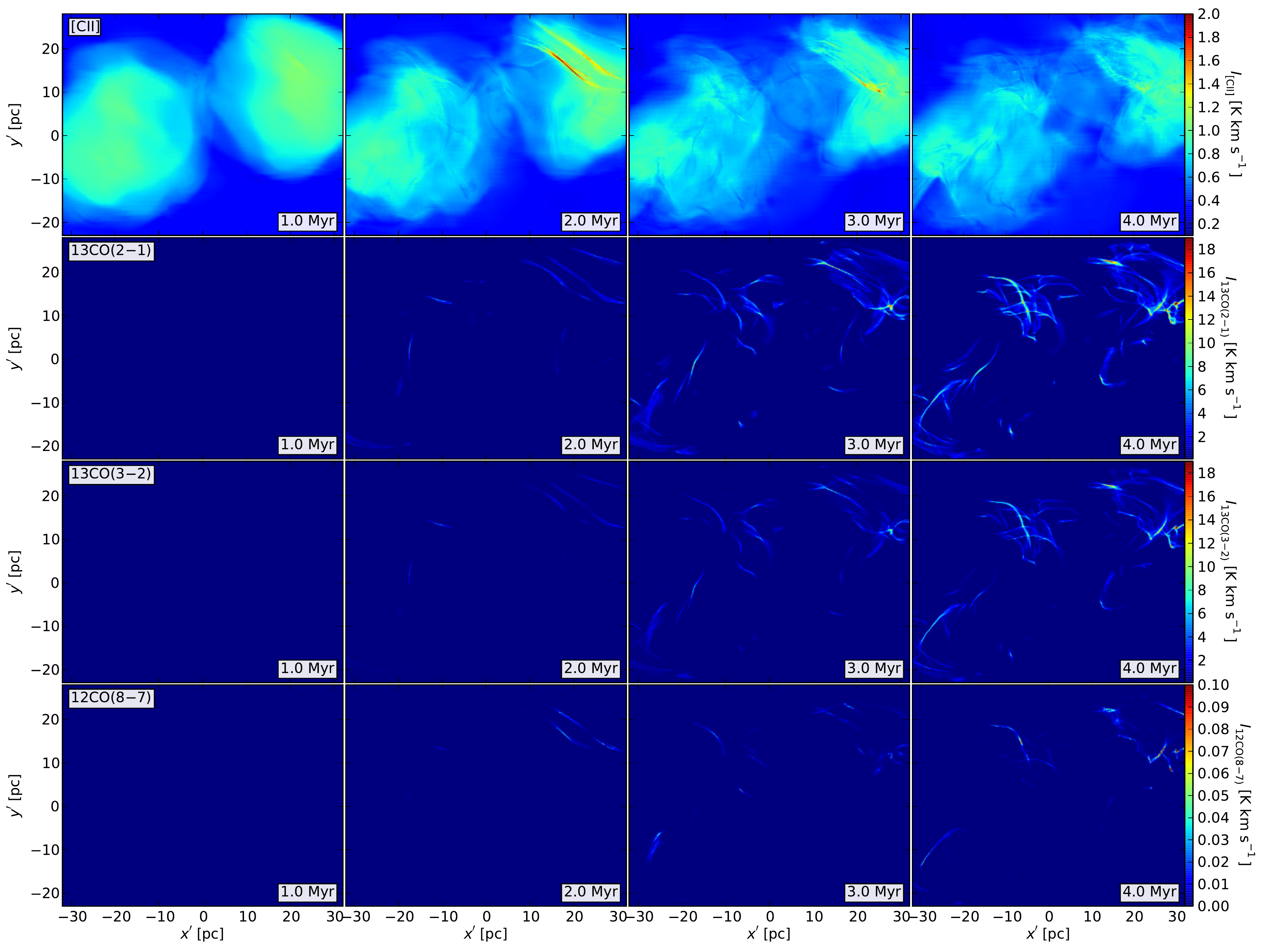}
  \figcaption{
Same as Fig.~\ref{fig:diagnostics_colliding} except for the
non-colliding case, simulating various line emissivities at 1.0, 2.0,
3.0, and 4.0 Myr. Integrated intensity maps derived from the PDR-based
cooling functions: Row 1: [CII].  Row 2: $^{13}$CO($J$=2-1).  Row 3:
$^{13}$CO($J$=3-2).  Row 4: $^{12}$CO($J$=8-7).
  \label{fig:diagnostics_noncolliding}}
\end{figure*}

Figures~\ref{fig:diagnostics_colliding} and
~\ref{fig:diagnostics_noncolliding} show the time-evolution of maps of
[CII], $^{13}$CO($J$=2-1), $^{13}$CO($J$=3-2), and $^{12}$CO($J$=8-7)
integrated intensity for the fiducial colliding and non-colliding 
cases, respectively. 

[CII] acts as a probe for the lower density, PDR gas enveloping GMCs.
This region contains gas transitioning to the molecular phase and
joining the GMC material. Our synthetic maps of [CII] show emission 
in extended regions surrounding the denser gas. The colliding case
exhibits higher [CII] intensities, but over a smaller volume
concentrated about the converging flows. The original GMCs show
$I_{\rm [CII]}\sim 1\:{\rm K~km/s}$, with subsequent evolution
reaching up to $\sim 4\:{\rm K~km/s}$. The non-colliding case 
remains at $\sim 1\:{\rm K~km/s}$ throughout the evolution, keeping 
a fairly consistent distribution. The emission is extended and 
encompasses the denser molecular gas.

$^{13}$CO($J$=2-1),
$^{13}$CO($J$=3-2) are seen to be good tracers of cold, dense gas. As
both colliding and non-colliding clouds evolve, dense filaments form
and become traceable by these low-$J$ CO lines. Noting the differences
in integrated intensity scales between the two models, the densities
in the colliding case reach significantly higher levels at earlier
times compared to the non-colliding case and can be traced through
CO. The morphologies of the structures differ, as one primary dense
filamentary region can be seen being formed at the interface of the
colliding flows, while distinct, distributed filaments are formed for
the non-colliding case. The primary filamentary structure in the
colliding case exhibits dense clumps reaching $I\approx 80\:{\rm
  K~km/s}$, while the separate filaments evolving in the non-colliding
case reach values of $I\approx 20\:{\rm K~km/s}$ for both
$^{13}$CO($J$=2-1) and $^{13}$CO($J$=3-2).

Stark differences, however, can be seen in $^{12}$CO($J$=8-7), where
high intensities are produced later in the evolution of the fiducial
colliding case, as the dense filaments in both GMCs collide and
merge. These begin to become visible at $t\approx3\:{\rm Myr}$ and
reach levels of $\sim 10^{3}\:{\rm K~km/s}$. In the non-colliding
case, there is almost no emission at this rotational level, indicating
a lack of strong shocks. By the $t\approx4\:{\rm Myr}$ mark, only
$I_{\rm 12CO(J=8-7)}\sim 0.2\:{\rm K~km/s}$ can be detected.

\subsection{Kinematics}
\label{sec:results-kinematics}

Synthetic spectra were created to gain more quantitative comparisons
between the various emission lines as well as to understand the
kinematics of the models. Additionally, line-of-sight velocity spectra
can be directly compared to those measured from observed clouds.

The majority of cloud collision candidates have relied primarily on
multiple velocity components deduced from CO spectra in conjunction
with coherent density structures and/or young stars as evidence for
detection. The current study offers a unique method of directly
reproducing various CO spectra for clouds undergoing collisions and
comparing them with non-colliding scenarios.

\begin{figure*}
  \centering
  \includegraphics[width=2\columnwidth]{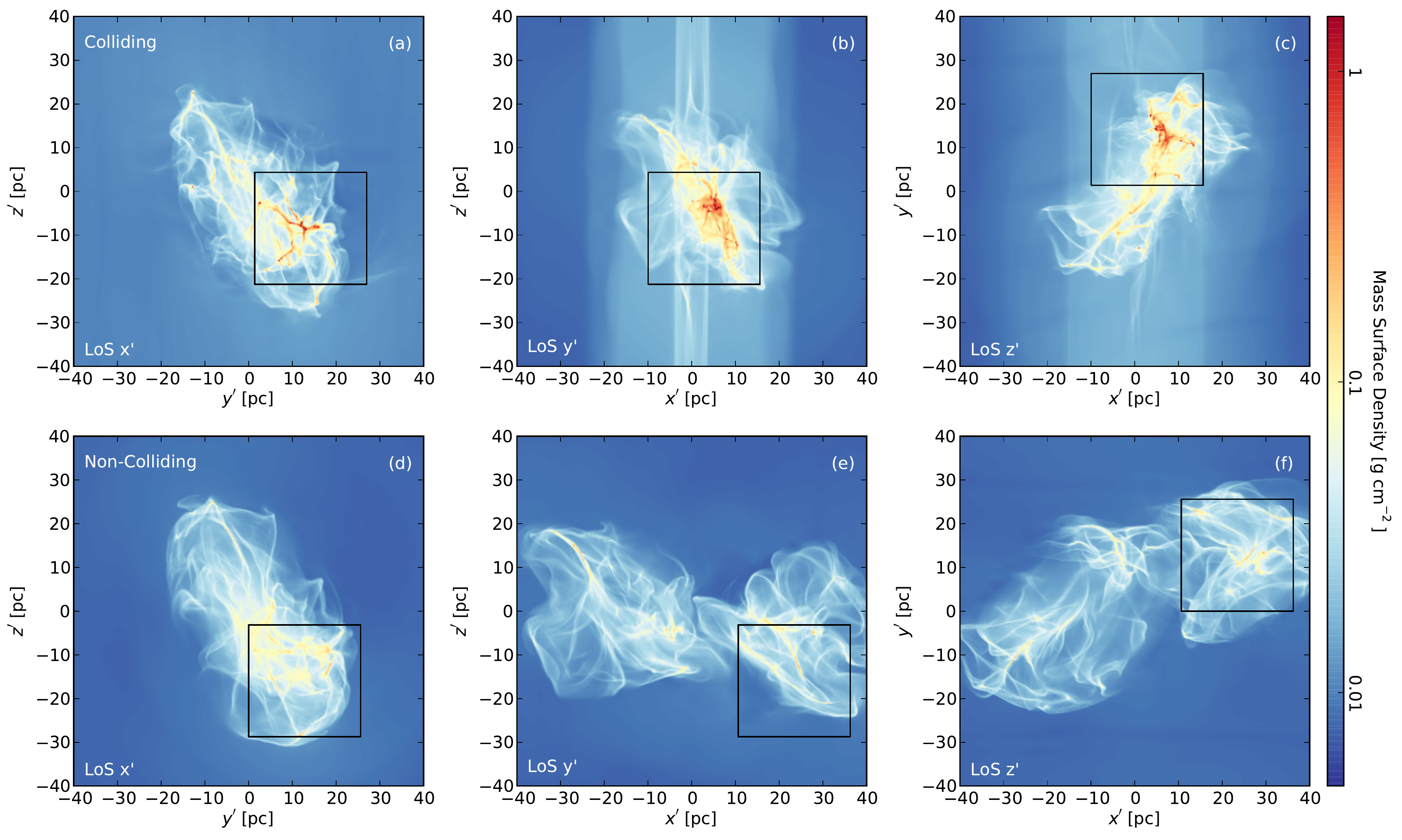}
  \includegraphics[width=2\columnwidth]{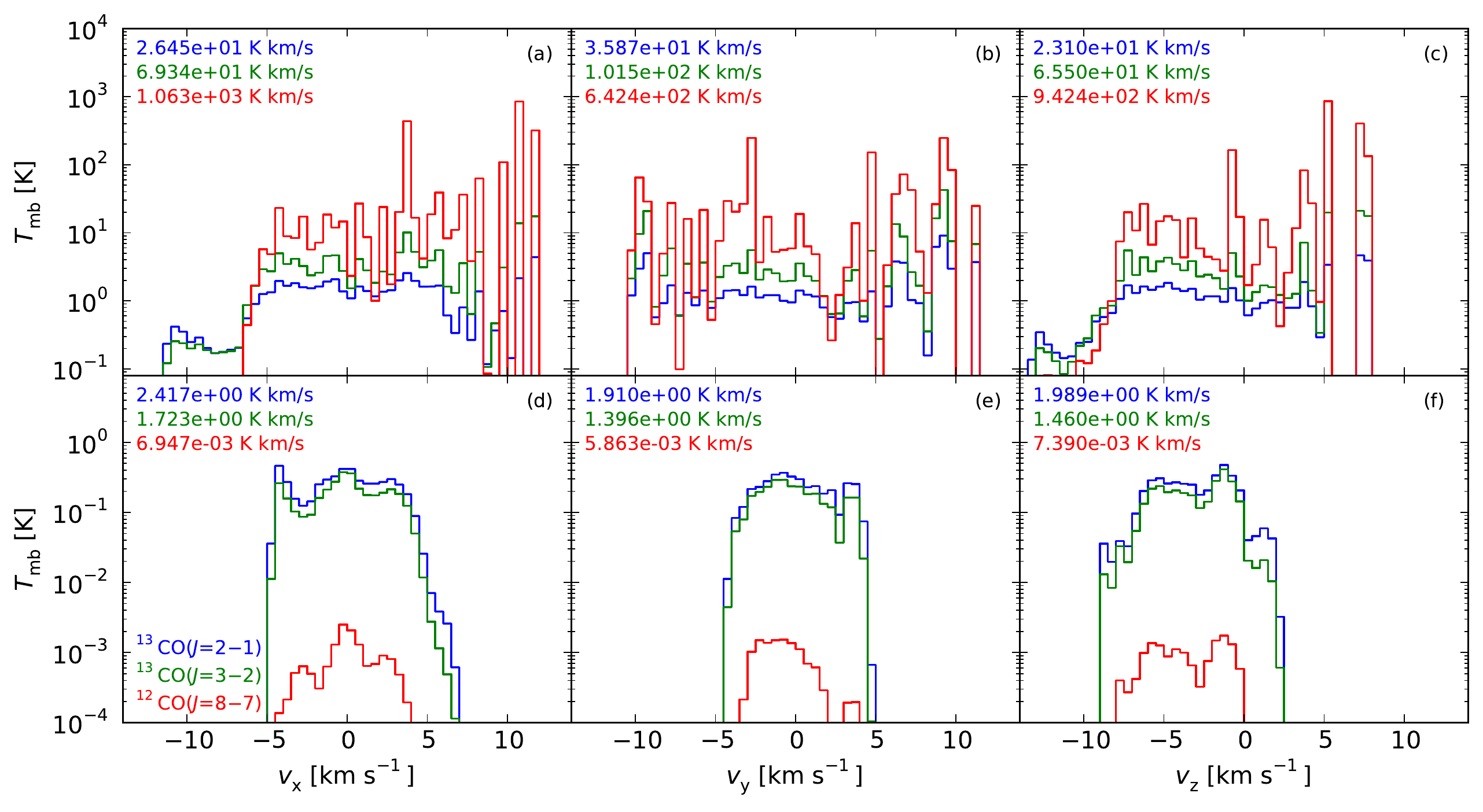}
  \figcaption{
Top: Mass surface density maps are shown for various lines of sight
directed through the fiducial colliding case (top row:(a)-(c)) and
non-colliding case (bottom row:(d)-(f)). The black boxes bound equal-volume 
regions containing the primary filament in each simulation at $t$=4.0 Myr. 
Bottom: Synthetic velocity spectra for $^{13}$CO(2-1), $^{13}$CO(3-2),
$^{12}$CO(8-7) from the respective selected regions shown in the upper
figures. Note large differences in $^{12}$CO(8-7) integrated
intensities relative to $^{13}$CO(2-1) and $^{13}$CO(3-2) between the
colliding and non-colliding models.
  \label{fig:spectra}}
\end{figure*}

\subsubsection{Spectra}

In Figure~\ref{fig:spectra}, we have created spectra of
$^{13}$CO($J$=2-1), $^{13}$CO($J$=3-2), and $^{12}$CO($J$=8-7) (same
as the integrated intensity maps) through square patches of area
$(25.6\:{\rm pc})^{2}$ projected through the x, y, and z lines of
sight for both the fiducial colliding and non-colliding cases. Each
spectrum corresponds to the respective mass surface density map, on
which the projected patch is indicated. The patches center on the
highest mass surface density regions for both cases.

The first main difference seen in the spectra between the colliding
and non-colliding cases is the width of the velocity ranges. The
non-colliding case exhibits fairly narrow ($\Delta v \lesssim 10\:{\rm
  km/s}$) velocity widths for each line of sight. The colliding case,
on the other hand, has broader ($\Delta v \sim 15-20\:{\rm km/s}$)
velocity widths and what may be interpreted as multiple components, at
least for the $v_{x}$ and $v_{y}$ directions, but to a lesser extent
$v_{z}$ as well.

Another key result is the relative strength of the various CO
lines. Throughout each of the non-colliding lines of sight, the
strength of the integrated intensity follows the trend of
\begin{equation}
I_{\rm 13CO(2-1)}>I_{\rm 13CO(3-2)}>I_{\rm 12CO(8-7)}
\end{equation}
The magnitudes are of the order $\sim 2$, $\sim 1$, and $\sim
2\times10^{-3}$ K km/s, respectively. For the colliding case, the
exact opposite trend is seen:
\begin{equation}
I_{\rm 13CO(2-1)}<I_{\rm 13CO(3-2)}<I_{\rm 12CO(8-7)}
\end{equation}
with intensities of order $\sim 10$, $\sim 30$, and $\sim
4\times10^{3}$ K km/s, respectively. Thus, the
$^{12}$CO($J$=8-7)/$^{13}$CO($J$=2-1) line intensity ratio is $\sim
10^{-3}$ for the non-colliding case and $\sim 10^{2}$ to $10^{3}$ for
the colliding case.

As a result, measurement of CO spectra, especially the
$^{12}$CO($J$=8-7)/$^{13}$CO($J$=2-1) line intensity ratio, is another
potentially strong diagnostic of cloud collisions. From our models,
both the velocity range and especially the values of integrated
intensities are differentiators between colliding and non-colliding
GMCs and both appear to be generally independent of line of sight.

\subsubsection{Velocity Gradients}

We can determine velocity dispersions and gradients of dense
structures within our simulations using synthetic $^{13}$CO($J$=1-0)
line intensity maps and $p-v$ diagrams. Our goal is to use similar
methods in determining these quantities as those used observationally
in order to directly compare with GMCs and IRDCs in the Galaxy \citep[see e.g.,][hereafter HT15]{Hernandez_Tan_2015}.

\begin{figure*}
  \centering
  \includegraphics[width=1\columnwidth]{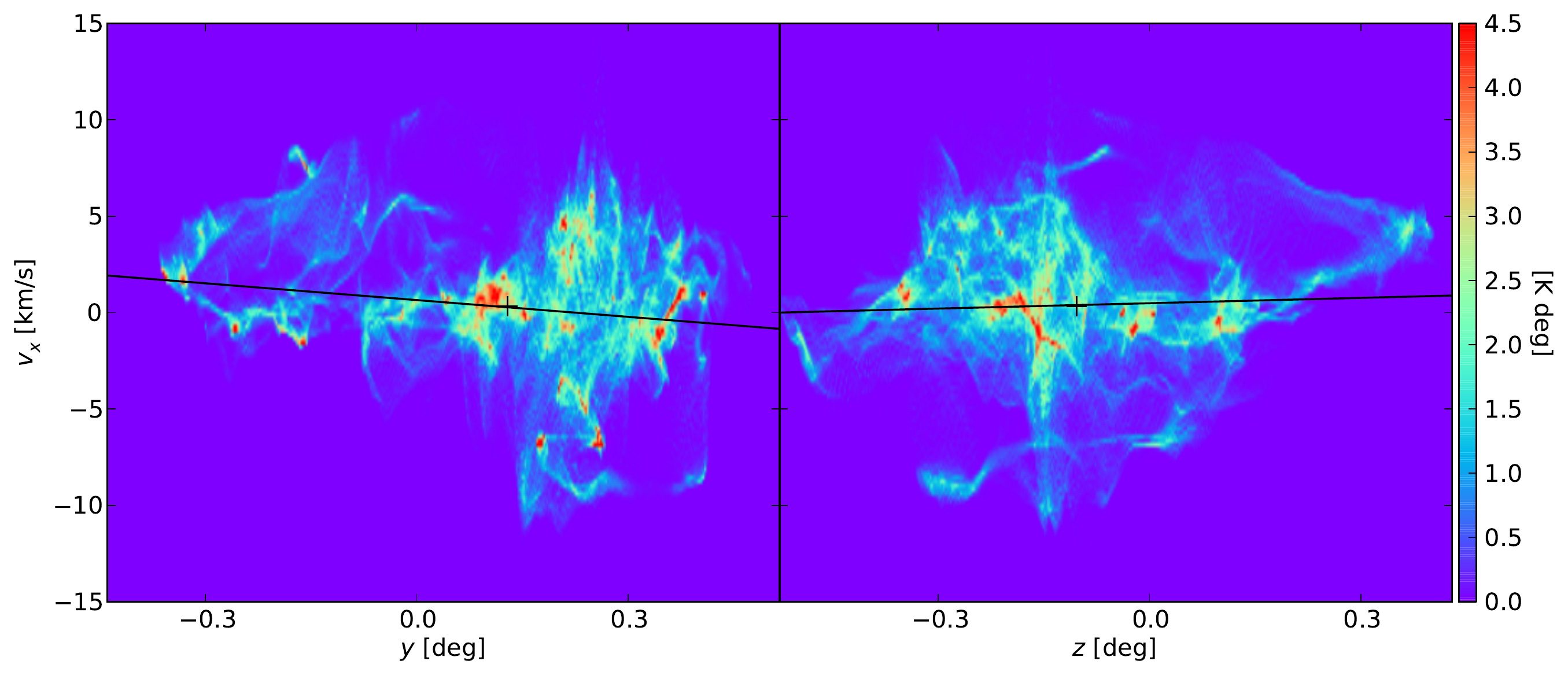}
  \includegraphics[width=1\columnwidth]{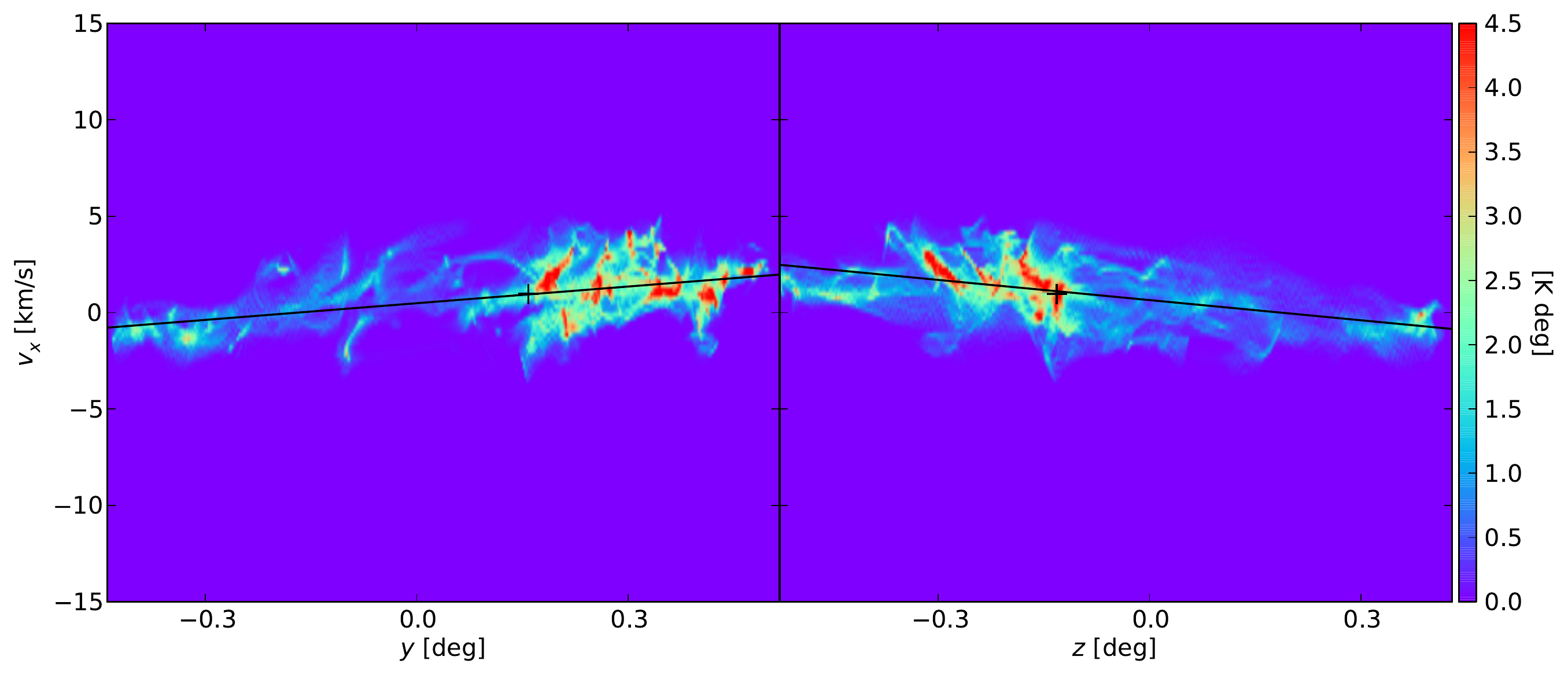}
  \includegraphics[width=1\columnwidth]{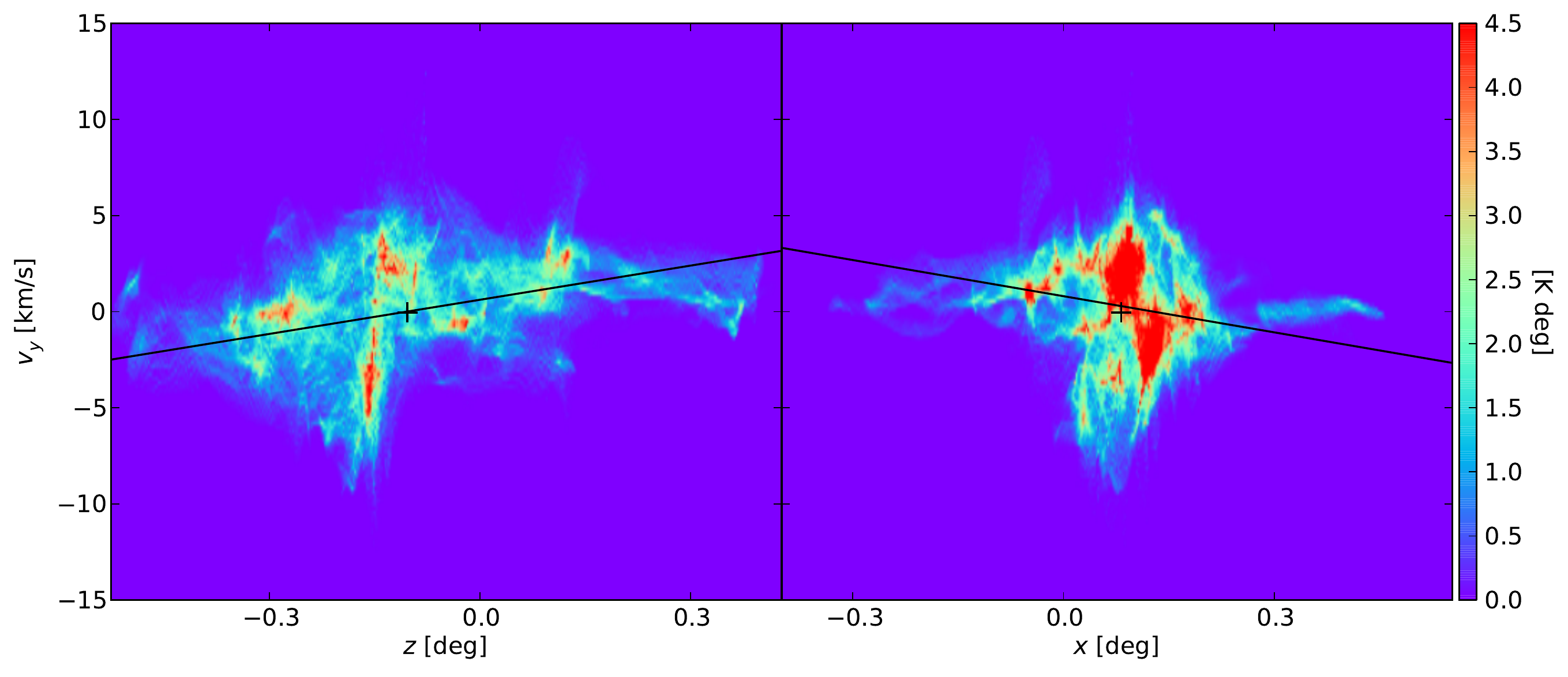}
  \includegraphics[width=1\columnwidth]{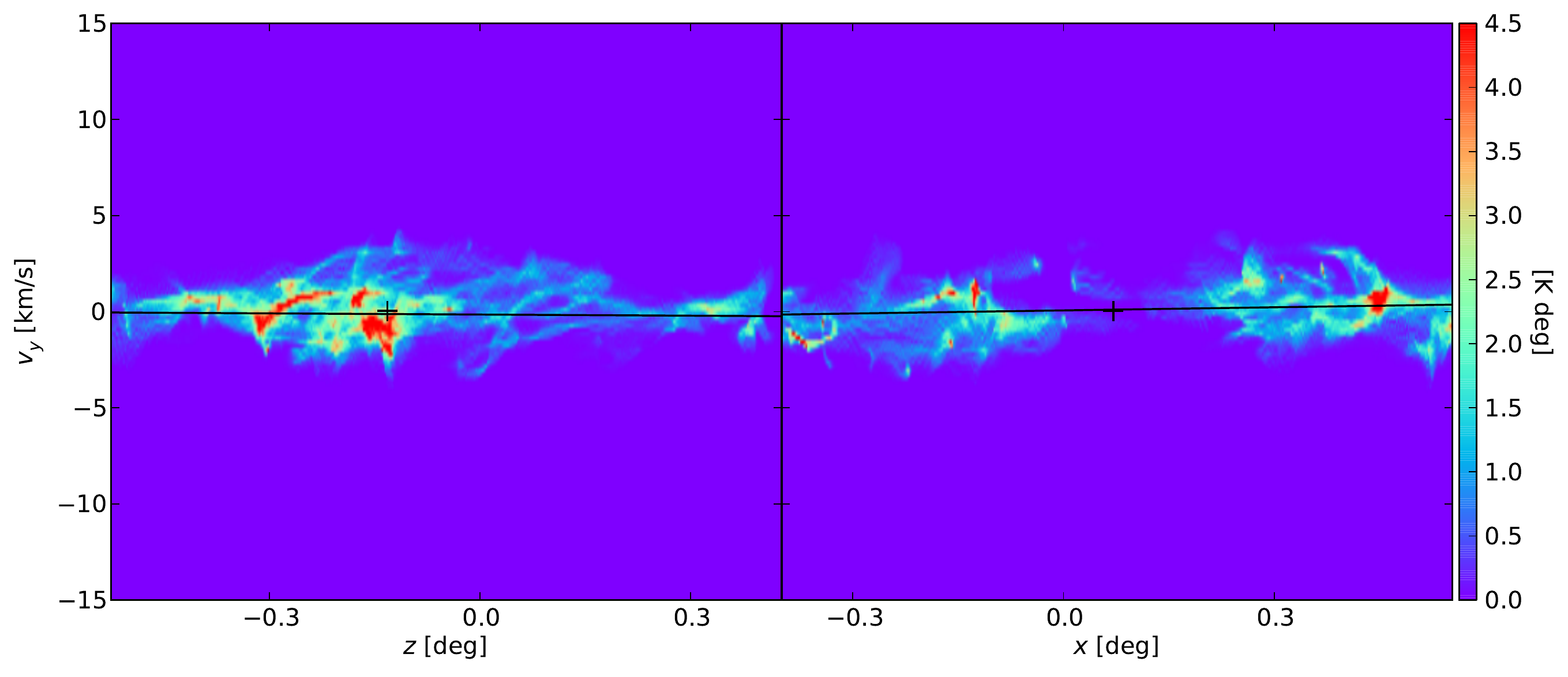}
  \includegraphics[width=1\columnwidth]{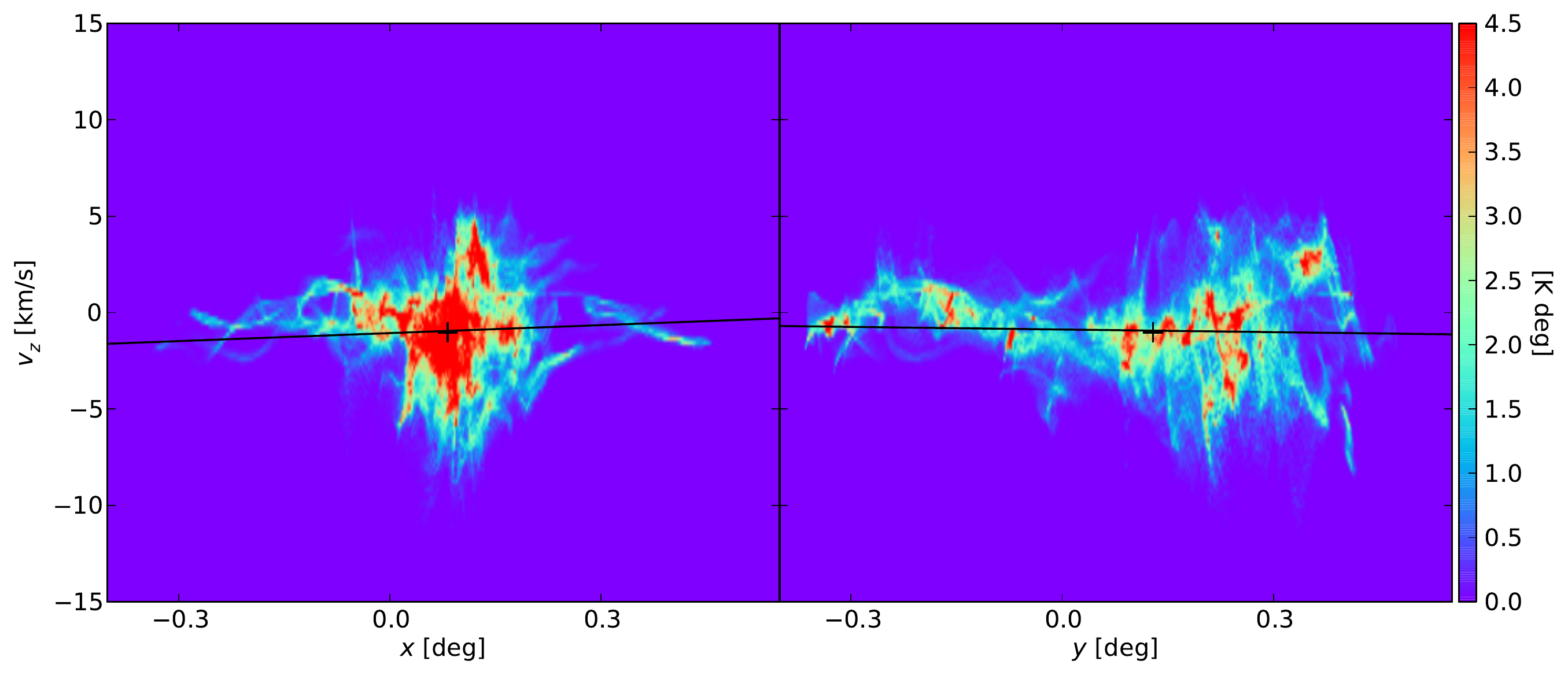}
  \includegraphics[width=1\columnwidth]{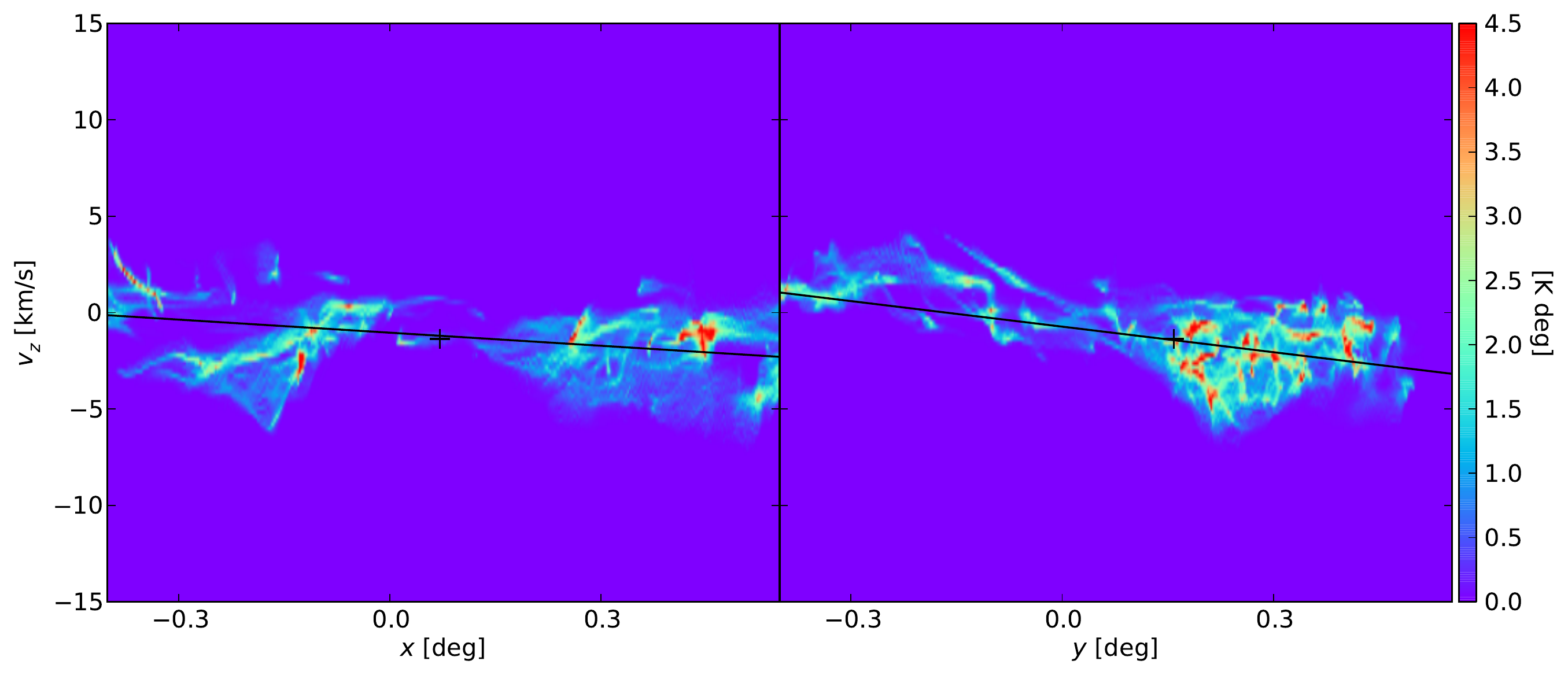}
  \figcaption{
Position-velocity diagrams for the fiducial colliding case (left column)
and non-colliding case (right column). The scaling is derived from
synthetic $^{13}$CO($J$=1-0) line intensities through velocity bins of 
$\Delta v=0.212\:{\rm km\:s^{-1}}$. The black cross indicates the position
of the center of mass and the solid black line shows the intensity-weighted 
linear velocity gradient ($dv_{\rm los}/ds$) across each cloud.
  \label{fig:ppv}}
\end{figure*}

We investigate the fiducial colliding and non-colliding cases 
transformed from our 3D spatial data to $p$-$p$-$v$-space for each of 
the $x$,$y$,and $z$ lines of sight (see Fig.~\ref{fig:ppv}).
The velocity dispersion was defined using the intensity-weighted 
rms 1D velocity dispersion of the corresponding region.
Velocity gradients were calculated along each spatial direction for 
coordinate axes orthogonal to the chosen line of sight (e.g., for 
$\frac{dv_{z}}{dx}$, the best linear fit was determined through each 
intensity-weighted cell in $(x,v_{z})$ space). 
Table~\ref{tab:velgrad} summarizes the velocity information for the 
fiducial colliding and non-colliding cases, for each line of sight.

\begin{table}
\caption{Velocity Gradients}
\label{tab:velgrad}
\begin{center}
\begin{tabular}{lccc}
\hline\hline
Case           & LoS  & $\sigma$ (dens)           & $\frac{dv_{\rm los}}{ds}$     \\
               &      & ($\rm{km\:s^{-1}}$) & ($\rm{km\:s^{-1}\:pc^{-1}}$)  \\
\hline
Colliding      & $x$  & 3.6588  & 0.0581 \\
               & $y$  & 2.7611  & 0.1648 \\
               & $z$  & 2.4760  & 0.0278 \\
Non-Colliding  & $x$  & 1.4926  & 0.0864 \\
               & $y$  & 1.1996  & 0.0110 \\
               & $z$  & 1.9649  & 0.0948 \\
\hline
\end{tabular}
\end{center}
\end{table}

Strong differences between the two models are revealed through the 
velocity dispersion, with the colliding case showing indications of 
much greater dispersion. The largest velocity dispersion of 
$\sim 3.7$~km/s is seen along the collision axis, while the orthogonal 
directions also experience greater dispersion relative to the 
non-colliding case. The RMS velocity dispersion over the 3 lines of 
sight is 3.01~km/s for the colliding case and 1.58~km/s for the 
non-colliding case.

The velocity gradients reveal differences as well. The largest
velocity gradient occurs when looking in the direction of the
collisional impact parameter, at $\sim 0.16$~km/s/pc. However, the
gradients along the remaining directions are similar in magnitude 
and even somewhat smaller when compared with the non-colliding case.
The RMS velocity gradient over the 3 lines of sight is 0.1022~km/s/pc
for the colliding case and 0.0743~km/s/pc for the 
non-colliding case.

Overall, the kinematics measured in the fiducial colliding case are in
rough agreement with the ten observed IRDCs and associated GMCs from
HT15, in which velocity dispersions of order $\sim$few km/s and
velocity gradients generally at $\sim$0.1 (but upwards of
$\sim$0.6-0.7) km/s/pc were found,
though these results do not necessarily preclude the non-colliding case. 
However, the kinematics of observed IRDCs, especially those with higher 
measured values of velocity gradient and dispersion, may suggest a more 
dynamic formation scenario with compression of GMC material.

\subsection{Dynamics}
\label{sec:results-dynamics}

\begin{figure*}
  \centering
  \includegraphics[width=2\columnwidth]{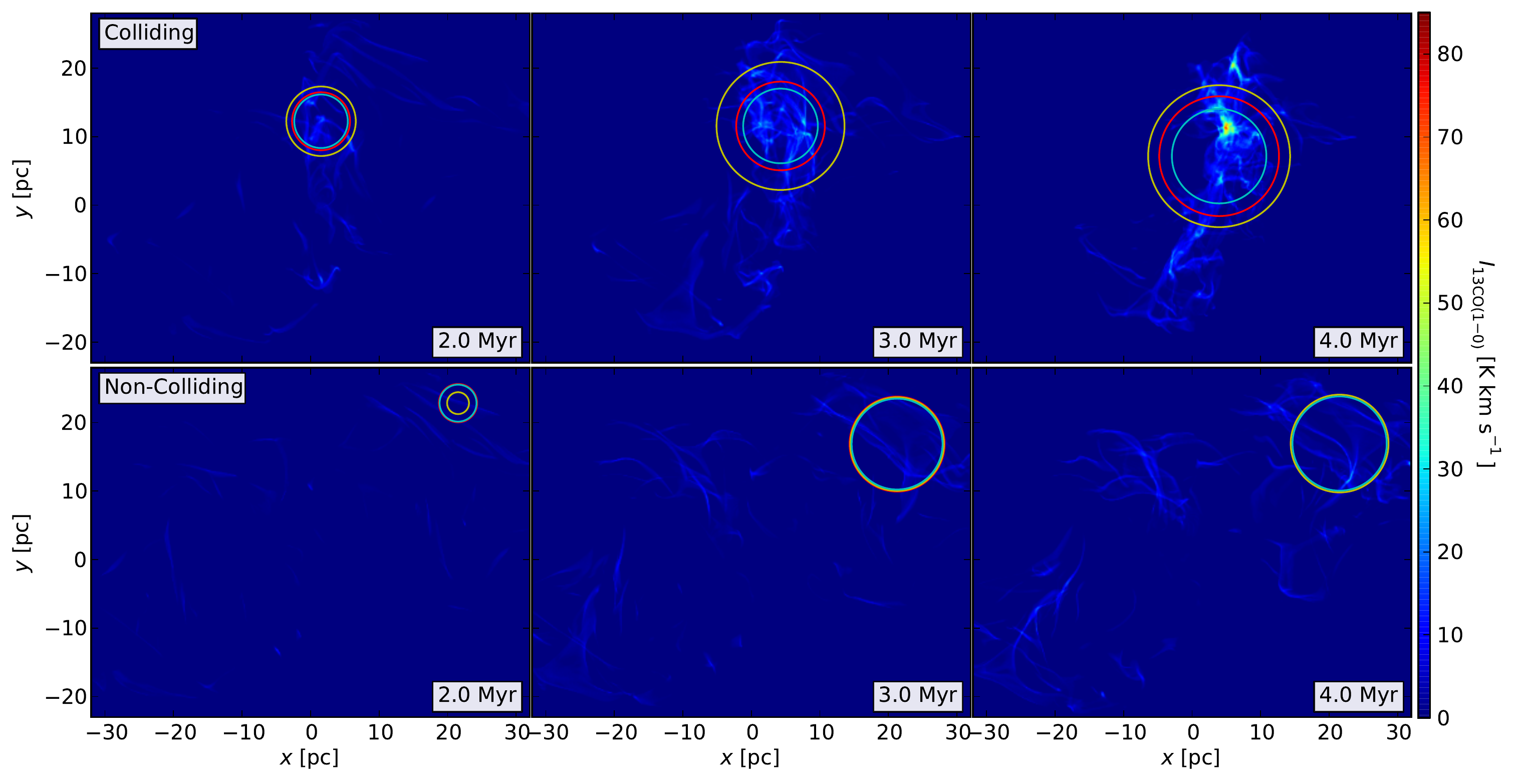}
  \figcaption{
Time evolution maps of $^{13}$CO($J$=1-0) integrated intensity for the
fiducial colliding case at 2.0, 3.0, and 4.0 Myr. The different effective
radii calculated for the virial analysis are plotted as colored circles 
with radii defined by $R_{\rm M}$ (blue), $R_{\rm A}$ (green), and 
$R_{1/2}$ (red).
  \label{fig:virial1}}
\end{figure*}

\begin{figure*}
  \centering
  \includegraphics[width=2\columnwidth]{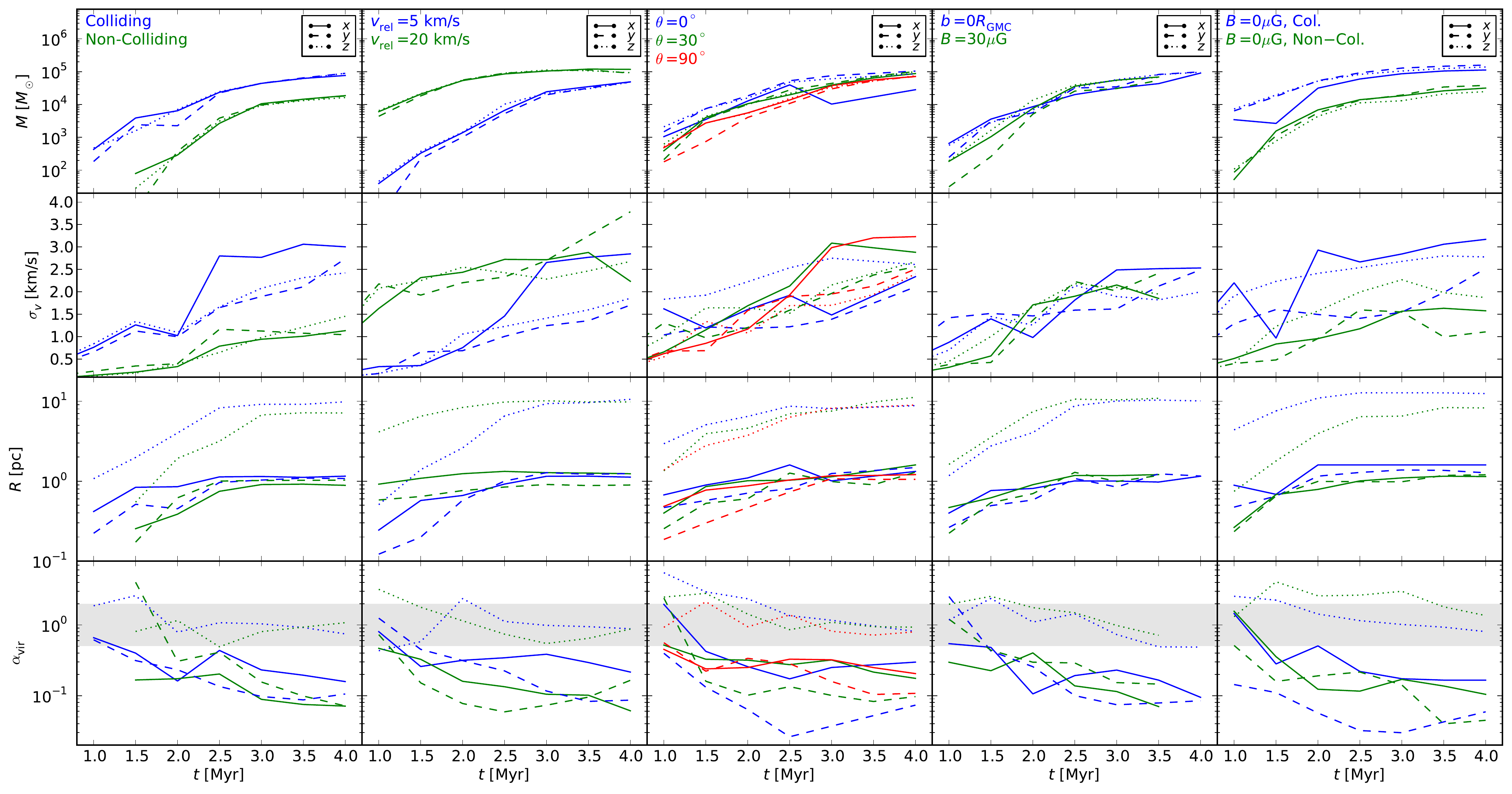}
  \figcaption{
From top to bottom rows: Time evolution of the total mass, velocity 
dispersion, virial radii as defined by the areal radius $R_{\rm A}$,
and the corresponding virial
parameter. 
The columns compare the various models, as indicated by the respective
color and label. The three primary lines of sight are also
investigated for each model, denoted as solid ($x$), dashed ($y$), 
and dotted ($z$) lines.
The shaded region is centered on $\alpha_{\rm vir}=1$ with a factor of 
2 to each side, roughly the range seen by HT15. 
  \label{fig:virial2}}
\end{figure*}

Virial analysis of clouds compares the relative importance of
self-gravity with internal motions and can reveal dynamical properties
of the material and, in turn, provide evidence for recent kinematic
history.  HT15 performed virial analysis on ten observed IRDCs and
associated GMCs based on $^{13}$CO($J$=1-0) emission and found that
IRDCs have moderately enhanced velocity dispersions and virial
parameters relative to GMCs, potentially indicating more disturbed
kinematics of the densest gas. If GMC collisions indeed trigger the
formation of IRDCs and then star clusters, virial analysis may be
another important diagnostic for the products of cloud collisions.

We follow the ``simple extraction (SE)'' and ``connected extraction
(CE)'' techniques detailed in HT15, applying them to our fiducial
colliding and non-colliding models. First, we calculate the cloud
center of mass in $p$-$p$-$v$-space based on $^{13}$CO($J$=1-0)
intensity. SE selects all voxels with $^{13}$CO(1-0) emission out to
radii of $R$=5,10,20, and 30~pc and within a $v_{0}\pm 15\:{\rm
  km\:s^{-1}}$ line-of-sight velocity range.  CE, on the other hand,
selects voxels directly connected face-wise in $p$-$p$-$v$-space. Each
must exceed the same $^{13}$CO(1-0) intensity threshold of 
$T_{B,v}\geq 1.35$~K as in HT15, i.e., the Galactic Ring Survey (GRS)
\citep{Jackson_ea_2006} $5\sigma_{\rm rms}$ level. The
connected voxel must also lie within a 30~pc radius and $\pm
15\:{\rm km\:s^{-1}}$ velocity. All connected structures in the 
$p$-$p$-$v$ domain are found via the established graph theory method 
of connected components of undirected graphs, with cells meeting the 
above-mentioned criteria acting as the nodes. The subgraph with the 
largest number of nodes is designated as the connected extraction, and 
further analysis is performed on this subset of voxels. 

For CE, three different radii are calculated, based on various
definitions: the mass-weighted radius ($R_{\rm M}$; the mean projected
radial distance of cloud mass from the center of mass), areal radius
($R_{\rm A}$; from the total projected area $A=\pi
R_{A}^{2}=N_{p}A_{p}$, where $N_{p}$ and $A_{p}$ are the pixel number
and area, respectively, of the defined cloud), and half-mass radius
($R_{\rm 1/2}$; the radius from the center of mass that contains half
the total cloud mass).

To study virialization of the cloud, we use the dimensionless virial
parameter $\alpha_{\rm vir}$ from \citet{Bertoldi_McKee_1992}
\begin{equation}
    \alpha_{\rm vir} = \frac{5 \sigma^{2} R}{GM},
\end{equation}
where $\sigma$ is the mass-averaged line-of-sight velocity dispersion. 

Figure~\ref{fig:virial1} shows the time evolution of the
$^{13}$CO($J$=1-0) integrated intensity maps for the fiducial
colliding and non-colliding cases and the corresponding virial radii. 
For both models, the $^{13}$CO($J$=1-0) structures grow in extent and 
encompass more material, leading to increasing effective radii. A 
central dominant filamentary structure forms in the colliding case, 
whereas the non-colliding case contains a number of smaller, more 
spatially separated filaments. The $^{13}$CO emission is generally
weaker and more dispersed in independent structures in the non-colliding
case. The chosen method for extraction successfully tracks the same 
singlelargest filamentary structure over time as it evolves in both 
cases.

The virial parameter and constituent variables for all models for the 
three $x$, $y$, and $z$ lines of sight are displayed in 
Fig.~\ref{fig:virial2}. These variables within the CE show distinctive
trends over time as well as systematic differences between various
models.

The total mass of the main connected $^{13}$CO-defined structure grows 
steadily over time.
The fiducial colliding case produces structures that grow from 
$10^{3}$ to just under $10^{5}\:M_{\odot}$ over 3 Myr. The non-colliding 
case grows at a similar rate, but generally contains $\sim$10 times less 
mass. The $v_{\rm rel}=20$~km/s model creates higher-mass structures at 
earlier times, but converges to just over $10^{5}\:M_{\odot}$ by
$t=4.0$~Myr. The $v_{\rm rel}=5$~km/s case follows an intermediate growth 
evolution. The $\theta=0^{\circ}$,$30^{\circ}$, and $90^{\circ}$ cases
have similar mass evolution, with slightly smaller masses corresponding 
to increasing values of $\theta$. The total structure mass for the 
$b=0R_{\rm GMC}$ and $B=30~{\rm \mu G}$ cases grow in a similar manner.
The non-magnetized colliding and non-colliding cases follow similar 
evolution as the fiducial colliding and non-colliding cases, respectively,
but do grow to slightly larger masses in general due to the lack of
magnetic pressure support.

The velocity dispersions of the $^{13}$CO emitting structures are found to 
grow throughout the time evolution for all cases, generally starting near  
$\lessapprox1$~km/s and reaching 2-3~km/s. The colliding cases in general 
show distinctly higher velocity dispersions, especially when viewing along 
the collision axis ($x$). Faster collision velocities result in larger
velocity dispersions, while there does not seem to be much dependence
on initial magnetic field direction. Stronger magnetic fields appear to
dampen the collision, resulting in slightly smaller values of $\sigma$.

The measured areal radii generally grow in a similar manner as the mass,
although there is a strong dependence on viewing direction. Specifically,
the $z$-directed line-of-sight, in which the plane-of-sky is sensitive to 
both the collision and impact parameter axes, shows a much greater radius 
in all cases. Along this direction, the radii are measured to increase 
from approximately 1~pc to 10~pc in all cases, though colliding cases in 
general created larger structures by a few pc. The higher velocity 
collisions grow much faster initially, but reach similar final spatial 
extents. Along the other lines of sight, there are similar trends,
although the initial and final radii are approximately a factor of 
10 smaller in these directions.

The general trend for all models is for the virial parameter to decrease 
over time, which appears to be mostly driven by the accumulation of more 
and more mass into the structures. The calculated radii of the structures 
have a strong dependence on viewing direction, as described above, thus
affecting $\alpha_{\rm vir}$ as well. In the $z$ line-of-sight, where
more extended structure is detected, the virial parameter values begin
moderately super-virial but evolve to approach those expected of virial 
equilibrium, i.e., $\alpha_{\rm vir}\sim 1$ (recall $\alpha_{\rm vir}<2$ 
implies a gravitationally bound structure, ignoring surface pressure and 
magnetic pressure effects). For other viewing directions, $\alpha_{\rm vir}$
of the structure is generally smaller, often already sub-virial. 
Systematic differences in $\alpha_{\rm vir}$ between models are less
distinct than from viewing direction, with virial parameters
decreasing by factors of a few over time. Despite the small
differences, the smallest values of $\alpha_{\rm vir}$ are present in
the non-magnetized cases. Overall, some of these structures them may
be undergoing rapid global collapse, but more likely in the magnetized
cases the $B$-fields are providing support that may keep them closer
to virial equilbrium.
We expect that: (1) the structures will continue to accumulate mass and 
become even more gravitationally bound; (2) they are likely to contain 
highly gravitationally unstable substructures, e.g., the dense filaments 
and clumps that appear from 3 to 4 Myr in the fiducial colliding case.

Results from the 10 IRDCs/GMCs studied in HT15 show relatively large 
variation of derived virial parameter depending on the analysis method: 
in particular, the most relevant method for comparison with our analysis 
is “CE,$\tau$”, i.e., connected extraction of a structure where an optical 
depth correction has been assessed, and where the velocity dispersion is 
measured directly from the second moment of the spectrum. 
This method finds values of $\alpha_{\rm vir}\sim 1$, but with
significant dispersion of about a factor of two. Still these values
are somewhat larger than those seen in most of our simulations at
$t\sim4$~Myr. In the context of the GMC-GMC collision scenario, this
may indicate that the relevant timescale for comparison is at earlier
times, e.g., $t\sim1$ to 2~Myr, or that the typical line of sight to
GMCs is in a direction that includes a significant component of the
collision velocity axis (which is likely for collisions mediated by
shear in the Galactic disk).

While the values of $\alpha_{\rm vir}$ are similar between all of the
simulations, ranging from slightly to strongly gravitationally bound 
objects, the total masses and velocity dispersion are notably larger for 
the colliding cases.
Thus we conclude that, in comparison to the $^{13}$CO emitting 
structures formed in non-colliding simulations, those formed via GMC 
collisions are more likely to lead to the conditions necessary for massive 
star cluster formation.

\section{Discussion and Conclusions}
\label{sec:conclusion}

We have investigated physical properties associated with and potential
observational signatures of magnetized, turbulent GMCs collisions. Our
method has utilized PDR-based heating and cooling functions, developed
in our previous study with 2D simulations, to allow our new 3D
simulations, with resolution of 0.125~pc, to follow the multi-phase,
non-equilibrium, thermal evolution of the clouds, including their
shock structures. We have explored the parameter space of GMC
collisions, including the effects of collision velocity, impact
parameter, magnetic field orientation and strength. We have also
carried out detailed comparisons of the results of otherwise identical
colliding and non-colliding clouds.

We have found that the relative orientations between magnetic fields
and mass surface density structures may be used to diagnose a cloud
collision.  HROs and subsequent histogram shape parameter analysis
reveal distinguishing behavior resulting from cloud collisions
compared with non-colliding clouds. In particular, the collision
velocity appears to have a strong effect on the HRO shape parameter.
The dependence on line of sight is fairly low, strengthening the
$\psi$ vs. $\Sigma$ diagnostic.

The $|B|$ vs. $n_{\rm H}$ relation found in our models reveals somewhat
stronger magnetic field strength when compared to the ``Crutcher 
relation'', although the general trend appears to follow $B_{\rm max}
\propto (n_{\rm H})^{2/3}$ at higher densities while staying near roughly 
constant $|B|$ at lower densities. This behavior is likely sensitive
to our choices of initial conditions, but may be representative of 
regions of slightly higher mean field strength compared to the
relatively nearby objects which comprise the ``Crutcher 
relation''. 

Area and mass-weighted $\Sigma$-PDFs show large differences among our
models, with strong distinguishing factors between colliding and
non-colliding cases. Although it is just a single case, a comparison 
with the $\Sigma$-PDF of an observed IRDC finds that the evolved GMC 
collision cases have more similar $\Sigma$-PDFs than the results of 
non-colliding simulations.

Intensity mapping of CO spectra, especially the
$^{12}$CO($J$=8-7)/$^{13}$CO($J$=2-1) line intensity ratio, is another
potentially strong diagnostic of cloud collisions. From synthetic
spectra of our models, the integrated intensities, as well as the
velocity spread, are differentiators between colliding and
non-colliding GMCs and both appear to be generally independent of line
of sight orientation.

Kinematically, the velocity dispersion of the colliding case was found
to be much higher than that of the non-colliding case, at almost a
factor of 2 higher, reaching $\sigma>3.5\:{\rm km\:s^{-1}}$ when
measured along the collision axis. Velocity gradients are also
enhanced due to collisions, with the highest values in the colliding
case measured when viewing orientation is along the same direction that 
the clouds are offset via the impact parameter, 
at $dv_{\rm los}/ds=0.20\:{\rm km\:s^{-1}\:pc^{-1}}$.

Finally, study of the $^{13}$CO-defined structures formed in the
colliding and non-colliding scenarios are quite different. In all of
the colliding cases, they are much more massive with generally larger
velocity dispersion. Both colliding and non-colliding cases are
gravitationally bound. This suggests a potential role for GMC
collisions in the triggering of massive star cluster formation.

\acknowledgments Computations described in this work were performed
using the publicly-available \texttt{Enzo} code
(http://enzo-project.org). 
This research also made use of the yt-project (http://yt-project.org/), 
a toolkit for analyzing and visualizing quantitative data 
\citep{Turk_ea_2011}. These are products of collaborative efforts
of many independent scientists from numerous institutions around the 
world. Their commitment to open science has helped make this work 
possible. 
The authors acknowledge University of Florida Research Computing
(http://researchcomputing.ufl.edu) for providing computational resources 
and support that have contributed to the research results reported in 
this publication. 
BW acknowledges the NASA Florida Space Grant 
Consortium Dissertation and Thesis Improvement Fellowship for its
support.

\software{Enzo (Bryan et al. 2014), Grackle (Bryan et al. 2014; Kim et al. 2014), PyPDR, and yt (Turk et al. 2011)}


\begin{thebibliography}{}

\bibitem[{{Anathpindika}(2009)}]{Anathpindika_2009}
{Anathpindika}, S. 2009, \aap, 504, 437

\bibitem[{{Balfour} {et~al.}(2015){Balfour}, {Whitworth}, {Hubber}, \&
  {Jaffa}}]{Balfour_ea_2015}
{Balfour}, S.~K., {Whitworth}, A.~P., {Hubber}, D.~A., \& {Jaffa}, S.~E. 2015,
  \mnras, 453, 2471

\bibitem[{{Beck}(2001)}]{Beck_2001}
{Beck}, R. 2001, \ssr, 99, 243

\bibitem[{{Bertoldi} \& {McKee}(1992)}]{Bertoldi_McKee_1992}
{Bertoldi}, F., \& {McKee}, C.~F. 1992, \apj, 395, 140

\bibitem[{{Bryan} {et~al.}(2014){Bryan}, {Norman}, {O'Shea}, {Abel}, {Wise},
  {Turk}, {Reynolds}, {Collins}, {Wang}, {Skillman}, {Smith}, {Harkness},
  {Bordner}, {Kim}, {Kuhlen}, {Xu}, {Goldbaum}, {Hummels}, {Kritsuk}, {Tasker},
  {Skory}, {Simpson}, {Hahn}, {Oishi}, {So}, {Zhao}, {Cen}, {Li}, \& {The Enzo
  Collaboration}}]{Bryan_ea_2014}
{Bryan}, G.~L., {Norman}, M.~L., {O'Shea}, B.~W., {et~al.} 2014, \apjs, 211, 19

\bibitem[{{Butler} {et~al.}(2014){Butler}, {Tan}, \&
  {Kainulainen}}]{Butler_ea_2014}
{Butler}, M.~J., {Tan}, J.~C., \& {Kainulainen}, J. 2014, \apjl, 782, L30

\bibitem[{{Cabral} \& {Leedom}(1993)}]{Cabral_Leedom_1993}
{Cabral}, B., \& {Leedom}, L.~C. 1993, in Special Interest Group on GRAPHics and 
Interactive Techniques Proceedings, 263

\bibitem[{{Chen} {et~al.}(2016){Chen}, {King}, \& {Li}}]{Chen_ea_2016arXiv}
{Chen}, C.-Y., {King}, P.~K., \& {Li}, Z.-Y. 2016, ArXiv e-prints,
  arXiv:1605.00648

\bibitem[{{Crutcher}(2012)}]{Crutcher_2012}
{Crutcher}, R.~M. 2012, \araa, 50, 29

\bibitem[{{Crutcher} {et~al.}(2010){Crutcher}, {Wandelt}, {Heiles},
  {Falgarone}, \& {Troland}}]{Crutcher_ea_2010}
{Crutcher}, R.~M., {Wandelt}, B., {Heiles}, C., {Falgarone}, E., \& {Troland},
  T.~H. 2010, \apj, 725, 466

\bibitem[{{Da Rio} {et~al.}(2014){Da Rio}, {Tan}, \& {Jaehnig}}]{DaRio_ea_2014}
{Da Rio}, N., {Tan}, J.~C., \& {Jaehnig}, K. 2014, \apj, 795, 55

\bibitem[{{Dedner} {et~al.}(2002){Dedner}, {Kemm}, {Kr{\"o}ner}, {Munz},
  {Schnitzer}, \& {Wesenberg}}]{Dedner_ea_2002}
{Dedner}, A., {Kemm}, F., {Kr{\"o}ner}, D., {et~al.} 2002, Journal of
  Computational Physics, 175, 645

\bibitem[{{Dickel} {et~al.}(1978){Dickel}, {Dickel}, \&
  {Wilson}}]{Dickel_ea_1978}
{Dickel}, J.~R., {Dickel}, H.~R., \& {Wilson}, W.~J. 1978, \apj, 223, 840

\bibitem[{{Dobashi} {et~al.}(2014){Dobashi}, {Matsumoto}, {Shimoikura},
  {Saito}, {Akisato}, {Ohashi}, \& {Nakagomi}}]{Dobashi_ea_2014}
{Dobashi}, K., {Matsumoto}, T., {Shimoikura}, T., {et~al.} 2014, \apj, 797, 58

\bibitem[{{Dobbs}(2008)}]{Dobbs_2008}
{Dobbs}, C.~L. 2008, \mnras, 391, 844

\bibitem[{{Dobbs} {et~al.}(2015){Dobbs}, {Pringle}, \&
  {Duarte-Cabral}}]{Dobbs_ea_2015}
{Dobbs}, C.~L., {Pringle}, J.~E., \& {Duarte-Cabral}, A. 2015, \mnras, 446,
  3608

\bibitem[{{Federrath} \& {Klessen}(2012)}]{Federrath_Klessen_2012}
{Federrath}, C., \& {Klessen}, R.~S. 2012, \apj, 761, 156

\bibitem[{{Federrath} \& {Klessen}(2013)}]{Federrath_Klessen_2013}
---. 2013, \apj, 763, 51

\bibitem[{{Federrath} {et~al.}(2011){Federrath}, {Sur}, {Schleicher},
  {Banerjee}, \& {Klessen}}]{Federrath_ea_2011}
{Federrath}, C., {Sur}, S., {Schleicher}, D.~R.~G., {Banerjee}, R., \&
  {Klessen}, R.~S. 2011, \apj, 731, 62

\bibitem[{{Fiege} \& {Pudritz}(2000)}]{Fiege_Pudritz_2000}
{Fiege}, J.~D., \& {Pudritz}, R.~E. 2000, \apj, 544, 830

\bibitem[{{Fukui} {et~al.}(2015){Fukui}, {Harada}, {Tokuda}, {Morioka},
  {Onishi}, {Torii}, {Ohama}, {Hattori}, {Nayak}, {Meixner}, {Sewi{\l}o},
  {Indebetouw}, {Kawamura}, {Saigo}, {Yamamoto}, {Tachihara}, {Minamidani},
  {Inoue}, {Madden}, {Galametz}, {Lebouteiller}, {Mizuno}, \&
  {Chen}}]{Fukui_ea_2015}
{Fukui}, Y., {Harada}, R., {Tokuda}, K., {et~al.} 2015, \apjl, 807, L4

\bibitem[{{Furukawa} {et~al.}(2009){Furukawa}, {Dawson}, {Ohama}, {Kawamura},
  {Mizuno}, {Onishi}, \& {Fukui}}]{Furukawa_ea_2009}
{Furukawa}, N., {Dawson}, J.~R., {Ohama}, A., {et~al.} 2009, \apjl, 696, L115

\bibitem[{{Gammie} {et~al.}(1991){Gammie}, {Ostriker}, \&
  {Jog}}]{Gammie_ea_1991}
{Gammie}, C.~F., {Ostriker}, J.~P., \& {Jog}, C.~J. 1991, \apj, 378, 565

\bibitem[{{Gutermuth} {et~al.}(2009){Gutermuth}, {Megeath}, {Myers}, {Allen},
  {Pipher}, \& {Fazio}}]{Gutermuth_ea_2009}
{Gutermuth}, R.~A., {Megeath}, S.~T., {Myers}, P.~C., {et~al.} 2009, \apjs,
  184, 18

\bibitem[{{Habe} \& {Ohta}(1992)}]{Habe_Ohta_1992}
{Habe}, A., \& {Ohta}, K. 1992, \pasj, 44, 203

\bibitem[{{Haworth} {et~al.}(2015{\natexlab{a}}){Haworth}, {Shima}, {Tasker},
  {Fukui}, {Torii}, {Dale}, {Takahira}, \& {Habe}}]{Haworth_ea_2015a}
{Haworth}, T.~J., {Shima}, K., {Tasker}, E.~J., {et~al.} 2015{\natexlab{a}},
  \mnras, 454, 1634

\bibitem[{{Haworth} {et~al.}(2015{\natexlab{b}}){Haworth}, {Tasker}, {Fukui},
  {Torii}, {Dale}, {Shima}, {Takahira}, {Habe}, \&
  {Hasegawa}}]{Haworth_ea_2015b}
{Haworth}, T.~J., {Tasker}, E.~J., {Fukui}, Y., {et~al.} 2015{\natexlab{b}},
  \mnras, 450, 10

\bibitem[{{Hernandez} \& {Tan}(2015)}]{Hernandez_Tan_2015}
{Hernandez}, A.~K., \& {Tan}, J.~C. 2015, \apj, 809, 154

\bibitem[{{Heyer} \& {Brunt}(2004)}]{Heyer_Brunt_2004}
{Heyer}, M.~H., \& {Brunt}, C.~M. 2004, \apjl, 615, L45

\bibitem[{{Jackson} {et~al.}(2010){Jackson}, {Finn}, {Chambers}, {Rathborne},
  \& {Simon}}]{Jackson_ea_2010}
{Jackson}, J.~M., {Finn}, S.~C., {Chambers}, E.~T., {Rathborne}, J.~M., \&
  {Simon}, R. 2010, \apjl, 719, L185

\bibitem[{{Jackson} {et~al.}(2006){Jackson}, {Rathborne}, {Shah}, {Simon},
  {Bania}, {Clemens}, {Chambers}, {Johnson}, {Dormody}, {Lavoie}, \&
  {Heyer}}]{Jackson_ea_2006}
{Jackson}, J.~M., {Rathborne}, J.~M., {Shah}, R.~Y., {et~al.} 2006, \apjs, 163,
  145

\bibitem[{{Kainulainen} {et~al.}(2009){Kainulainen}, {Beuther}, {Henning}, \&
  {Plume}}]{Kainulainen_ea_2009}
{Kainulainen}, J., {Beuther}, H., {Henning}, T., \& {Plume}, R. 2009, \aap,
  508, L35

\bibitem[{{Kainulainen} \& {Tan}(2013)}]{Kainulainen_Tan_2013}
{Kainulainen}, J., \& {Tan}, J.~C. 2013, \aap, 549, A53

\bibitem[{{Kataoka} {et~al.}(2012){Kataoka}, {Machida}, \&
  {Tomisaka}}]{Kataoka_ea_2012}
{Kataoka}, A., {Machida}, M.~N., \& {Tomisaka}, K. 2012, \apj, 761, 40

\bibitem[{{Kauffmann} {et~al.}(2013){Kauffmann}, {Pillai}, \&
  {Goldsmith}}]{Kauffmann_ea_2013}
{Kauffmann}, J., {Pillai}, T., \& {Goldsmith}, P.~F. 2013, \apj, 779, 185

\bibitem[{{Kennicutt}(1998)}]{Kennicutt_1998}
{Kennicutt}, Jr., R.~C. 1998, \apj, 498, 541

\bibitem[{{Kim} {et~al.}(2014){Kim}, {Abel}, {Agertz}, {Bryan}, {Ceverino},
  {Christensen}, {Conroy}, {Dekel}, {Gnedin}, {Goldbaum}, {Guedes}, {Hahn},
  {Hobbs}, {Hopkins}, {Hummels}, {Iannuzzi}, {Keres}, {Klypin}, {Kravtsov},
  {Krumholz}, {Kuhlen}, {Leitner}, {Madau}, {Mayer}, {Moody}, {Nagamine},
  {Norman}, {Onorbe}, {O'Shea}, {Pillepich}, {Primack}, {Quinn}, {Read},
  {Robertson}, {Rocha}, {Rudd}, {Shen}, {Smith}, {Szalay}, {Teyssier},
  {Thompson}, {Todoroki}, {Turk}, {Wadsley}, {Wise}, {Zolotov}, \& {AGORA
  Collaboration29}}]{Kim_ea_2014}
{Kim}, J.-h., {Abel}, T., {Agertz}, O., {et~al.} 2014, \apjs, 210, 14

\bibitem[{{Klein} \& {Woods}(1998)}]{Klein_Woods_1998}
{Klein}, R.~I., \& {Woods}, D.~T. 1998, \apj, 497, 777

\bibitem[{{Krumholz} \& {McKee}(2005)}]{Krumholz_McKee_2005}
{Krumholz}, M.~R., \& {McKee}, C.~F. 2005, \apj, 630, 250

\bibitem[{{Krumholz} \& {Tan}(2007)}]{Krumholz_Tan_2007}
{Krumholz}, M.~R., \& {Tan}, J.~C. 2007, \apj, 654, 304

\bibitem[{{Lada} \& {Lada}(2003)}]{Lada_Lada_2003}
{Lada}, C.~J., \& {Lada}, E.~A. 2003, \araa, 41, 57

\bibitem[{{Lee} {et~al.}(2015){Lee}, {Chang}, \& {Murray}}]{Lee_ea_2015}
{Lee}, E.~J., {Chang}, P., \& {Murray}, N. 2015, \apj, 800, 49

\bibitem[{{Lee} \& {Draine}(1985)}]{Lee_Draine_1985}
{Lee}, H.~M., \& {Draine}, B.~T. 1985, \apj, 290, 211

\bibitem[{{Leroy} {et~al.}(2008){Leroy}, {Walter}, {Brinks}, {Bigiel}, {de
  Blok}, {Madore}, \& {Thornley}}]{Leroy_ea_2008}
{Leroy}, A.~K., {Walter}, F., {Brinks}, E., {et~al.} 2008, \aj, 136, 2782

\bibitem[{{Li} {et~al.}(2014){Li}, {Goodman}, {Sridharan}, {Houde}, {Li},
  {Novak}, \& {Tang}}]{Li_ea_2014}
{Li}, H.-B., {Goodman}, A., {Sridharan}, T.~K., {et~al.} 2014, Protostars and
  Planets VI, 101

\bibitem[{{Lim} {et~al.}(2016){Lim}, {Tan}, {Kainulainen}, {Ma}, \&
  {Butler}}]{Lim_ea_2016arXiv}
{Lim}, W., {Tan}, J.~C., {Kainulainen}, J., {Ma}, B., \& {Butler}, M.~J. 2016,
  ArXiv e-prints, arXiv:1605.09320

\bibitem[{{Liszt} {et~al.}(1984){Liszt}, {Burton}, \& {Xiang}}]{Liszt_ea_1984}
{Liszt}, H.~S., {Burton}, W.~B., \& {Xiang}, D.-L. 1984, \aap, 140, 303

\bibitem[{{Loren}(1976)}]{Loren_1976}
{Loren}, R.~B. 1976, \apj, 209, 466

\bibitem[{{Loren}(1977)}]{Loren_1977}
---. 1977, \apj, 218, 716

\bibitem[{{McKee} \& {Ostriker}(2007)}]{McKee_Ostriker_2007}
{McKee}, C.~F., \& {Ostriker}, E.~C. 2007, \araa, 45, 565

\bibitem[{{Mouschovias}(2001)}]{Mouschovias_2001}
{Mouschovias}, T. 2001, in Astronomical Society of the Pacific Conference
  Series, Vol. 248, Magnetic Fields Across the Hertzsprung-Russell Diagram, ed.
  G.~{Mathys}, S.~K. {Solanki}, \& D.~T. {Wickramasinghe}, 515

\bibitem[{{Murray} {et~al.}(2010){Murray}, {Quataert}, \&
  {Thompson}}]{Murray_ea_2010}
{Murray}, N., {Quataert}, E., \& {Thompson}, T.~A. 2010, \apj, 709, 191

\bibitem[{{Odenwald} {et~al.}(1992){Odenwald}, {Fischer}, {Lockman}, \&
  {Stemwedel}}]{Odenwald_ea_1992}
{Odenwald}, S., {Fischer}, J., {Lockman}, F.~J., \& {Stemwedel}, S. 1992, \apj,
  397, 174

\bibitem[{{Ohama} {et~al.}(2010){Ohama}, {Dawson}, {Furukawa}, {Kawamura},
  {Moribe}, {Yamamoto}, {Okuda}, {Mizuno}, {Onishi}, {Maezawa}, {Minamidani},
  {Mizuno}, \& {Fukui}}]{Ohama_ea_2010}
{Ohama}, A., {Dawson}, J.~R., {Furukawa}, N., {et~al.} 2010, \apj, 709, 975

\bibitem[{{Ossenkopf} \& {Mac Low}(2002)}]{Ossenkopf_MacLow_2002}
{Ossenkopf}, V., \& {Mac Low}, M.-M. 2002, \aap, 390, 307

\bibitem[{{Padoan} {et~al.}(2014){Padoan}, {Federrath}, {Chabrier}, {Evans},
  {Johnstone}, {J{\o}rgensen}, {McKee}, \& {Nordlund}}]{Padoan_ea_2014}
{Padoan}, P., {Federrath}, C., {Chabrier}, G., {et~al.} 2014, Protostars and
  Planets VI, 77

\bibitem[{{Padoan} \& {Nordlund}(2011)}]{Padoan_Nordlund_2011}
{Padoan}, P., \& {Nordlund}, {\AA}. 2011, \apj, 730, 40

\bibitem[{{Planck Collaboration} {et~al.}(2015){Planck Collaboration}, {Ade},
  {Aghanim}, {Alina}, {Alves}, {Aniano}, {Armitage-Caplan}, {Arnaud},
  {Arzoumanian}, {Ashdown}, {Atrio-Barandela}, {Aumont}, {Baccigalupi},
  {Banday}, {Barreiro}, {Battaner}, {Benabed}, {Benoit-L{\'e}vy}, {Bernard},
  {Bersanelli}, {Bielewicz}, {Bond}, {Borrill}, {Bouchet}, {Boulanger},
  {Bracco}, {Burigana}, {Cardoso}, {Catalano}, {Chamballu}, {Chiang},
  {Christensen}, {Colombi}, {Colombo}, {Combet}, {Couchot}, {Coulais}, {Crill},
  {Curto}, {Cuttaia}, {Danese}, {Davies}, {Davis}, {de Bernardis}, {de Rosa},
  {de Zotti}, {Delabrouille}, {Dickinson}, {Diego}, {Donzelli}, {Dor{\'e}},
  {Douspis}, {Dupac}, {Efstathiou}, {En{\ss}lin}, {Eriksen}, {Falgarone},
  {Fanciullo}, {Ferri{\`e}re}, {Finelli}, {Forni}, {Frailis}, {Fraisse},
  {Franceschi}, {Galeotta}, {Ganga}, {Ghosh}, {Giard}, {Giraud-H{\'e}raud},
  {Gonz{\'a}lez-Nuevo}, {G{\'o}rski}, {Gregorio}, {Gruppuso}, {Guillet},
  {Hansen}, {Harrison}, {Helou}, {Hern{\'a}ndez-Monteagudo}, {Hildebrandt},
  {Hivon}, {Hobson}, {Holmes}, {Hornstrup}, {Huffenberger}, {Jaffe}, {Jaffe},
  {Jones}, {Juvela}, {Keih{\"a}nen}, {Keskitalo}, {Kisner}, {Kneissl},
  {Knoche}, {Kunz}, {Kurki-Suonio}, {Lagache}, {Lamarre}, {Lasenby},
  {Lawrence}, {Leonardi}, {Levrier}, {Liguori}, {Lilje}, {Linden-V{\o}rnle},
  {L{\'o}pez-Caniego}, {Lubin}, {Mac{\'{\i}}as-P{\'e}rez}, {Maino},
  {Mandolesi}, {Maris}, {Marshall}, {Martin}, {Mart{\'{\i}}nez-Gonz{\'a}lez},
  {Masi}, {Matarrese}, {Mazzotta}, {Melchiorri}, {Mendes}, {Mennella},
  {Migliaccio}, {Miville-Desch{\^e}nes}, {Moneti}, {Montier}, {Morgante},
  {Mortlock}, {Munshi}, {Murphy}, {Naselsky}, {Nati}, {Natoli}, {Netterfield},
  {Noviello}, {Novikov}, {Novikov}, {Oxborrow}, {Pagano}, {Pajot}, {Paoletti},
  {Pasian}, {Pelkonen}, {Perdereau}, {Perotto}, {Perrotta}, {Piacentini},
  {Piat}, {Pietrobon}, {Plaszczynski}, {Pointecouteau}, {Polenta}, {Popa},
  {Pratt}, {Prunet}, {Puget}, {Rachen}, {Reinecke}, {Remazeilles}, {Renault},
  {Ricciardi}, {Riller}, {Ristorcelli}, {Rocha}, {Rosset}, {Roudier},
  {Rusholme}, {Sandri}, {Scott}, {Soler}, {Spencer}, {Stolyarov}, {Stompor},
  {Sudiwala}, {Sutton}, {Suur-Uski}, {Sygnet}, {Tauber}, {Terenzi},
  {Toffolatti}, {Tomasi}, {Tristram}, {Tucci}, {Umana}, {Valenziano},
  {Valiviita}, {Van Tent}, {Vielva}, {Villa}, {Wade}, {Wandelt}, \&
  {Zonca}}]{PlanckXX_2015}
{Planck Collaboration}, {Ade}, P.~A.~R., {Aghanim}, N., {et~al.} 2015, \aap,
  576, A105

\bibitem[{{Planck Collaboration} {et~al.}(2016){Planck Collaboration}, {Ade},
  {Aghanim}, {Alves}, {Arnaud}, {Arzoumanian}, {Ashdown}, {Aumont},
  {Baccigalupi}, {Banday}, {Barreiro}, {Bartolo}, {Battaner}, {Benabed},
  {Beno{\^i}t}, {Benoit-L{\'e}vy}, {Bernard}, {Bersanelli}, {Bielewicz},
  {Bock}, {Bonavera}, {Bond}, {Borrill}, {Bouchet}, {Boulanger}, {Bracco},
  {Burigana}, {Calabrese}, {Cardoso}, {Catalano}, {Chiang}, {Christensen},
  {Colombo}, {Combet}, {Couchot}, {Crill}, {Curto}, {Cuttaia}, {Danese},
  {Davies}, {Davis}, {de Bernardis}, {de Rosa}, {de Zotti}, {Delabrouille},
  {Dickinson}, {Diego}, {Dole}, {Donzelli}, {Dor{\'e}}, {Douspis}, {Ducout},
  {Dupac}, {Efstathiou}, {Elsner}, {En{\ss}lin}, {Eriksen}, {Falceta-Gon{\c
  c}alves}, {Falgarone}, {Ferri{\`e}re}, {Finelli}, {Forni}, {Frailis},
  {Fraisse}, {Franceschi}, {Frejsel}, {Galeotta}, {Galli}, {Ganga}, {Ghosh},
  {Giard}, {Gjerl{\o}w}, {Gonz{\'a}lez-Nuevo}, {G{\'o}rski}, {Gregorio},
  {Gruppuso}, {Gudmundsson}, {Guillet}, {Harrison}, {Helou}, {Hennebelle},
  {Henrot-Versill{\'e}}, {Hern{\'a}ndez-Monteagudo}, {Herranz}, {Hildebrandt},
  {Hivon}, {Holmes}, {Hornstrup}, {Huffenberger}, {Hurier}, {Jaffe}, {Jaffe},
  {Jones}, {Juvela}, {Keih{\"a}nen}, {Keskitalo}, {Kisner}, {Knoche}, {Kunz},
  {Kurki-Suonio}, {Lagache}, {Lamarre}, {Lasenby}, {Lattanzi}, {Lawrence},
  {Leonardi}, {Levrier}, {Liguori}, {Lilje}, {Linden-V{\o}rnle},
  {L{\'o}pez-Caniego}, {Lubin}, {Mac{\'{\i}}as-P{\'e}rez}, {Maino},
  {Mandolesi}, {Mangilli}, {Maris}, {Martin}, {Mart{\'{\i}}nez-Gonz{\'a}lez},
  {Masi}, {Matarrese}, {Melchiorri}, {Mendes}, {Mennella}, {Migliaccio},
  {Miville-Desch{\^e}nes}, {Moneti}, {Montier}, {Morgante}, {Mortlock},
  {Munshi}, {Murphy}, {Naselsky}, {Nati}, {Netterfield}, {Noviello}, {Novikov},
  {Novikov}, {Oppermann}, {Oxborrow}, {Pagano}, {Pajot}, {Paladini},
  {Paoletti}, {Pasian}, {Perotto}, {Pettorino}, {Piacentini}, {Piat},
  {Pierpaoli}, {Pietrobon}, {Plaszczynski}, {Pointecouteau}, {Polenta},
  {Ponthieu}, {Pratt}, {Prunet}, {Puget}, {Rachen}, {Reinecke}, {Remazeilles},
  {Renault}, {Renzi}, {Ristorcelli}, {Rocha}, {Rossetti}, {Roudier},
  {Rubi{\~n}o-Mart{\'{\i}}n}, {Rusholme}, {Sandri}, {Santos}, {Savelainen},
  {Savini}, {Scott}, {Soler}, {Stolyarov}, {Sudiwala}, {Sutton}, {Suur-Uski},
  {Sygnet}, {Tauber}, {Terenzi}, {Toffolatti}, {Tomasi}, {Tristram}, {Tucci},
  {Umana}, {Valenziano}, {Valiviita}, {Van Tent}, {Vielva}, {Villa}, {Wade},
  {Wandelt}, {Wehus}, {Ysard}, {Yvon}, \& {Zonca}}]{PlanckXXXV_2016}
---. 2016, \aap, 586, A138

\bibitem[{{Pon} {et~al.}(2015){Pon}, {Caselli}, {Johnstone}, {Kaufman},
  {Butler}, {Fontani}, {Jim{\'e}nez-Serra}, \& {Tan}}]{Pon_ea_2015}
{Pon}, A., {Caselli}, P., {Johnstone}, D., {et~al.} 2015, \aap, 577, A75

\bibitem[{{Ragan} {et~al.}(2014){Ragan}, {Henning}, {Tackenberg}, {Beuther},
  {Johnston}, {Kainulainen}, \& {Linz}}]{Ragan_ea_2014}
{Ragan}, S.~E., {Henning}, T., {Tackenberg}, J., {et~al.} 2014, \aap, 568, A73

\bibitem[{{Roman-Duval} {et~al.}(2010){Roman-Duval}, {Jackson}, {Heyer},
  {Rathborne}, \& {Simon}}]{Roman-Duval_ea_2010}
{Roman-Duval}, J., {Jackson}, J.~M., {Heyer}, M., {Rathborne}, J., \& {Simon},
  R. 2010, \apj, 723, 492

\bibitem[{{Scoville} {et~al.}(1986){Scoville}, {Sanders}, \&
  {Clemens}}]{Scoville_Sanders_Clemens_1986}
{Scoville}, N.~Z., {Sanders}, D.~B., \& {Clemens}, D.~P. 1986, \apjl, 310, L77

\bibitem[{{Soler} {et~al.}(2013){Soler}, {Hennebelle}, {Martin},
  {Miville-Desch{\^e}nes}, {Netterfield}, \& {Fissel}}]{Soler_ea_2013}
{Soler}, J.~D., {Hennebelle}, P., {Martin}, P.~G., {et~al.} 2013, \apj, 774,
  128

\bibitem[{{Solomon} {et~al.}(1987){Solomon}, {Rivolo}, {Barrett}, \&
  {Yahil}}]{Solomon_ea_1987}
{Solomon}, P.~M., {Rivolo}, A.~R., {Barrett}, J., \& {Yahil}, A. 1987, \apj,
  319, 730

\bibitem[{{Stark}(1984)}]{Stark_1984}
{Stark}, A.~A. 1984, \apj, 281, 624

\bibitem[{{Suwannajak} {et~al.}(2014){Suwannajak}, {Tan}, \&
  {Leroy}}]{Suwannajak_ea_2014}
{Suwannajak}, C., {Tan}, J.~C., \& {Leroy}, A.~K. 2014, \apj, 787, 68

\bibitem[{{Takahira} {et~al.}(2014){Takahira}, {Tasker}, \&
  {Habe}}]{Takahira_ea_2014}
{Takahira}, K., {Tasker}, E.~J., \& {Habe}, A. 2014, \apj, 792, 63

\bibitem[{{Tan}(2000)}]{Tan_2000}
{Tan}, J.~C. 2000, \apj, 536, 173

\bibitem[{{Tan}(2010)}]{Tan_2010}
---. 2010, \apjl, 710, L88

\bibitem[{{Tan} {et~al.}(2013){Tan}, {Shaske}, \& {Van
  Loo}}]{Tan_Shaske_Van_Loo_2013}
{Tan}, J.~C., {Shaske}, S.~N., \& {Van Loo}, S. 2013, in IAU Symposium, Vol.
  292, IAU Symposium, ed. T.~{Wong} \& J.~{Ott}, 19--28

\bibitem[{{Tasker} \& {Tan}(2009)}]{Tasker_Tan_2009}
{Tasker}, E.~J., \& {Tan}, J.~C. 2009, \apj, 700, 358

\bibitem[{{Torii} {et~al.}(2011){Torii}, {Enokiya}, {Sano}, {Yoshiike},
  {Hanaoka}, {Ohama}, {Furukawa}, {Dawson}, {Moribe}, {Oishi}, {Nakashima},
  {Okuda}, {Yamamoto}, {Kawamura}, {Mizuno}, {Maezawa}, {Onishi}, {Mizuno}, \&
  {Fukui}}]{Torii_ea_2011}
{Torii}, K., {Enokiya}, R., {Sano}, H., {et~al.} 2011, \apj, 738, 46

\bibitem[{{Truelove} {et~al.}(1997){Truelove}, {Klein}, {McKee}, {Holliman},
  {Howell}, \& {Greenough}}]{Truelove_ea_1997}
{Truelove}, J.~K., {Klein}, R.~I., {McKee}, C.~F., {et~al.} 1997, \apjl, 489,
  L179

\bibitem[{{Turk} {et~al.}(2011){Turk}, {Smith}, {Oishi}, {Skory}, {Skillman},
  {Abel}, \& {Norman}}]{Turk_ea_2011}
{Turk}, M.~J., {Smith}, B.~D., {Oishi}, J.~S., {et~al.} 2011, \apjs, 192, 9

\bibitem[{{Wang} \& {Abel}(2008)}]{Wang_Abel_2008}
{Wang}, P., \& {Abel}, T. 2008, \apj, 672, 752

\bibitem[{{Wu} {et~al.}(2015){Wu}, {Van Loo}, {Tan}, \&
  {Bruderer}}]{Wu_ea_2015}
{Wu}, B., {Van Loo}, S., {Tan}, J.~C., \& {Bruderer}, S. 2015, \apj, 811, 56

\bibitem[{{Zuckerman} \& {Evans}(1974)}]{Zuckerman_Evans_1974}
{Zuckerman}, B., \& {Evans}, II, N.~J. 1974, \apjl, 192, L149

\end{thebibliography}

\clearpage

\end{document}